\documentclass[a4paper,11pt]{article}
\usepackage{jcappub}
\usepackage{lineno}
\usepackage[T1]{fontenc}
\usepackage[utf8]{inputenc}
\usepackage{amsmath}
\usepackage{amssymb}
\usepackage{graphicx}
\usepackage{xcolor}
\usepackage{xspace}
\usepackage{booktabs}
\usepackage{multirow}
\usepackage{siunitx}
\usepackage{mleftright}
\usepackage{enumitem}
\usepackage{wick}
\usepackage[many]{tcolorbox}
\usepackage{acro}
\usepackage{cleveref}

\DeclareAcronym{2LPT}{short=2LPT, long=second-order Lagrangian perturbation theory}
\DeclareAcronym{2PCF}{short=2PCF, long=two-point correlation function}
\DeclareAcronym{3PCF}{short=3PCF, long=three-point correlation function}
\DeclareAcronym{4PCF}{short=4PCF, long=four-point correlation function}
\DeclareAcronym{8PCF}{short=8PCF, long=eight-point correlation function}
\DeclareAcronym{BAO}{short=BAO, long=baryon acoustic sscillations}
\DeclareAcronym{BOSS}{short=BOSS, long=Baryon Oscillation Spectroscopic Survey}
\DeclareAcronym{CL}{short=CL, long=confidence levels}
\DeclareAcronym{CMASS}{short=CMASS, long=``Constant Stellar Mass''}
\DeclareAcronym{CMB}{short=CMB, long=cosmic microwave background}
\DeclareAcronym{DEC}{short=DEC, long=declination}
\DeclareAcronym{DESI}{short=DESI, long=Dark Energy Spectroscopic Instrument}
\DeclareAcronym{FKP}{short=FKP, long=Feldman--Kaiser--Peacock}
\DeclareAcronym{LSS}{short=LSS, long=large-scale structure}
\DeclareAcronym{NGC}{short=NGC, long=North Galactic Cap}
\DeclareAcronym{PDF}{short=PDF, long=probability density function}
\DeclareAcronym{RA}{short=RA, long=right ascension}
\DeclareAcronym{SGC}{short=SGC, long=South Galactic Cap}

\makeatletter
\input{aas_macros.sty}
\let\jnl@style=\relax
\makeatother

\title{No evidence for parity violation in BOSS}

\author[a,b,c]{Alex Krolewski,}
\author[c,d]{Simon May,}
\author[c]{Kendrick Smith,}
\author[a,c]{\\Hans Hopkins}
\affiliation[a]{Department of Physics and Astronomy, University of Waterloo\\
    200 University Avenue West, Waterloo, ON N2L 3G1, Canada}
\affiliation[b]{Waterloo Centre for Astrophysics, University of Waterloo\\
    200 University Avenue West, Waterloo, ON N2L 3G1, Canada}
\affiliation[c]{Perimeter Institute for Theoretical Physics\\
    31 Caroline Street North, Waterloo, ON N2L 2Y5, Canada} 
\affiliation[d]{Department of Physics, North Carolina State University\\
    Raleigh, NC, 27695-8202, USA}

\emailAdd{akrolews@uwaterloo.ca}
\emailAdd{simon.may@pitp.ca}
\emailAdd{kmsmith@pitp.ca}
\emailAdd{hshopkins@uwaterloo.ca}

\DeclareSIUnit{\hHubble}{\ensuremath{\mathnormal h}}
\DeclareSIUnit{\pc}{pc}
\DeclareSIUnit{\kpc}{\kilo\pc}
\DeclareSIUnit{\Mpc}{\mega\pc}
\DeclareSIUnit{\Gpc}{\giga\pc}
\crefname{figure}{figure}{figures}

\newcommand{\bx}{{\mathbf{x}}}
\newcommand{\br}{{\mathbf{r}}}
\newcommand{\bk}{{\mathbf{k}}}
\newcommand{\hr}{{\hat{\mathbf{r}}}}
\newcommand{\hk}{{\hat{\mathbf{k}}}}
\newcommand{\hE}{\widehat{\mathcal{E}}}
\newcommand{\bE}{\bar{\mathcal{E}}}
\newcommand{\tW}{\widetilde{W}}
\newcommand{\Var}{\mathrm{Var}}
\newcommand{\Cov}{\mathrm{Cov}}
\newcommand{\Tr}{\mathrm{Tr}}
\newcommand{\N}{\mathcal{N}}
\newcommand{\bd}{\bar{d}}
\newcommand{\classpt}{\textsc{class-pt}}
\newcommand{\nn}{\nonumber}

\newcommand{\cmassnorth}{\ac{CMASS}-\ac{NGC}\xspace}
\newcommand{\cmasssouth}{CMASS\-LOWZTOT-\ac{SGC}\xspace}
\newcommand{\cmassnorthacs}{\acs*{CMASS}-\acs*{NGC}\xspace}
\newcommand{\cmasssouthacs}{CMASS\-LOWZTOT-\acs*{SGC}\xspace}

\newcommand{\SigNorthTot}{$7.3\sigma$\xspace}
\newcommand{\SigSouthTot}{$6.9\sigma$\xspace}
\newcommand{\SigNorthCross}{$1.8\sigma$\xspace}
\newcommand{\SigSouthCross}{$1.7\sigma$\xspace}
\newcommand{\SigNorthCrossTot}{$5.9\sigma$\xspace}
\newcommand{\SigSouthCrossTot}{$9.4\sigma$\xspace}
\newcommand{\SigNorthNull}{$5.6\sigma$\xspace}
\newcommand{\SigSouthNull}{$6.8\sigma$\xspace}

\abstract{%
    Recent studies have found evidence for parity violation in the BOSS spectroscopic galaxy survey, with statistical significance as high as $7\sigma$.
    These analyses assess the significance of the parity-odd four-point correlation function (4PCF) with a statistic called $\chi^2$. 
    This statistic is biased if the \emph{parity-even} eight-point correlation function (8PCF) of the data differs from the mock catalogs.
    We construct new statistics $\chi^2_\times$, $\chi^2_{\mathrm{null}}$ that separate the parity violation signal from the 8PCF bias term, allowing them to be jointly constrained.
    Applying these statistics to BOSS, we find that the parity violation signal ranges from $0$ to $2.5\sigma$ depending on analysis choices, whereas the 8PCF bias term is $\sim 6\sigma$.
    We conclude that there is no compelling evidence for parity violation in BOSS.
    Our new statistics can be used to search for parity violation in future surveys, such as DESI, without 8PCF biases.%
}

\begin{document}

\maketitle
\flushbottom

\section{Introduction}
\label{sec:intro}

\subsection{Background: Parity violation in BOSS?}

\enlargethispage{\baselineskip}

Recently, two groups \cite{hou_parity,oliver_parity} reported evidence for parity violation in the \ac{BOSS}, following the proposal of \cite{Cahn:2021ltp} and using methods developed in \cite{hou_covariance}.
In \cite{hou_parity}, statistical significance as high as $7.1\sigma$ was reported, and in \cite{oliver_parity} statistical significance as high as $2.9\sigma$ was reported.\footnote{%
    Refs.\ \cite{hou_parity,oliver_parity} report different statistical significances mainly because a key analysis parameter $N_\beta$ (number of radial bins in the $\chi^2$ estimator) is chosen differently.
    In \cite{hou_parity,oliver_parity}, $N_\beta$ is chosen to be 18, 10 respectively.
    (Note that \cite{hou_parity} also presents results with $N_\beta = 10$, and these agree qualitatively with \cite{oliver_parity}.)
    The parameter $N_\beta$ is defined precisely in \cref{ssec:hE}.%
}

Cosmological parity violation, if confirmed, would have profound implications for fundamental physics, and so the results of \cite{hou_parity,oliver_parity} have attracted a great deal of interest.\footnote{%
    E.\,g.\ three recent workshops were devoted to parity violation: 
    \url{https://events.asiaa.sinica.edu.tw/workshop/20231204/},
    \url{https://inspirehep.net/seminars/2170834},
    \url{https://parity.cosmodiscussion.com/}.%
}
A variety of models were proposed which generate parity violation on cosmological scales (e.\,g.\ \cite{Cabass:2022rhr,Creque-Sarbinowski:2023wmb,Jazayeri:2023kji,Fujita:2023inz,Vanzan:2023cze,Callister:2023tws} and references therein).

On the observational side, the situation has been puzzling.
Follow-up searches in \ac{BOSS} for specific parity-violating models of inflation produced null results \cite{Cabass:2022oap}.
A re-analysis of \ac{BOSS} using a different set of mock catalogs shifted the detection significance of parity violation by around $2\sigma$ \cite{Philcox:2024mmz}, suggesting that the analysis may not be very robust to the choice of mocks.
In the \ac{CMB}, some analyses have found tentative evidence for parity violation (e.\,g.\ \cite{Minami:2020odp,Diego-Palazuelos:2022dsq,Eskilt:2022cff} and references therein) whereas others have found null results \cite{Philcox:2023ffy}.

These follow-up studies may suggest that the original detection of parity violation is spurious.
On the other hand, no follow-up study has directly refuted the $7\sigma$ detection from \cite{hou_parity}, so the current observational situation is unclear.
(It is also unclear how to interpret results from future datasets, e.\,g.\ the \ac{DESI} survey \cite{DESI:2024mwx}, until the $7\sigma$ excess in \ac{BOSS} is diagnosed.)

One may ask, is there a statistical procedure which will unambiguously determine whether the $7\sigma$ detection of parity violation in \ac{BOSS} is spurious or not?
In this paper, we develop such a procedure.
We construct improved statistics (denoted $\chi^2_\times$, $\chi^2_{\mathrm{null}}$) which separate the $7\sigma$ detection from \cite{hou_parity} into two contributions: a parity violation contribution, and a ``data--mock mismatch'' contribution which is nonzero if the \emph{parity-even} \ac{8PCF} of the data differs from the mock catalogs.

We apply these statistics to \ac{BOSS} data, and find that the parity violation signal is not statistically significant (significance varies between $0$ to $2.5\sigma$ depending on analysis choices, see \cref{fig:2d,fig:ngc_sgc}), whereas the data--mock mismatch signal is $\sim 6\sigma$.
Our interpretation is that there is not compelling evidence for parity violation in \ac{BOSS}.

Our new statistics $\chi^2_\times$, $\chi^2_{\mathrm{null}}$ are conceptually simple, but the details are complicated, since algebraically messy objects arise, e.\,g.\ the parity-odd \ac{4PCF}, \cref{eq:Edef}, and its analytic covariance (\cref{app:analytic_covariance}).
In the rest of this extended introduction (\crefrange{ssec:intro_mm}{ssec:new_statistics}), we present the main results of the paper in streamlined form.
Details and derivations will be given in later sections (\crefrange{sec:data}{sec:discussion}).

\subsection{The \texorpdfstring{$\chi^2$}{χ²} statistic and the difficulty of modeling the 4PCF covariance}
\label{ssec:intro_mm}

\begin{figure}
    \centerline{\includegraphics[width=8cm]{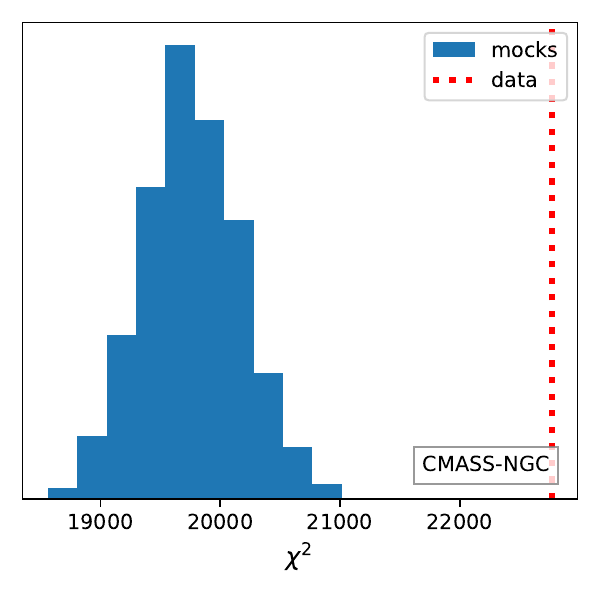} \includegraphics[width=8cm]{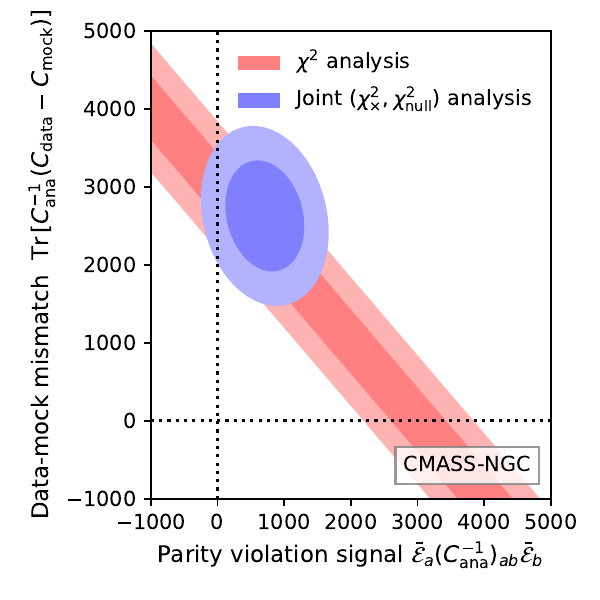}}
    \caption{%
        Analysis of parity violation in the \ac{BOSS} \cmassnorthacs dataset. 
        (Results for \cmasssouthacs are qualitatively similar and shown in \cref{sec:new_statistics}.)
        \textbf{\textit{Left panel.}} When the $\chi^2$ statistic defined in \cref{eq:chi2_def_intro} is evaluated on \ac{BOSS} data (dashed vertical line), the result is a $\sim7\sigma$ outlier relative to mock catalogs (solid histogram). 
        This reproduces the main result from \cite{hou_parity}.
        \textbf{\textit{Right panel.}} We interpret this $7\sigma$ signal as a sum of parity violation and ``data--mock mismatch'' contributions (\cref{eq:intro_2terms}).
        If only $\chi^2$ is used, these contributions are perfectly degenerate (red regions). The new statistics $\chi^2_\times, \chi^2_{\mathrm{null}}$ defined in \cref{eq:chi2_times_def_intro,eq:chi2_null_def_intro} break this degeneracy (blue regions). 
        We see that the parity violation signal drops to $<2\sigma$, while the data-mismatch signal remains at high significance.
        Throughout this panel, statistical errors are assumed Gaussian, with covariance estimated from mock catalogs.
        Light/dark regions are \SI{68}{\percent} and \SI{95}{\percent} \acf{CL}.%
    }
    \label{fig:intro}
\end{figure}
\acuse{CL}

The analysis in \cite{hou_parity,oliver_parity} is based on a particular statistic applied to galaxy catalogs, denoted $\chi^2$ and defined below, which is sensitive to parity violation.
Statistical significance is assigned by computing $\chi^2$ on the \ac{BOSS} data, and comparing it to $\chi^2$ values from an ensemble of \ac{BOSS} mock galaxy catalogs (as first proposed for the two-point correlation function in \cite{Xu13}, applied to the three-point function in \cite{Slepian:2015qza}, and the parity-odd four-point function in \cite{hou_parity}).
Following \cite{hou_parity,oliver_parity}, we have used the MultiDark-PATCHY \ac{BOSS} mock catalogs \cite{Kitaura:2015uqa,Rodriguez-Torres:2015vqa} (or ``PATCHY mocks'' for short) throughout this paper.
We have reproduced the result from \cite{hou_parity} in the left panel of \cref{fig:intro}.
We find that $\chi^2_{\mathrm{data}}$ is indeed a $7\sigma$ outlier, relative to a histogram of $\chi^2_{\mathrm{mock}}$ values.
Here are three possible interpretations of this $7\sigma$ result:
\begin{enumerate}
    \item \textit{Parity-violating new physics:} The spatial clustering of galaxies in the universe is not parity-invariant.
    
    \item \textit{Parity-violating systematics:} \ac{BOSS} has undiagnosed systematics which are not parity-invariant.
    
    \item \textit{Data--mock mismatch:} There is no evidence for parity violation (either physical or systematic) in \ac{BOSS}, but the mocks do not perfectly model the \emph{parity-even} higher $N$-point functions of the data.
    (More precisely, if the parity-even \ac{8PCF} of the mock catalogs differs from the data, then the mocks may underpredict the covariance of the \ac{4PCF}, leading to a biased $\chi^2$.)\footnote{%
        This possibility is emphasized throughout \cite{hou_parity,oliver_parity}, where it is described as underestimating the covariance (or noise) of the statistic $\smash{\hE_a}$ (defined later in the paper).
        E.\,g.\ the abstract of \cite{hou_parity} reads ``Underestimation of the noise could also lead to a spurious detection'', and \cite{oliver_parity} writes ``A spurious detection of parity-violation could be caused by the simulations underestimating the true covariance''.
        \cite{hou_parity} estimated the impact of a wrong covariance matrix on the detection significance in their section~7.%
    }
\end{enumerate}
To explore the ``data--mock mismatch'' possibility in more detail, we explain how $\chi^2$ is defined.
The steps are (schematically):
\begin{equation}
    \Big( \text{Galaxy catalog} \Big) \rightarrow 
    \Big( \text{Parity-odd \ac{4PCF} } \hE_a \Big) \rightarrow
    \Big( \text{Parity-even \ac{8PCF} } \chi^2 \Big)
\end{equation}
The new quantities $\hE_a$ and $(C_{\mathrm{ana}})_{ab}$ will be defined precisely later (\cref{sec:reproducing_7sigma}). 
In the introduction, the following qualitative descriptions of $\hE_a$ and $(C_{\mathrm{ana}})_{ab}$ will suffice:
\begin{itemize}
    \item Each component $a = 1, \dots, N_{\mathrm{dof}}$ of $\hE_a$ is a parity-odd four-point function in the galaxy catalog.
    ``Parity-odd'' means that if a spatial reflection is applied to the galaxy catalog, then $\hE_a$ transforms as $\hE_a \rightarrow -\hE_a$.
    This implies:
    \begin{equation}
        \bE_a = 0
          \hspace{1.5cm}
        \text{if the statistics of the galaxy field are parity-invariant}
    \end{equation}
    where $\bE_a$ denotes the true (i.\,e.\ cosmic average) parity-odd four-point function $\hE_a$.\footnote{%
        In most of the paper, we denote the parity-odd four-point function $\hE_a$ using a single index $a=1, \dots, N_{\mathrm{dof}}$. However, the ``natural'' definition of $\hE$ (\cref{ssec:hE}) is a six-index object $\smash{\hE^{\beta_1\beta_2\beta_3}_{\ell_1\ell_2\ell_3}}$, where $\beta_i$ denotes a radial bin and $0 \le l_i \le 4$ denotes an ``angular momentum'' (index of the spherical harmonics).
        When we use the compressed notation $\hE_a$, each value of the ``flattened'' index $a = 1, \dots, N_{\mathrm{dof}}$ represents a different six-tuple $((\beta_i), (l_i))$.
        The number of components $N_{\mathrm{dof}}$ can be large.
        In \cref{fig:intro}, we have used an ``18-bin'' setup with $N_{\mathrm{dof}} = 18768$.
        See \cref{ssec:hE} for details.%
    }

    \item The ``analytic'' covariance matrix $(C_{\mathrm{ana}})_{ab}$ is the $N_{\mathrm{dof}}$-by-$N_{\mathrm{dof}}$ matrix:
    \begin{equation}
        (C_{\mathrm{ana}})_{ab} = \big\langle \hE_a \hE_b \big\rangle
          \hspace{1.5cm} \text{assuming that the galaxy field $\delta_{\mathrm{g}}$ is Gaussian}
    \end{equation}
    The assumption that $\delta_{\mathrm{g}}$ is Gaussian is not intended to be an accurate approximation.
    It is only intended to give a calculable, well-motivated, invertible covariance matrix that can be used for data compression purposes when defining the $\chi^2$ statistic:
    \begin{equation}
        \chi^2 \equiv \hE_a (C^{-1}_{\mathrm{ana}})^{ab} \hE_b
        \label{eq:chi2_def_intro}
    \end{equation}
\end{itemize}
Note that the definition \eqref{eq:chi2_def_intro} of $\chi^2$ involves squaring the parity-odd \ac{4PCF} $\hE_a$.
Therefore, $\chi^2$ is a parity-even eight-point statistic, whereas $\hE_a$ is a parity-odd four-point statistic.
This makes the $\chi^2$ statistic more fragile: it can be biased by parity-even effects
(whereas many observational systematics cannot generate a parity-odd signal).
In particular, if the parity-even \ac{8PCF} of the mock catalogs does not agree with the data (``data--mock mismatch''), then there is no symmetry which protects the $\chi^2$ statistic from bias.
Quantitatively, a $\sim \SI{20}{\percent}$ discrepancy between the parity-even \ac{8PCF} of the mocks and data would explain the results of \cite{hou_parity,oliver_parity} without parity violation (either cosmological or systematic).

A priori, a $\sim \SI{20}{\percent}$ \ac{8PCF} discrepancy between mocks and data is entirely plausible.
The PATCHY mocks include free parameters (mostly pertaining to the galaxy--halo relation) which are adjusted so that the \ac{2PCF} of the mocks agrees with the data.
The mocks are not intended to model higher-point correlation functions precisely.\footnote{%
    This is a natural consequence of the fact that \ac{LSS} analyses have focused on the two-point function, and the massive catalogs of mocks necessary for this analysis have only been created for analyses of the large-scale power spectrum and correlation function.
    Large simulation suites devoted to higher-point statistics have only recently become available (e.\,g.\ Quijote \cite{Quijote}) and have not generally been used to create mocks for the \ac{BOSS} survey, partially because the \SI{1}{\per\hHubble\Gpc} boxes are smaller than the \ac{BOSS} survey volume.%
}
Indeed, the \ac{3PCF} of the PATCHY mocks generally agrees with the data \cite{Slepian:2016kfz,Kitaura:2015uqa}, but has some discrepancies \cite{Rodriguez-Torres:2015vqa}.
In section~4.2.3 of \cite{hou_parity} it is reported that the parity-even \acp{4PCF} disagree at $4.9\sigma$ (for some choices of binning).
Generally speaking, higher-point functions are sensitive to tails of distributions, and can magnify small modeling issues.
Therefore, it seems completely plausible that the \acp{8PCF} of the mocks and data could disagree by $\sim \SI{20}{\percent}$.

(As an aside, squaring $\hE_a$ seems to be necessary in an analysis where no model of the parity-odd \ac{4PCF} is assumed. On the other hand, if a model for $\hE_a$ is assumed, then the optimal statistic is linear in $\hE_a$, and \ac{8PCF} bias is not an issue. This may explain why model-based analyses have produced null results so far \cite{Cabass:2022oap}.)

\subsection{New statistics that distinguish parity violation and data--mock mismatch}
\label{ssec:new_statistics}

Now we can present the main idea of this paper.
So far, we have proposed data--mock mismatch as a possible explanation for the $7\sigma$ signal in \cref{fig:intro} (left panel), but we have not presented evidence for or against this possibility.
We will now construct new statistics, denoted $\chi^2_\times$ and $\chi^2_{\mathrm{null}}$, which cleanly separate parity violation from data--mock mismatch.

Our construction is based on the following simple idea.
If $\chi^2$ excess is due to parity violation, then the true parity-odd four-point function $\bE_a$ of the universe is nonzero.
In this case, we should see the same (within statistical errors) parity-odd four-point function $\hE_a$ in different parts of the sky.
On the other hand, if the $\chi^2$ excess is due to data--mock mismatch, then $\hE_a$ has mean zero, but the mocks underestimate the covariance $\langle \hE_a \hE_b \rangle$.
In this case, we should see uncorrelated (within statistical errors) parity-odd four-point functions $\hE_a$ in different parts of the sky.

To make this idea precise, we start by splitting the \ac{BOSS} survey area into $N_p$ patches of roughly equal area, where $N_p = 3$ for the \ac{BOSS} \ac{CMASS} \ac{NGC} dataset, denoted as \cmassnorth (which we focus on in this introduction), and $N_p = 2$ for the CMASSLOWZTOT \ac{SGC} dataset, denoted as \cmasssouth.\footnote{%
    We explain in \cref{sec:cmass_vs_cmasslowztot} why we consider two different samples in the northern and southern galactic caps.%
}
We separate patches by gaps of 5--10 degrees, to make the patches approximately statistically independent.
The patches are shown in \cref{ssec:patches} and \cref{fig:regions}.

We estimate the parity-odd \ac{4PCF} independently in each patch $\mu = 1, \cdots, N_p$, and denote the result by $\hE_a^\mu$, now with an extra index $\mu$. We then define new statistics:
\begin{align}
    \chi^2_\times &\propto \sum_{\mu \ne \nu} \hE_a^\mu (C_{\mathrm{ana}}^{-1})^{ab} \hE_b^\nu 
      \label{eq:chi2_times_def_intro} \\
    \chi^2_{\mathrm{null}} &\propto \sum_{\mu\ne\nu} \big( \hE_a^\mu - \hE_a^\nu \big) 
    (C_{\mathrm{ana}}^{-1})^{ab} \big( \hE_b^\mu - \hE_b^\nu \big) 
    \label{eq:chi2_null_def_intro}
\end{align} 
(The overall normalizations of $\chi^2_\times$ and $\chi^2_{\mathrm{null}}$ will be defined in \cref{sec:new_statistics}.)
At an intuitive level, we expect that $\chi_\times^2$ will only be sensitive to parity violation, and $\chi^2_{\mathrm{null}}$ will only be sensitive to data--mock mismatch, by the following argument:

\begin{itemize}
    \item The $\chi_\times^2$ statistic measures correlations between parity-odd four-point functions $\hE_a^\mu$  in different ($\mu\ne\nu$) patches of sky.
    Such correlations do not acquire expectation values from data--mock mismatch (which acts as ``noise'' that is uncorrelated between well-separated patches).
    On the other hand, if parity is violated, then $\bE_a$ is the same in all patches, leading to a nonzero expectation value $\langle \chi^2_\times \rangle \propto \bE_a (C^{-1}_{\mathrm{ana}})^{ab} \bE_b$.

    \item The $\chi^2_{\mathrm{null}}$ statistic defines a null test: it measures consistency between four-point functions in different parts of the sky.
    Parity violation does not contribute to $\chi^2_{\mathrm{null}}$, since we still expect consistent values of $\bE_a$ in different parts of the sky.
    However, systematics or data--mock mismatch will add ``noise'' to $\hE_a$, which does contribute to $\chi^2_{\mathrm{null}}$.
\end{itemize}
More formally, in \cref{sec:new_statistics} we will show that the new statistics $\chi^2_\times$ and $\chi^2_{\mathrm{null}}$ separate parity violation and data--mock mismatch, in the following precise sense.
Going back to the original $\chi^2$ statistic in \cref{eq:chi2_def_intro}, we calculate the expectation value $\langle \chi^2 \rangle$ relative to mocks, and find two terms:
\begin{equation}
    \big\langle \chi^2 \big\rangle_{\mathrm{data}} 
      - \big\langle \chi^2 \big\rangle_{\mathrm{mock}}
     = \underbrace{\bE_a \big( C_{\mathrm{ana}}^{-1} \big)^{ab} \bE_b}_{\text{parity violation}} \,\, + \,\,
        \underbrace{\Tr\big[ \big( C_{\mathrm{data}} - C_{\mathrm{mock}} \big) C_{\mathrm{ana}}^{-1} \big]}_{\text{data--mock mismatch}}
    \label{eq:intro_2terms}
\end{equation}
The ``parity violation'' term in \cref{eq:intro_2terms} is nonzero if the true parity-odd four-point function $\bE_a$ of the universe is nonzero, and the ``data--mock mismatch'' term is nonzero if the covariance matrix $C_{ab} = \langle \hE_a \hE_b \rangle$ of the mocks differs from the data.
We then show that:
\begin{align}
    \big\langle \chi^2_{\times} \big\rangle_{\mathrm{data}} 
      - \underbrace{\big\langle \chi^2_{\mathrm{\times}} \big\rangle_{\mathrm{mock}}}_{\text{$= 0$}}
     &= \underbrace{\bE_a \big( C_{\mathrm{ana}}^{-1} \big)^{ab} \bE_b}_{\text{parity violation}} \label{eq:intro_2terms_split1} \\
    \big\langle \chi^2_{\mathrm{null}} \big\rangle_{\mathrm{data}} 
      - \big\langle \chi^2_{\mathrm{null}} \big\rangle_{\mathrm{mock}}
     &= \underbrace{\Tr\big[ \big( C_{\mathrm{data}} - C_{\mathrm{mock}} \big) C_{\mathrm{ana}}^{-1} \big]}_{\text{data--mock mismatch}}
    \label{eq:intro_2terms_split2}
\end{align}
Comparing with \cref{eq:intro_2terms}, we see that the statistics $\chi^2_\times$ and $\chi^2_{\mathrm{null}}$ formally separate parity violation from data--mock mismatch, as argued intuitively above.

The idea of $\chi^2_\times$ was inspired by a standard trick from \ac{CMB} data analysis.
The most straightforward way to estimate a \ac{CMB} power spectrum $C_l$ would be to make a single \ac{CMB} map, take its auto power spectrum, and subtract the noise bias $N_l$.
In practice, this is not robust since the noise bias $N_l$ is difficult to model.
A more robust approach is to cross-correlate maps with independent noise realizations (e.\,g.\ made from data taken at different times), so that there is no noise bias $N_l$ to subtract.
Analogously, in this paper we obtain independent estimates of the parity-odd \ac{4PCF}, by computing $\hE_a$ in different parts of the sky.
By cross-correlating these measurements, we obtain a statistic $\chi^2_\times$ with no bias $\langle \chi^2_\times \rangle_{\mathrm{mock}}$ to subtract.

In the right panel of \cref{fig:intro}, we apply the statistics $\chi^2_\times$ and $\chi^2_{\mathrm{null}}$ to \ac{BOSS}.
The parity violation signal drops to $< 2\sigma$, while the data-mismatch signal remains at high significance.
Our interpretation is that there is not compelling evidence for parity violation in \ac{BOSS} (for the \ac{CMASS}-\ac{NGC} sample -- see \cref{fig:2d,fig:ngc_sgc} for \ac{SGC} and \ac{NGC}+\ac{SGC} results).

In this introduction, we explained the main ideas of our analysis, glossing over technical details.
In the rest of the paper, we will fill in the details.
In \cref{sec:data}, we describe the \ac{BOSS} data and mock catalogs, and details of our processing.
In \cref{sec:reproducing_7sigma} we define the $\chi^2$ statistic, and reproduce the $\sim 7\sigma$ and $\sim 3\sigma$ results from \cite{hou_parity,oliver_parity}.
In \cref{sec:new_statistics}, we define the new statistics $\chi^2_\times$ and $\chi^2_{\mathrm{null}}$, derive their main properties, and apply them to \ac{BOSS}.
We conclude in \cref{sec:discussion}.
We make our data products publicly available.\footnote{%
    Our data can be accessed here: \url{ https://doi.org/10.5281/zenodo.12537418}.%
}

\section{BOSS data}
\label{sec:data}

\subsection{Catalogs and weights}

We measure the parity-odd four-point function
using the publicly available \ac{BOSS} DR12 galaxy catalogs \cite{DR12,Eisenstein11,Dawson13}.\footnote{\url{https://data.sdss.org/sas/dr12/boss/lss/}}
We use the same fiducial cosmology as \ac{BOSS} to convert angles and redshifts into Cartesian coordinates \cite{Beutler17}: 
$\Omega_m = 0.31$, $\Omega_b h^2 = 0.022$, $h = 0.676$, $\sigma_8 = 0.8$, $n_s = 0.96$, and $\Sigma m_\nu = \SI{0.06}{\eV}$.
We compare the parity-odd signal in data to the parity-odd signal in 2048 MultiDark-Patchy mocks \cite{Kitaura:2015uqa,Rodriguez-Torres:2015vqa},\footnote{%
    Described at \url{https://www.skiesanduniverses.org/page/page-3/page-15/page-9/}.
    We use the mocks available at \url{https://data.sdss.org/sas/dr12/boss/lss/dr12_multidark_patchy_mocks/} for CMASSLOWZTOT, and the ``pre-reconstruction'' -- i.\,e.\ without \ac{BAO} reconstruction applied -- mocks (\texttt{Patchy\_prerecon.tar.gz}) available at \url{https://www.ub.edu/bispectrum/page11.html} for \ac{CMASS}.
} run with the parameter set $\Omega_{\mathrm{m}} = 0.3071$, $\Omega_{\mathrm{b}} h^2 = 0.02205$, $h = 0.6777$, $\sigma_8 = 0.8288$, $n_s = 0.96$, and $\Sigma m_\nu = \SI{0}{\eV}$.
The mocks were produced using an approximate (fast) simulation based on \ac{2LPT} on large scales, and spherical collapse on small scales \cite{Kitaura14}.
The parameters were tuned to match an abundance-matched $N$-body simulation, which itself matches the two-point \ac{BOSS} clustering \cite{Rodriguez-Torres:2015vqa}.
The mocks were then cut to the \ac{BOSS} survey geometry, and coordinates were converted to Cartesian using the \ac{BOSS} fiducial cosmology.
(We also noticed a previously unknown systematic in the PATCHY mocks, described in \cref{app:data_mock_investigations}, but we do not believe that it significantly affects the $\chi^2$ statistic.)

In the data catalogs, galaxies are assigned weights $w_{\mathrm{g,data}}$ to correct for imaging systematics $w_{\mathrm{sys}}$ (the product of weights correcting for dependence on stellar density and seeing); fiber collisions $w_{\mathrm{cp}}$; and redshift failures $w_{\mathrm{noz}}$ \cite{Reid2016}; as well as \ac{FKP} weights $w_{\mathrm{FKP}} = 1/(1 + n(z) P_0)$ \cite{FKP} using the observed comoving number density $n(z)$ and $P_0 = \SI{e4}{\per\hHubble\cubed\Mpc\cubed}$.
\begin{equation}
    w_{\mathrm{g,data}} = (w_{\mathrm{noz}} + w_{\mathrm{cp}} - 1) w_{\mathrm{sys}} w_{\mathrm{FKP}}
    \label{eq:wgD}
\end{equation}
$w_{\mathrm{noz}}$ and $w_{\mathrm{cp}}$ are both nearest-neighbor weights that transfer weights from redshift failures or fiber-collided galaxies to the nearest observed neighbor, with default values of 1; thus the proper way to combine them is $(w_{\mathrm{noz}} + w_{\mathrm{cp}} - 1)$ to ensure that observed galaxies are given weight 1.
Random catalogs are generated uniformly within the \ac{BOSS}
imaging region, without fluctuations due to imaging systematics or redshift-dependent effects; hence randoms are only weighted by the \ac{FKP} weights.
\begin{equation}
    w_{\mathrm{r,data}} = w_{\mathrm{FKP}} 
\end{equation}
The mock weighting scheme is slightly different; mock galaxies are assigned weights
\begin{equation}
    w_{\mathrm{g,mock}} = w_{\mathrm{cp}} w_{\mathrm{veto}} w_{\mathrm{FKP}}
\end{equation}
Fiber collisions are implemented in mocks following the approximate
method of \cite{Guo12} (randomly sub-sampling potentially collided galaxies, based on the number of tiles of coverage at a given point on the sky), rather than running the fiber assignment algorithm on the mocks.
Close-pair weights $w_{\mathrm{cp}}$ are applied in the same way to mocks as to data. Since the mocks do not have redshift failures,
there is no need for the $w_{\mathrm{noz}}-1$ term.
Also, the mocks come with a binary veto flag $w_{\mathrm{veto}}$ to remove objects within the veto mask, whereas the veto mask (i.\,e.\ bright stars, poor imaging, etc.) has already been applied to the publicly available data products.
Finally, the random weighting is slightly different in the mocks compared to the data: unlike the data randoms, the mock randoms are run through fiber assignment, and thus must also have close-pair weights applied
\begin{equation}
    w_{\mathrm{r,mock}} = w_{\mathrm{cp}} w_{\mathrm{FKP}} \label{eq:wrM}
\end{equation}

There are various choices of random catalogs of differing density.
We use randoms with $50\times$ the galaxy density: the ``\texttt{random0}'' catalog on data, the $50\times$ random catalog for Patchy CMASSLOWZTOT, and randomly subsampling the $100\times$ random catalog for Patchy \ac{CMASS}.
Choosing different random catalogs (i.\,e.\ using $50\times$ randoms for Patchy \ac{CMASS}, or $100\times$ randoms on mocks and combining ``\texttt{random0}'' and ``\texttt{random1}'' on data) changes our results by $\sim 0.5\sigma$.

\subsection{Sample definition: CMASS vs.\ CMASSLOWZTOT}
\label{sec:cmass_vs_cmasslowztot}

\ac{BOSS} galaxies are typically split in two ways \cite{Reid2016}: by targeting algorithm, LOWZ or \ac{CMASS}, or by redshift bin (after combining LOWZ and \ac{CMASS} into the ``CMASSLOWZTOT'' sample), dubbed \texttt{z1}, \texttt{z2}, and \texttt{z3} in \cite{2017MNRAS.464.1168R}, at $0.2 < z < 0.5$, $0.4 < z < 0.6$, and $0.5 < z < 0.75$, respectively.
LOWZ and \ac{CMASS} were targeted as distinct samples. This leads to differences in their geometric coverage.
In particular, a targeting error in early LOWZ data led to lower density in three regions in the North Galactic Cap (see appendix~A and figure~8 in \cite{Reid2016}; figure~8 shows the LOWZ region excluding these areas with lower density).
These early data are referred to as ``LOWZE2'' and ``LOWZE3'' (after the two different targeting algorithms used), and are included in the combined ``CMASSLOWZTOT'' sample, since the error is known and can be propagated into a reduced number of randoms and different \ac{FKP} weights.
Finally, the imaging systematic weights are different for LOWZ and \ac{CMASS}: no angular weights are applied to LOWZ, since the targets are brighter and do not show significant variation with seeing or stellar density.\footnote{%
    However, LOWZE3 does require angular weights depending on the seeing, due to the incorrect application of the \ac{CMASS} star-galaxy separation \cite{2017MNRAS.464.1168R}.%
}

The different possible choices for the samples introduce additional options into our analysis.
However, due to the associated computational cost, it is not straightforward to simply run all possibilities.
In the following, we will describe the rationale behind our choices.

For the analysis with 10 radial bins ($N_\beta = 10$, see \cref{ssec:hE}), we match \cite{oliver_parity} and use CMASSLOWZTOT for both \ac{NGC} and \ac{SGC}, restricted to $0.43 < z < 0.7$.\footnote{%
    The sample selection was described in more detail in the preceding paper on the parity-even four-point function \cite{ParityEvenPhilcox}.%
}
For the case of 18 radial bins ($N_\beta = 18$, see \cref{ssec:hE}),
we test the impact of varying the sample definition on the parity-odd detection, to test if it is affected by the targeting inhomogeneities in LOWZ, or by the different treatment of systematic weights in \ac{CMASS} versus LOWZ.

Since the targeting inhomogeneities are only present in the \ac{NGC}, our default 18-bin analysis uses \ac{CMASS} only in the \ac{NGC}, and CMASSLOWZTOT in the \ac{SGC}, where we can benefit from the higher number density without potentially adding systematics due to the LOWZ inhomogeneity.
Both are cut to the same redshift range $0.43 < z < 0.70$, matching the redshift cut in \cite{hou_parity,oliver_parity}. 
The sample selection in the \ac{NGC} thus matches that of \cite{hou_parity}, who use \ac{CMASS} in both hemispheres (J.~Hou and Z.~Slepian, priv.~comm).
This is a conservative choice, in the sense that \ac{BOSS} has released large-scale structure catalogs (and mocks) for CMASSLOWZTOT with randoms that correctly follow LOWZ's angular variation; however, the impacts of these choices have only been validated on the large-scale two-point functions \cite{2017MNRAS.464.1168R}, and the possibility remains that higher-point functions are more sensitive to these choices.
Nevertheless, we find consistent detections of the 18-bin parity-odd four-point functions between both \ac{CMASS} and CMASSLOWZTOT and between \ac{NGC} and \ac{SGC}, suggesting that these issues in sample construction are not responsible for the detection.

Summarizing, for the 18-bin case (analogous to \cite{hou_parity}) we will consider two galaxy samples \cmassnorth and \cmasssouth, restricted to $0.43 < z < 0.7$, and with per-object weights given by \crefrange{eq:wgD}{eq:wrM}.
For the 10-bin case (analogous to \cite{oliver_parity}), we will use CMASSLOWZTOT-\ac{NGC} and CMASSLOWZTOT-\ac{SGC}, again restricted to $0.43 < z < 0.7$, and with per-object weights given by \crefrange{eq:wgD}{eq:wrM}.

\section{Reproducing results from \texorpdfstring{\cite{hou_parity,oliver_parity}}{Hou et al., Philcox et al.}}
\label{sec:reproducing_7sigma}

In this section we will define the $\chi^2$ statistic, and reproduce the $\sim 7\sigma$ and $\sim 3\sigma$ results from \cite{hou_parity,oliver_parity}.
This section mostly reviews results from previous papers, especially \cite{hou_parity,oliver_parity,Cahn:2020axu,hou_covariance}, but is included to establish consistency between our pipeline and previous results, and to make our paper self-contained.

\subsection{The parity-odd four-point estimator \texorpdfstring{$\hE_a$}{ℰₐ}}
\label{ssec:hE}

As described in the introduction, the first step in our pipeline is a parity-odd four-point estimator $\hE$, which we will sometimes denote as a six-index object $\smash{\hE_{l_1l_2l_3}^{\beta_1\beta_2\beta_3}}$, and sometimes denote with a single ``flattened'' index $\smash{\hE_a}$. 
Each value of the flattened index $a=1,\dots,N_{\mathrm{dof}}$ corresponds to one six-tuple $((\beta_i), (l_i))$,
In this section, we will review the definition and key properties of $\hE$ from \cite{hou_parity,oliver_parity,Cahn:2020axu,hou_covariance}.

For simplicity, our discussion of $\hE$ in this section assumes a simplified box geometry with periodic boundary conditions, no window function, and no lightcone evolution.
To generalize to a realistic survey geometry, we use the edge correction procedure described in section~2 of \cite{hou_parity} or section~II.C of \cite{oliver_parity}.
This edge correction is implemented in the ENCORE software \cite{Philcox:2021bwo}, which we use to compute $\hE$ in our pipeline.

The input to $\hE$ is a galaxy field $\delta_{\mathrm{g}}(\bx)$ defined by the usual ``data minus randoms'' prescription \cite{LandySzalay93,SzapudiSzalay98,Philcox:2021bwo}:
\begin{equation}
    \delta_{\mathrm{g}}(\bx) = \Bigg( \frac{1}{n_{\mathrm{gal}}} \sum_{i \in \mathrm{gal}} \delta^3(\bx-\bx_i) \Bigg)
      - \Bigg( \frac{1}{n_{\mathrm{rand}}} \sum_{j \in \mathrm{rand}} \delta^3(\bx-\bx_j) \Bigg)
\end{equation}
To write down the definition of $\hE$, we will use six-index notation $\smash{\hE_{l_1l_2l_3}^{\beta_1\beta_2\beta_3}}$.
Here, each index $l_i \ge 0$ is an integer, and each index $\beta_i$ is a non-overlapping radial bin of the form $[R_{\mathrm{min}}, R_{\mathrm{max}}]$.
We define the function $W^\beta_{l m}(\br)$, where the interval $\beta = [R_{\mathrm{min}}, R_{\mathrm{max}}]$ is a radial bin, by:
\begin{equation}
    W^\beta_{l m}(\br) = \left\{ \begin{array}{cl}
      4\pi Y_{lm}^*(\hr)/V_\beta & \text{if } |r| \in \beta \\
      0 & \text{otherwise}
    \end{array} \right.
     \hspace{1.5cm}
    \text{where } V_\beta = \frac{4\pi}{3} \big( R_{\mathrm{max}}^3 - R_{\mathrm{min}}^3 \big)
    \label{eq:Wr_def}
\end{equation}
and $Y_{lm}$ are the spherical harmonics.
We define the estimator $\hE_{l_1l_2l_3}^{\beta_1\beta_2\beta_3}$ by:
\begin{equation}
    \hE_{l_1l_2l_3}^{\beta_1\beta_2\beta_3}
     = \frac{(-1)^{\sum l_i}}{V_{\mathrm{fid}}} \sum_{m_1m_2m_3} 
      \left( \begin{array}{ccc} l_1 & l_2 & l_3 \\ m_1 & m_2 & m_3 \end{array} \right)
     \int d^3\bx \,\, \delta_{\mathrm{g}}(\bx) \, 
     \left( \prod_{i=1}^3 \int d^3\br_i \, W^{\beta_i}_{l_im_i}(\br_i) \, \delta_{\mathrm{g}}(\bx+\br_i) \right)
    \label{eq:Edef}
\end{equation}
where $V_{\mathrm{fid}}$ is a fiducial survey volume (see \cref{tab:cov_ana_parameters}).
The estimator $\hE_{l_1l_2l_3}^{\beta_1\beta_2\beta_3}$ defined by \cref{eq:Edef} has the following properties:
\begin{itemize}
    \item $\hE_{l_1l_2l_3}^{\beta_1\beta_2\beta_3}$ is a four-point estimator in the galaxy field $\delta_{\mathrm{g}}(\bx)$.
    \item $\hE_{l_1l_2l_3}^{\beta_1\beta_2\beta_3}$ is invariant under rotations of the coordinate system.
    \item $\hE_{l_1l_2l_3}^{\beta_1\beta_2\beta_3}$ is nonzero if and only if the triangle inequality $|l_1-l_2| \le l_3 \le (l_1+l_2)$ is satisfied.
    \item $\hE_{l_1l_2l_3}^{\beta_1\beta_2\beta_3}$ is either parity-even or parity-odd, depending on whether $(l_1+l_2+l_3)$ is even or odd.
    \item $\hE_{l_1l_2l_3}^{\beta_1\beta_2\beta_3}$ is either real or imaginary, depending on whether $(l_1+l_2+l_3)$ is even or odd.
    \item $\hE_{l_1l_2l_3}^{\beta_1\beta_2\beta_3}$ is either symmetric or antisymmetric under permutations of $(\beta_i,l_i)$, depending on whether $(l_1+l_2+l_3)$ is even or odd:
    \begin{equation}
        \hE_{l_1l_2l_3}^{\beta_1\beta_2\beta_3} 
          = (-1)^{\sum l_i} \hE_{l_2l_1l_3}^{\beta_2\beta_1\beta_3}
          = (-1)^{\sum l_i} \hE_{l_1l_3l_2}^{\beta_1\beta_3\beta_2}
        \label{eq:hE_perm}
    \end{equation}
\end{itemize}
Since we are interested in the parity-odd case, we will only consider components of $\hE$ such that $\sum l_i$ is odd.
In view of the permutation symmetry \eqref{eq:hE_perm}, we will also assume $\beta_1 \le \beta_2 \le \beta_3$.

Additionally, in \cite{hou_parity} all three radial bins must be distinct $(\beta_1 < \beta_2 < \beta_3)$, and in \cite{oliver_parity} they must be separated by at least one bin $(\beta_1+2 < \beta_2+1 < \beta_3)$, in order to reduce dependence on nonlinear scales.
We will cover both cases by introducing a parameter $\Delta$, and requiring
\begin{equation}
    (\beta_1 + 2\Delta) < (\beta_2 + \Delta) < \beta_3
    \label{eq:beta_constraint}
\end{equation}
where $\Delta=0,1$ in refs.\ \cite{hou_parity,oliver_parity} respectively.

Summarizing, the parity-odd four-point function $\hE_{l_1l_2l_3}^{\beta_1\beta_2\beta_3}$ has one component for each six-tuple $((\beta_i), (l_i))$ such that $\sum l_i$ is odd, the triangle inequality $|l_1-l_2| \le l_3 \le (l_1+l_2)$ is satisfied, and the bin constraint \eqref{eq:beta_constraint} is satisfied.
A short calculation gives the following general formula for the number of components $N_{\mathrm{dof}}$, as a function of $(N_\beta,l_{\mathrm{max}},\Delta)$:
\begin{equation}
    N_{\mathrm{dof}} = \hspace{0.2cm}
    \underbrace{\left( \frac{(N_\beta-2\Delta)(N_\beta-2\Delta-1)(N_\beta-2\Delta-2)}{6} \right)}_{\substack{\text{number of triples } (\beta_1,\beta_2,\beta_3) \\ \text{such that } (\beta_1 + 2\Delta) \, < \, (\beta_2 + \Delta) \, < \, \beta_3 }}
     \times 
    \underbrace{\left\lfloor
     \frac{2l_{\mathrm{max}}^3 + 3l_{\mathrm{max}}^2 + 2l_{\mathrm{max}} + 4}{8}
      \right\rfloor}_{\substack{\text{number of triples } (l_1,l_2,l_3) \\ \text{ such that } l_1+l_2+l_3 \text{ is odd, and} \\ \text{triangle inequality is satisfied}}}
    \label{eq:ndof}
\end{equation}
where the notation $\lfloor x \rfloor$ means ``$x$ rounded down to an integer''.
In this paper, we will focus on the following specific cases:
\begin{itemize}
    \item A ``10-bin'' case following \cite{oliver_parity}:
    We use $N_\beta=10$ linearly spaced radial bins in the interval $[20, 160]$~\si{\per\hHubble\Mpc}, with $(l_{\mathrm{max}}, \Delta) = (4,1)$.
    Then \cref{eq:ndof} gives $N_{\mathrm{dof}}=1288$. (Note that our ``10-bin'' case is the same as the 10-bin analysis from \cite{oliver_parity} which uses $\Delta = 1$, but not the same as the 10-bin analysis from \cite{hou_parity} which uses $\Delta = 0$.)
    \item An ``18-bin'' case following \cite{hou_parity}:
    We use $N_\beta=18$ linearly spaced radial bins in the interval $[20, 160]$~\si{\per\hHubble\Mpc}, with $(l_{\mathrm{max}}, \Delta) = (4, 0)$.
    Then \cref{eq:ndof} gives $N_{\mathrm{dof}} = 18768$.
    (This differs slightly from the setup in \cite{hou_parity}, which uses bins in $[20, 164]$~\si{\per\hHubble\Mpc}.)
\end{itemize}

To compute $\hE$, we run the public ENCORE software \cite{Philcox:2021bwo} adapting the shell script available in the GitHub repository.\footnote{%
    \url{https://github.com/oliverphilcox/encore/blob/master/run_npcf.csh}%
}
The strategy is to reduce the runtime by splitting the randoms into 32 equal files (after first randomizing their order to ensure that each random subset covers the full area).
Then we compute counts of data powers $D^N$ (with $N = 4$ for the four-point function), random powers $R^N$ for the first random subset (for later use in edge correction), and data minus random powers $(D - R)^N$ for all 32 subsets.
Note that the random catalogs are identical for each of the 2048 Patchy mocks, and hence we only generate them for the first Patchy mock. The random weights are automatically balanced to ensure that $\sum(D - R) = 0$ in the $(D - R)^N$ step; however, this re-weighting must also be applied to the first random subset used to calculate $R^N$ counts.
Hence, we re-create the first random subset for each Patchy mock, but with the weights rescaled to ensure that they match the sum of the corresponding Patchy mock's data weights. 

After the counts are combined into the two-point, three-point, and (parity-odd) four-point functions, edge correction is performed.
Before and during edge correction, we keep both parity-even and parity-odd \acp{4PCF} with $l_{\mathrm{max}} = 5$.
After edge correction, we keep only parity-odd \acp{4PCF} with $l_{\mathrm{max}} = 4$.

\subsection{Computing the analytic covariance \texorpdfstring{$C_{\mathrm{ana}}$}{Cₐₙₐ}}
\label{ssec:analytic_covariance}

Consider the covariance matrix $\Cov(\hE_a, \hE_b)$, where $\hE_a$ is the parity-odd four-point estimator defined in the previous section.
In \cite{hou_parity}, it was shown that this covariance matrix can be computed analytically under two simplifying approximations: the galaxy field $\delta_{\mathrm{g}}(\bx)$ is a Gaussian field, and the survey geometry is a 3D periodic box with volume $V_{\mathrm{fid}}$ (i.\,e.\ neglecting lightcone evolution and anisotropic power spectra from redshift space distortions).

Throughout this paper, we define the ``analytic'' covariance matrix $(C_{\mathrm{ana}})_{ab}$ to be the covariance computed under these approximations.
Note that these approximations are not particularly good, and $C_{\mathrm{ana}}$ is not necessarily a good approximation to the true covariance $C_{\mathrm{data}}$.
This is not a problem, since in this paper we only use $C_{\mathrm{ana}}$ for data compression purposes, when defining  statistics such as $\smash{\chi^2 = \hE_a (C_{\mathrm{ana}}^{-1})^{ab} \hE_b}$.

\begin{table}
    \centering
    \begin{tabular}{cclcc}
        \toprule
        \multicolumn{1}{c}{Reference} & Bins & \multicolumn{1}{c}{Dataset} & $P_0$ $/$ \si{\per\hHubble\cubed\Mpc\cubed} & $V_{\mathrm{fid}}$ $/$ \si{\per\hHubble\cubed\Gpc\cubed}
        \\
        \midrule
        \multirow{2}{*}{\cite{hou_parity}} & \multirow{2}{*}{18} & \cmassnorth & $1/(\num{1.40e-4})$ & 2.50\phantom{0}
        \\
        && \cmasssouth & $1/(\num{1.40e-4})$ & 0.790
        \\
        \multirow{2}{*}{\cite{oliver_parity}} & \multirow{2}{*}{10} & CMASSLOWZTOT-\ac{NGC} & $1/(\num{3.19e-4})$ & 1.90\phantom{0}
        \\
        && CMASSLOWZTOT-\ac{SGC} & $1/(\num{3.16e-4})$ & 0.766
        \\
        \bottomrule
    \end{tabular}
    \caption{%
        Input parameters for the analytic covariance (\cref{ssec:analytic_covariance}) in different cases, following \cite{hou_parity,oliver_parity}.
        We note that the values of $P_0$ and $V_{\mathrm{fid}}$ are taken directly from \cite{hou_parity,oliver_parity}, where they are determined from fits of the analytic covariance matrix to that of the mocks.
        They do \emph{not} therefore represent the actual number density or volume of the \ac{CMASS} and CMASSLOWZTOT samples.%
    }
    \label{tab:cov_ana_parameters}
\end{table}

In \cref{app:analytic_covariance}, we give a self-contained derivation of the analytic covariance (re-deriving results from \cite{hou_covariance}).
The input parameters and main computational steps can be summarized as follows:
\begin{itemize}
    \item Following \cite{hou_parity,oliver_parity}, the galaxy power spectrum $P_{\mathrm{g}}(k)$ is modelled using {\classpt} \cite{Chudaykin:2020aoj} at redshift $z = 0.57$, and a Poisson noise term $P_0$ is added by hand.\footnote{%
        The power spectrum $P_{\mathrm{g},\text{\classpt}}(k)$ was provided to us by Oliver Philcox, and was also used in \cite{oliver_parity}.
        The CLASS-PT bias parameters were obtained by fitting the measured \ac{CMASS} galaxy power spectra.
        This power spectrum file is included in our public code repository at \url{https://gitlab.com/Socob/analytic4pc}.%
    }
    The specific values of $P_0$ used vary with each considered case, and match those given in \cite{hou_parity,oliver_parity} (where they were obtained by fitting the analytic covariance to the mock covariance).
    These values are summarized in \cref{tab:cov_ana_parameters} and do not necessarily match the number density of each sample.

    \item Following \cite{hou_covariance}, we apply a Gaussian damping to the power spectrum:
    \begin{equation}
        P_{\mathrm{g}}(k) = \big(P_{\mathrm{g},\text{\classpt}}(k) + P_0\big) \exp\mleft({-}(k/k_0)^2\mright)
        \qquad
        \text{where $k_0 = \SI{1}{\hHubble\per\Mpc}$.}
        \label{eq:pg_dampening}
    \end{equation}

    \item The fiducial effective survey volume appears in $C_{\mathrm{ana}}$ via an overall prefactor $1/V_{\mathrm{fid}}$. As with $P_0$, the values of this volume are different in each considered case, matching those from the fits in \cite{hou_parity,oliver_parity}, and are summarized in \cref{tab:cov_ana_parameters}, again, not necessarily matching
    the actual survey volume.

    \item The first computational step is computing a large number of correlation functions, for example the standard galaxy correlation function
    \begin{equation}
        \xi(r) = \int_0^\infty dk \, \frac{k^2}{2\pi^2} P_{\mathrm{g}}(k) j_0(kr)
    \end{equation}
    and additionally some non-standard correlation functions $H^{\beta}_l(r)$, $F^{\beta\beta'}_{ll'L}(r)$ defined using similar $k$-integrals (see \cref{eq:H_def,eq:F_def}).
    Following section~5.1 of \cite{hou_covariance}, we do these $k$-integrals by summing over $N_k = 5000$ linearly spaced $k$-values with $k_{\mathrm{min}} = \SI{e-4}{\hHubble\per\Mpc}$, $k_{\mathrm{max}} = \SI{5}{\hHubble\per\Mpc}$.
    \footnote{These settings were derived from convergence tests performed on the analytic three-point function covariance in Section 6.5 of \cite{Slepian:2015qza}.}

    \item The matrix elements of $C_{\mathrm{ana}}$ are given by integrals involving  correlation functions computed in the previous step.
    For example, \cref{eq:T_final} contains an integral of the form:
    \begin{equation}
        \int_0^\infty dr \, r^2 \xi(r) \,
             F^{\beta_1\beta_1'}_{l_1l_1'L_1}(r) \,
             F^{\beta_2\beta_2'}_{l_2l_2'L_2}(r) \,
             F^{\beta_3\beta_3'}_{l_3l_3'L_3}(r)
    \end{equation}
    Following section~5.1 of \cite{hou_covariance}, we do these $r$-integrals by
    summing over $N_r = 4100$ linearly spaced $r$-values with $r_{\mathrm{min}} = \SI{e-5}{\Mpc\per\hHubble}$, $r_{\mathrm{max}} = \SI{1000}{\Mpc\per\hHubble}$.
\end{itemize}

\par\noindent
We wrote a publicly available Julia software package to compute $C_{\mathrm{ana}}$ given the above inputs.\footnote{%
    Our code is available at \url{https://gitlab.com/Socob/analytic4pc}.%
}
Although some of the code used in \cite{hou_covariance} is also publicly available,\footnote{%
    \url{https://github.com/Moctobers/npcf_cov}%
} we opted to create an independent implementation for several reasons:
\begin{itemize}
    \item The public version of the code from \cite{hou_covariance} is only applicable to the calculation of the parity-even part of the analytic covariance (see \cref{app:analytic_covariance} for details), so it could not directly be used in the parity-odd analysis.
    \item In order to rule out any issues with the analysis related to the computation of $C_{\mathrm{ana}}$, an independent implementation is very valuable, especially considering the complexity of the calculation and the mathematical objects involved.
    \item Since the calculation is quite computationally intensive for larger numbers of bins, there was a large motivation to make full use of parallel computing resources in order to improve run time.
    \item In order to perform various tests of numerical stability and convergence, we wanted to create a flexible code that allows easily varying the numerical parameters (e.\,g.\ $N_k$, $k_{\mathrm{max}}$) without modifying the code itself. 
\end{itemize}

\par\noindent
In principle, the calculation of $C_{\mathrm{ana}}$ is an ``embarrassingly parallel'' problem, since each matrix element can be computed independently.
However, an important part of the calculation involves caching the values of intermediate quantities, namely the ordinary and binned Bessel functions, $j_l(x)$ and $B^\beta_l(k)$, as well as the correlation functions $\xi(r)$, $H^{\beta}_l(r)$, $F^{\beta\beta'}_{ll'L}(r)$, and combinatorial factors.
Since these same quantities appear many times even for different covariance matrix elements, it would be immensely inefficient to re-compute them every time.
This complicates parallelization, as naïvely running a serial code in parallel for different matrix elements would not only duplicate the cost of computing the cache for each instance, but also run into memory limitations since the size of these objects is not insignificant.

Our code makes use of Julia's built-in distributed computing module Distributed.jl, which allows it to run in parallel on an arbitrary number of computing cores across several machines without being limited to a single shared-memory node, and without requiring the installation of additional libraries or software.
The cache of intermediate quantities is implemented within each shared-memory machine via shared process memory in order to avoid storing multiple copies of the same data.

We have performed cross-checks of our code's output with the public code from \cite{hou_covariance} and the related 10-bin covariance matrix used in \cite{oliver_parity}, and have found excellent agreement approaching numerical precision.
Further details of investigations concerning the numerical accuracy of the analytic covariance computation are laid out in \cref{app:analytic_covariance}.
For instance, the integration parameters $(N_k, k_{\mathrm{max}}, N_r, r_{\mathrm{max}})$ should be chosen conservatively enough that the integrals converge.
In \cref{app:convergence_tests}, we present an ``end-to-end'' check which shows that all matrix entries of $C_{\mathrm{ana}}$ have converged, if we use the parameter choices above.
In \cref{app:cov_mat_properties}, we show that $C_{\mathrm{ana}}$ can be poorly conditioned, but this does not seem to be an issue for the $\chi^2$ analysis.

\subsection{Computing \texorpdfstring{$\chi^2$}{χ²}}
\label{ssec:computing_chi2}

Following \cite{hou_parity,oliver_parity}, we define the $\chi^2$ statistic by:
\begin{equation}
    \chi^2 \equiv \hE_a (C_{\mathrm{ana}}^{-1})^{ab} \, \hE_b
    \label{eq:chi2_def}
\end{equation}
The purpose of the $\chi^2$ statistic is to provide data  compression: it reduces the many-component parity-odd \ac{4PCF} $\hE_a$ to a scalar quantity $\chi^2$.\footnote{%
    Refs.\ \cite{hou_parity,oliver_parity} also explore other forms of data compression, such as limiting the number of eigenvalues $N_{\mathrm{eig}}$ in the covariance, or using the mock covariance instead of the analytic covariance.
    In this paper, we choose to only use the $\chi^2$ statistic defined in \cref{eq:chi2_def}, since it leads to the highest detection significance in \cite{hou_parity,oliver_parity}.%
}

\begin{figure}[p!]
    \centerline{\includegraphics[width=8cm]{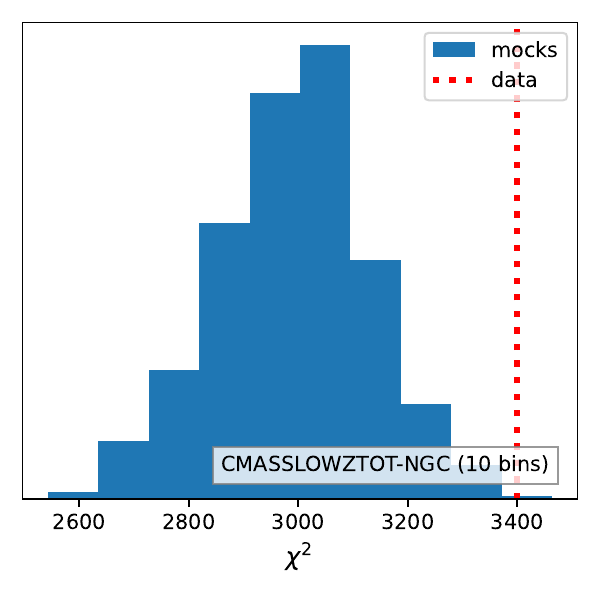} \includegraphics[width=8cm]{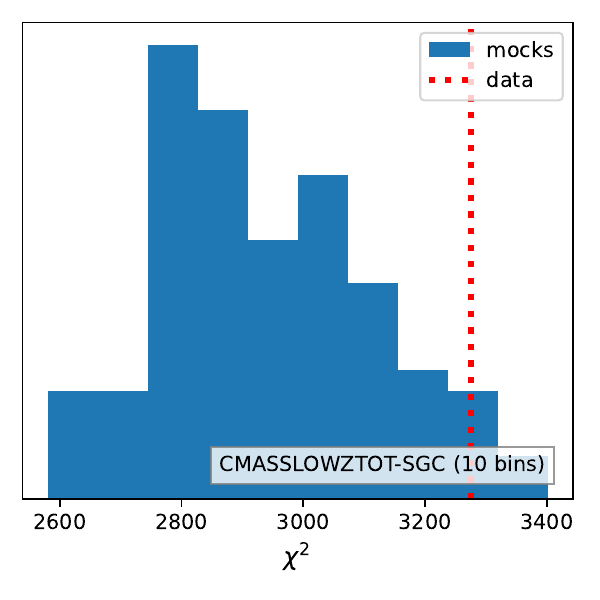}}
    \centerline{\includegraphics[width=8cm]{figures/north_chi2_tot.pdf} \includegraphics[width=8cm]{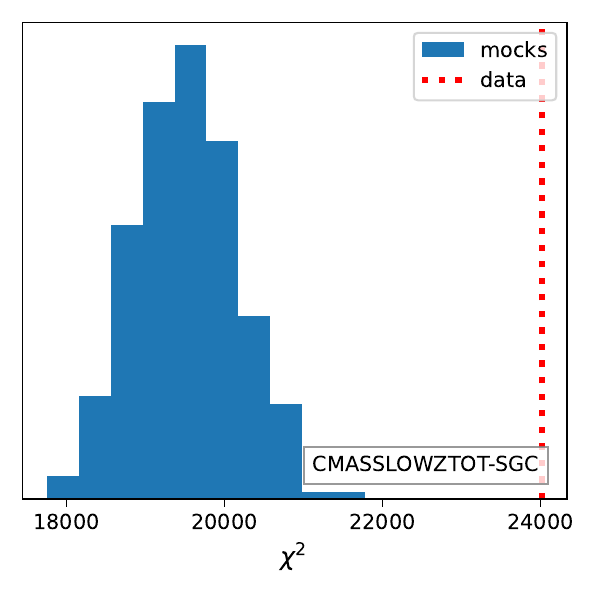}}
    \caption{%
        Reproducing results from \cite{hou_parity,oliver_parity}, by evaluating $\chi^2$ on \ac{BOSS} data (vertical dashed lines) and mocks (solid histograms).
        \textbf{\textit{Top row.}}
        A 10-bin analysis following \cite{oliver_parity}. We find $2.8\sigma$ for CMASSLOWZTOT-\ac{NGC} (top left) and $1.9\sigma$ for \cmasssouth (top right).
        \textbf{\textit{Bottom row.}}
        An 18-bin analysis following \cite{hou_parity}. We find \SigNorthTot for \cmassnorth (bottom left) and \SigSouthTot for \cmasssouth (bottom right).
    }
    \label{fig:chi2}
\end{figure}

Our goal in this section is to reproduce the main results of \cite{hou_parity,oliver_parity}, by evaluating $\chi^2$ on \ac{BOSS} data and mocks, before defining new statistics $\chi^2_\times$, $\chi^2_{\mathrm{null}}$ in later sections.
Our main results are shown in \cref{fig:chi2} and the rest of this section will be devoted to interpretation and discussion.

We note that the overall normalization of $\chi^2$ is arbitrary, since $\chi^2$ is proportional to the fiducial survey volume $V_{\mathrm{fid}}$, and we chose $V_{\mathrm{fid}}$ (\cref{tab:cov_ana_parameters}) without a precise $\chi^2$ normalization in mind.
On a related note, $\chi^2$ is not precisely $\chi^2$-distributed, for two reasons: $\hE_a$ is not a multivariate Gaussian, and $C_{\mathrm{ana}}$ does not perfectly model $C_{\mathrm{mock}}$.
(Similar comments apply to the new statistics $\chi^2_\times$, $\chi^2_{\mathrm{null}}$ in the next section.)

Therefore, throughout the paper, we assess statistical significance by comparing to a histogram of mocks.
This procedure does not assume an analytic distribution for $\chi^2$, and is independent of the overall normalization of $\chi^2$.
When we report statistical significance as a ``number of sigmas'' $S\sigma$, we report
$S = (\chi^2_{\mathrm{data}} - \langle \chi^2 \rangle_{\mathrm{mock}}) / \mbox{Var}(\chi^2)_{\mathrm{mock}}^{1/2}$, 
rather than reporting the $p$-value of the data within the mock distribution.
This makes the approximation that the mock distribution is Gaussian, but lets us quantify large outliers (e.\,g.\ $7\sigma$).

\begin{table}
    \centering
    \begin{tabular}{lccc}
        \toprule
        Analysis choice & 10-bin, \ac{NGC} & 18-bin, \ac{NGC} & 18-bin, \ac{SGC} \\
        \midrule
        This work, baseline & 2.8 & 7.3 & 6.9 \\
        Previous work: \cite{hou_parity,oliver_parity} & 2.4 & 4.7 & 5.4 \\
        \midrule
        Swap \ac{CMASS} and CMASSLOWZTOT & -- & 5.4 & 9.2 \\
        Double $\bar{n}$ in analytic covariance & -- & -- & 8.1 \\
        Halve $\bar{n}$ in analytic covariance & -- & -- & 5.5 \\
        Turn off $w_{\textrm{noz}}$ & -- & 4.3 & -- \\
        Turn off $w_{\textrm{cp}}$ & -- & 8.9 & -- \\
        Turn off $w_{\textrm{sys}}$ & 3.0 & -- & -- \\
        \bottomrule
    \end{tabular}
    \caption{Detection significance of the parity-odd four point function using $\chi^2$, in number of Gaussian $\sigma$. Top rows compare our baseline results to those of \cite{hou_parity,oliver_parity}, and subsequent rows consider swapping the galaxy samples used in the 18-bin case between \ac{CMASS} and CMASSLOWZTOT; changing the number density in the analytic covariance; and turning off various systematic weights. We perform these tests on one hemisphere and choice of $N_\beta$ to illustrate their effect, which is expected to be similar for all three choices of $N_\beta$ and hemisphere.}
    \label{tab:analysis_choices}
\end{table}

We start by attempting to reproduce the results of \cite{hou_parity,oliver_parity}.
First we consider the 10-bin CMASSLOWZTOT-\ac{NGC} results from \cite{oliver_parity} (top left in \cref{fig:chi2}).
We find $\chi^2 = 3399$ on the data, compared to a mean $\chi^2 = 2991$ and standard deviation of 144 on the mocks. 
The data has higher $\chi^2$ by $2.84\sigma$, or two-tailed $p$-value of 0.996 (equivalent to $2.9\sigma$) by the non-parametric rank test, with 9 simulations having $\chi^2 > 3399$.
Ref.\ \cite{oliver_parity} finds $\chi^2 = 3382$ on the data compared to $\chi^2 = 2999 \pm 162$ on the mocks, for a detection of $2.36\sigma$ or $p = 0.988$ ($2.5\sigma$) in the non-parametric rank test.\footnote{Following the publicly available notebook in \url{https://github.com/oliverphilcox/Parity-Odd-4PCF/blob/main/BOSS\%20Odd-Parity\%204PCF\%20(CS\%20template).ipynb}.}
The individual mock parity odd four-point correlation functions
have $\chi^2 \equiv \Delta\hE_a (C_{\mathrm{ana}}^{-1})^{ab} \, \Delta\hE_b$ $\sim 60$, where $\Delta\hE_a$ is the change in parity-odd \ac{4PCF} between our work and \cite{oliver_parity}.
Overall, the level of agreement with \cite{oliver_parity} is excellent.

Our baseline agreement with the 18-bin results of \cite{hou_parity} is less good (bottom row of \cref{fig:chi2}).
Our baseline detection significance is $7.31\sigma$ for \cmassnorth and $6.90\sigma$ for \cmasssouth, compared to $4.7\sigma$ in \ac{NGC} and $5.4\sigma$ in \ac{SGC} in \cite{hou_parity}.
Here, and throughout the paper when discussing the 18-bin case, we define the detection significance by dividing the difference in $\chi^2$ by the standard deviation from the mocks; the non-parametric rank test from \cite{oliver_parity} is not useful because all mocks have $\chi^2$ lower than the data.

Our reproduction of the 18-bin result is quite sensitive to various analysis choices.
In \cref{tab:analysis_choices}, we summarize the detection significance of $\chi^2$ under various analysis choices, also including the baseline choices.

If we instead use CMASSLOWZTOT-\ac{NGC} and \ac{CMASS}-\ac{SGC}, we find detection significance of $5.35\sigma$ in \ac{NGC} and $9.16\sigma$ in \ac{SGC}.
Hence, CMASS ($7.31\sigma$ and $9.16\sigma$) gives consistently higher detection significance than CMASSLOWZTOT ($5.35\sigma$ and $6.90\sigma$), driven by changes in $\chi^2_{\mathrm{data}}$ rather
than changes in $\langle \chi^2_{\mathrm{mock}} \rangle$.
This suggests that the parity-odd four point detection is not driven by inhomogeneous sample selection issues in CMASSLOWZTOT, either created by the complex LOWZ selection function or an improper combination of \ac{CMASS} and LOWZ with their slightly different angular weights and redshift distributions.

Our results also depend on the number density $\bar{n} = 1/P_0$ used to calculate the analytic covariance (\cref{tab:cov_ana_parameters}).
Doubling the number density increases the CMASSLOWZTOT-\ac{SGC} significance to $8.13\sigma$ whereas halving it decreases the significance to $5.50\sigma$.
This suggests that the detection is weighted towards small scales.
Increasing the number density reduces shot noise and pushes the range of cosmic variance-limited modes to higher $k$, effectively up-weighting smaller scales.

The results are even slightly sensitive to the number of randoms used.
If we double the number density of randoms on mocks and data, the significance decreases by $0.4\sigma$.
If we use the $50\times$ randoms on mocks, rather than randomly subsampling the $100\times$ randoms, the significance increases by $0.6\sigma$.
We measured these changes using the change in $\chi^2_{\mathrm{data}}$ and $\chi^2_{\mathrm{mock}}$ on 50 mocks; the scatter contributed by the limited number of mocks is therefore $1/\sqrt{50} = 0.15\sigma$.

Finally, we tested the impact of removing some of the weights applied to the data.
The mocks do not have redshift failures and hence do not have $w_{\mathrm{noz}}$.
This leads to a significant difference in the tails of the distribution of $w_{\mathrm{g,data}}$ and $w_{\mathrm{g,mock}}$, and an even larger difference in the third and fourth moments of the weights $\langle w_{\mathrm{g}}^3 \rangle$, $\langle w_{\mathrm{g}}^4 \rangle$.
Turning off $w_{\mathrm{noz}}$ in the data drops the \cmassnorth detection significance from $7.32\sigma$ to $4.26\sigma$. 

Fiber collisions are added to the mocks in only an approximate way -- and the impact of fiber collisions in the mocks is sensitive to the modeling of sub-\si{\Mpc} scales, which does not match the data perfectly \cite{Kitaura:2015uqa}.
Turning off $w_{\mathrm{cp}}$ changes $\chi^2_{\mathrm{data}}$ substantially, from 22772 to 19747; it also substantially reduces $\langle \chi^2_{\mathrm{mock}}\rangle$, from 19599 to 18344, and $\sigma_{\chi^2}$, from 405 to 158.
Thus, the detection significance goes up, from $7.3\sigma$ to $8.9\sigma$, if we did not correct for fiber collisions in either the data or the mocks.

We also tried turning off the imaging systematics weights on data: again, the mocks do not have imaging systematics and hence do not have this weight, although turning off the weight will add an extra source of systematics in the data.
In the 10-bin case, we find that this increases $\chi^2$ from 3399 to 3581, increasing the detection significance from $2.65\sigma$ to $3.0\sigma$.

Summarizing, we agree nearly perfectly with the 10-bin results from \cite{oliver_parity}, and agree qualitatively with the 18-bin results from \cite{hou_parity}.
In the 18-bin case, the level of disagreement is comparable to the effect of varying analysis choices.
This is good enough agreement that we are confident that we have reproduced the essential features of the analysis in \cite{hou_parity,oliver_parity}.
We find that the $\chi^2$ detection significance remains high even after
removing some of the systematics corrections from the data (though with $\sim3\sigma$ changes), in agreement with the extensive set of validation tests in \cite{hou_parity}.

\section{The new statistics \texorpdfstring{$\chi^2_\times$}{χ²-×} and \texorpdfstring{$\chi^2_{\mathrm{null}}$}{χ²-null}}
\label{sec:new_statistics}

Throughout this section, we only consider the 18-bin case from \cite{hou_parity} (rather than the 10-bin case from \cite{oliver_parity} -- see \cref{ssec:hE} for more discussion).

\subsection{Splitting BOSS into patches}
\label{ssec:patches}

\begin{figure}
    \centerline{\includegraphics[width=8cm]{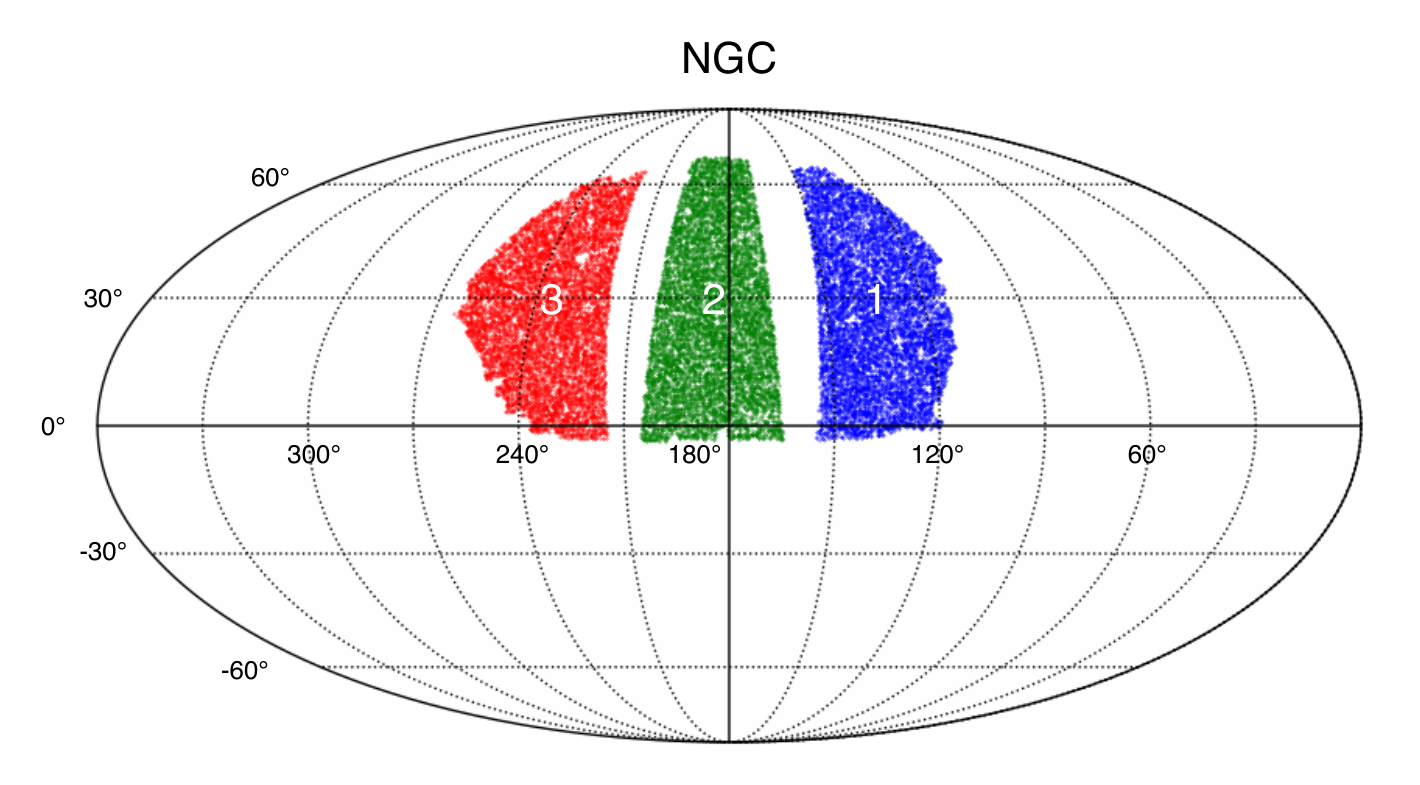} \includegraphics[width=8cm]{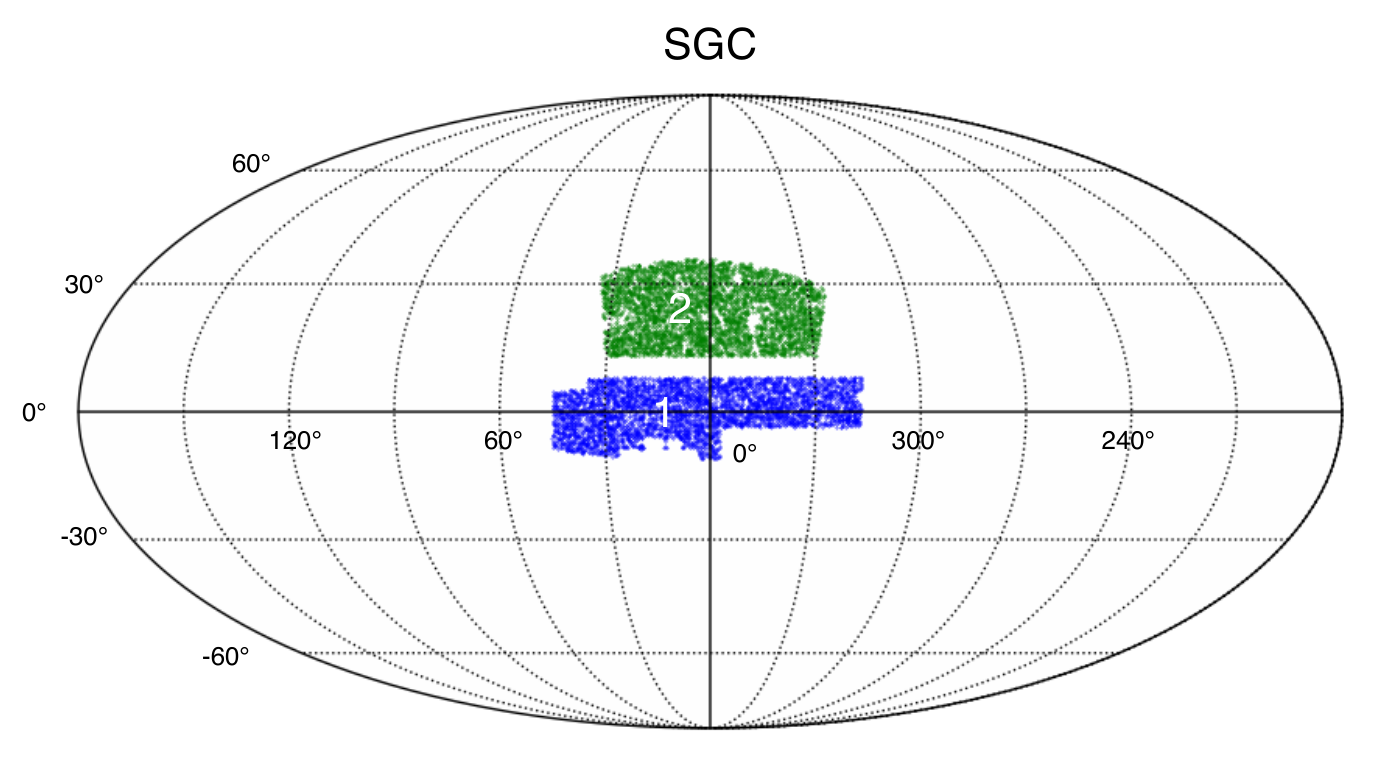}
    }
    \centerline{\includegraphics[width=8cm]{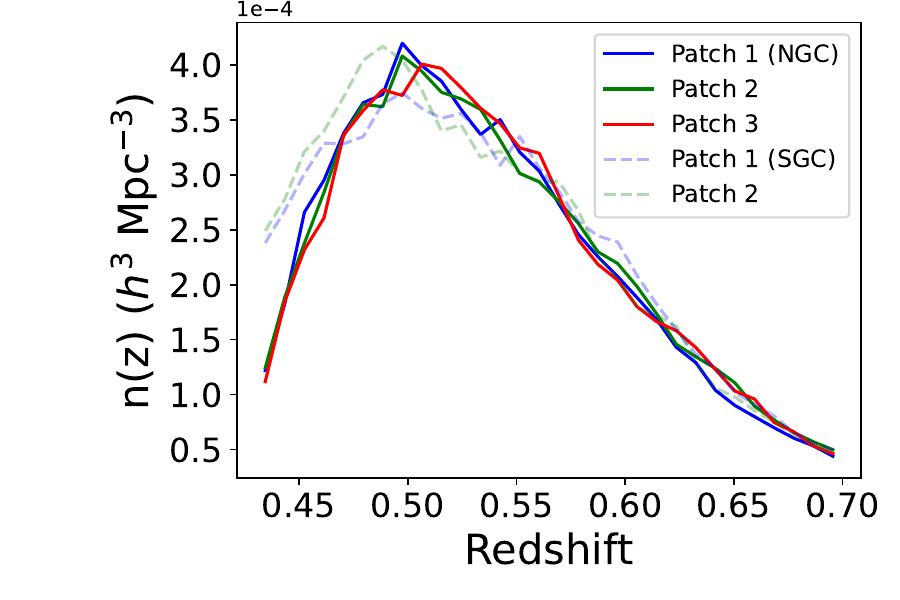}}
    \caption{%
        Patches considered for \ac{NGC} (top left) and \ac{SGC} (top right) to compute $\chi^2_\times$ and $\chi^2_{\mathrm{null}}$.
        Note that the sky has been rotated between the \ac{NGC} and \ac{SGC} plots.
        Bottom panel shows the redshift distribution of all five patches, using the default choices, \ac{CMASS} in the \ac{NGC} and CMASSLOWZTOT in the \ac{SGC}.%
    }
    \label{fig:regions}
\end{figure}

We divide each \ac{BOSS} survey (either \cmassnorth or \cmasssouth) into $N_p$ patches, separated by $\SI{10}{\degree}$ ($\SI{5}{\degree}$) in the \ac{NGC} (\ac{SGC}) -- corresponding to 140 (70) \si{\per\hHubble\Mpc} at $z = 0.4$ -- in order to ensure statistical independence.
We use $N_p = 3$ patches for \cmassnorth, and $N_p = 2$ patches for \cmasssouth.
The patches are defined with the following simple cuts in \ac{RA} and \ac{DEC}, for \ac{NGC}:
\begin{align}
     \mathrm{RA} & < \SI{155}{\degree} - \frac{1}{12} \mathrm{DEC}
    & \text{(\ac{NGC} patch 1)} \nn \\
     \SI{165}{\degree} - \frac{1}{12} \mathrm{DEC} < \mathrm{RA} & < \SI{205}{\degree} - \frac{1}{12} \mathrm{DEC}
    & \text{(\ac{NGC} patch 2)} \nn \\
     \SI{215}{\degree} - \frac{1}{12} \mathrm{DEC} < \mathrm{RA} &
    &  \text{(\ac{NGC} patch 3)}
\intertext{and for \ac{SGC}:}
     \mathrm{DEC} &< \SI{8}{\degree} & \text{(\ac{SGC} patch 1)} \nn \\
     \mathrm{DEC} &> \SI{13}{\degree} & \text{(\ac{SGC} patch 2)}
\end{align}
The sky and redshift distributions of the patches are shown in \cref{fig:regions}.
The cuts remove \SI{22}{\percent} (\SI{12}{\percent}) of the survey area in the \ac{NGC} (\ac{SGC}), respectively.
The mean redshifts are $\bar{z} = \numlist{0.5417; 0.5477; 0.5475; 0.5461; 0.5424}$ for patches 1--3 in \cmassnorth and patches 1--2 in \cmasssouth, respectively.
The standard error of the mean is 0.00017 for the \ac{NGC} patches and 0.00021 for the \ac{SGC} patches, though since galaxies are correlated, the standard error of the mean underestimates the true error on $\bar{z}$ as it neglects the cosmic variance contribution.

\subsection{Definitions and notation}

We use the term ``survey'' to mean either \cmassnorth or \cmasssouth, the term ``patch'' to mean one of the patches $\mu = 1, \dots, N_p$ which comprise a survey, and the term ``region'' to mean either a survey or a patch.

We denote the parity-odd \ac{4PCF} for a survey by $\hE_a$, and the parity-odd \ac{4PCF} for a patch by $\hE_a^\mu$, where $\mu = 1, \dots, N_p$.
We define the following covariance matrices:
\begin{align}
    (C_{\mathrm{data}})_{ab} &= \text{Covariance matrix $\langle \hE_a \hE_b \rangle$ of \ac{BOSS} galaxy field (not directly observable)}
    \\
    (C_{\mathrm{mock}})_{ab} &= \text{Covariance matrix $\langle \hE_a \hE_b \rangle$ of \ac{BOSS} mocks (in the limit $N_{\mathrm{mocks}} \rightarrow \infty$)}
    \label{eq:Cmock}
    \\
    (C_{\mathrm{ana}})_{ab} &= \text{Analytic covariance matrix from \cref{ssec:analytic_covariance}}
\end{align}
In principle, $C_{\mathrm{mock}}$ could be computed by brute force with a large number of mocks ($N_{\mathrm{mocks}} \gg N_{\mathrm{dof}})$.
In practice, we don't have enough mocks, so we can't estimate every entry of the dense matrix $C_{\mathrm{mock}}$.
(For example, this is why we define $\chi^2$ as $\hE_a (C_{\mathrm{ana}}^{-1})^{ab} \hE_b$ rather than $\hE_a (C_{\mathrm{mock}}^{-1})^{ab} \hE_b$.)

Let $(C_{\mathrm{mock}}^\mu)_{ab}$ and $(C_{\mathrm{data}}^\mu)_{ab}$ denote the \ac{4PCF} covariance in a single patch (where $\mu=1,\dots,N_p$).
We expect the covariance to scale with survey volume roughly as $1/V$, so a single-patch covariance $C^\mu_{ab}$ will be larger than the corresponding full-survey covariance $C_{ab}$.
To quantify this, we define the ``effective volume'' $V_{\mathrm{eff}}$ of the full survey and each patch $\mu$ by:
\begin{equation}
    V_{\mathrm{eff}} \equiv \frac{V_{\mathrm{fid}} N_{\mathrm{dof}}}{\Tr(C_{\mathrm{ana}}^{-1} \, C_{\mathrm{mock}})} 
      \hspace{1.5cm}
    V_{\mathrm{eff}}^\mu \equiv \frac{V_{\mathrm{fid}} N_{\mathrm{dof}}}{\Tr(C_{\mathrm{ana}}^{-1} \, C_{\mathrm{mock}}^\mu)} 
    \label{eq:Veff}
\end{equation}
The traces in the denominators can be computed by Monte Carlo, e.\,g.\ $\Tr(C_{\mathrm{ana}}^{-1} C_{\mathrm{mock}}) = \langle \hE_a (C_{\mathrm{ana}}^{-1})^{ab} \hE_b \rangle_{\mathrm{mock}}$.
In \cref{tab:Veff}, we show values of $V_{\mathrm{eff}}, V_{\mathrm{eff}}^\mu$ computed using this method.

\begin{table}
    \centering
    \begin{tabular}{lc}
        \toprule
        Region & Effective volume \\
        \midrule
        \rule{0pt}{2.7ex}    
        \cmassnorth & \\
        \hspace{0.3cm} Full survey
           & $V_{\mathrm{eff}}=2.37$ $h^{-3}$ Gpc$^3$ \\
        \hspace{0.3cm} Patch $\mu=1$
           & $V_{\mathrm{eff}}^\mu=0.52$ $h^{-3}$ Gpc$^3$ \\
        \hspace{0.3cm} Patch $\mu=2$
           & $V_{\mathrm{eff}}^\mu=0.58$ $h^{-3}$ Gpc$^3$ \\
        \hspace{0.3cm} Patch $\mu=3$
           & $V_{\mathrm{eff}}^\mu=0.53$ $h^{-3}$ Gpc$^3$ \\
        \rule{0pt}{2.7ex}
        \cmasssouth & \\
        \hspace{0.3cm} Full survey
           & $V_{\mathrm{eff}}=0.76$ $h^{-3}$ Gpc$^3$ \\
        \hspace{0.3cm} Patch $\mu=1$
           & $V_{\mathrm{eff}}^\mu=0.29$ $h^{-3}$ Gpc$^3$ \\
        \hspace{0.3cm} Patch $\mu=2$
           & $V_{\mathrm{eff}}^\mu=0.30$ $h^{-3}$ Gpc$^3$ \\
        \bottomrule
    \end{tabular}
    \caption{Effective volumes $V_{\mathrm{eff}}$ and $V_{\mathrm{eff}}^\mu$ defined in \cref{eq:Veff}.}
    \label{tab:Veff}
\end{table}

We assume that $\hE_a$ is an unbiased estimator of $\bE_a$ (the true parity-odd four-point function of the universe):
\begin{equation}
    \big\langle \hE_a \big\rangle_{\mathrm{data}} = \bE_a
     \hspace{1.5cm}
    \big\langle \hE_a \big\rangle_{\mathrm{mock}} = 0
    \label{eq:E_survey}
\end{equation}
Then, by definition of $C_{\mathrm{data}}$ and $C_{\mathrm{mock}}$, we have:
\begin{equation}
    \big\langle \hE_a \hE_b \big\rangle_{\mathrm{data}} 
      = \bE_a \bE_b + (C_{\mathrm{data}})_{ab}
    \hspace{1.5cm}
    \big\langle \hE_a \hE_b \big\rangle_{\mathrm{mock}} 
      = (C_{\mathrm{mock}})_{ab}
    \label{eq:EE_survey}
\end{equation}
In \cref{sec:intro}, we stated without proof that the quantity $\big\langle \chi^2 \big\rangle_{\mathrm{data}} - \big\langle \chi^2 \big\rangle_{\mathrm{mock}}$ (i.\,e.\ the ``excess'' $\chi^2$) was the sum of parity violation and data--mock mismatch terms (\cref{eq:intro_2terms}).
Now we can prove this formally as follows:
\begin{align}
\big\langle \chi^2 \big\rangle_{\mathrm{data}} - \big\langle \chi^2 \big\rangle_{\mathrm{mock}}
 &= \big\langle \hE_a (C_{\mathrm{ana}}^{-1})^{ab} \hE_b \rangle_{\mathrm{data}}
      - \big\langle \hE_a (C_{\mathrm{ana}}^{-1})^{ab} \hE_b \rangle_{\mathrm{mock}} \nn \\
 &= (C_{\mathrm{ana}}^{-1})^{ab} \big( \bE_a \bE_b + (C_{\mathrm{data}})_{ab} \big)
    - (C_{\mathrm{ana}}^{-1})^{ab} (C_{\mathrm{mock}})_{ab} \nn \\
 &= \underbrace{\bE_a \big( C_{\mathrm{ana}}^{-1} \big)^{ab} \bE_b}_{\text{parity\ violation}} \,\, + \,\,
    \underbrace{\Tr\big[ \big( C_{\mathrm{data}} - C_{\mathrm{mock}} \big) C_{\mathrm{ana}}^{-1} \big]}_{\text{data--mock mismatch}}
  \label{eq:chi2_tot_ev}
\end{align}
In the first line, we used the definition \eqref{eq:chi2_def} of $\chi^2$, and in the second line we used \cref{eq:EE_survey}.

\subsection{The statistic \texorpdfstring{$\chi^2_\times$}{χ²-×}}
\label{ssec:chi2_cross}

We define the new statistic $\chi^2_\times$ by:
\begin{equation}
    \chi^2_\times \equiv \frac{1}{N_p(N_p-1)} \sum_{\mu \ne \nu} \hE_a^\mu (C_{\mathrm{ana}}^{-1})^{ab} \hE_b^\nu
    \label{eq:chi2_times_def}
\end{equation}
In \cref{sec:intro}, we argued that $\chi^2_\times$ is sensitive to parity violation but not data--mock mismatch.
Intuitively, this is because $\chi^2_\times$ is constructed from cross-correlations between patches, and cross-correlations are not sensitive to data--mock mismatch (which acts as ``noise'' which is uncorrelated between patches).
We will prove this formally shortly, but first we want to state our assumptions explicitly:
\begin{itemize}
    \item \textit{Assumption 1.}
    When we estimate the parity-odd \ac{4PCF} in a single patch, rather than a full survey, the estimator $\hE_a^\mu$ is still unbiased:
    \begin{equation}
        \big\langle \hE_a^\mu \big\rangle_{\mathrm{data}} = \bE_a
         \hspace{1.5cm}
        \big\langle \hE_a^\mu \big\rangle_{\mathrm{mock}} = 0
        \label{eq:E_patch}
    \end{equation}
    (Note that this equation is the same as \cref{eq:E_survey}, but with a $\mu$ index added.)

    \item \textit{Assumption 2.}
    For patches $\mu\ne\nu$, the estimators $\hE_a^\mu$ and $\hE_b^\nu$ are uncorrelated:
    \begin{equation}
        \big\langle \hE_a^\mu \hE_b^\nu \big\rangle_{\mathrm{data}} = \bE_a \bE_b + (C_{\mathrm{data}}^\mu)_{ab} \, \delta^{\mu\nu}
         \hspace{1.5cm}
        \big\langle \hE_a^\mu \hE_b^\nu \big\rangle_{\mathrm{mock}} = (C_{\mathrm{mock}}^\mu)_{ab} \, \delta^{\mu\nu}
        \label{eq:EE_patch}
    \end{equation}
    where the $\delta^{\mu\nu}$ factors follow from the assumption of uncorrelated patches, and the rest of \cref{eq:EE_patch} follows from \cref{eq:E_patch} and the definitions of $C_{\mathrm{data}}^\mu$ and $C_{\mathrm{mock}}^\mu$. (Note the similarity between \cref{eq:EE_survey,eq:EE_patch}.)

    As a test of assumption 2, we verified that for each patch pair $\mu\ne\nu$, the quantity $\langle \hE^\mu_a (C_{\mathrm{ana}}^{-1})^{ab} \hE^\nu_b \rangle_{\mathrm{mock}}$ is zero, within statistical errors from the finite number of mocks.
    (If the spatial ``padding'' between sky patches in \cref{fig:regions} were reduced, we expect that this test would eventually fail.)
\end{itemize}
Now we give a formal proof that $\chi^2_\times$ is sensitive to parity violation but not data--mock mismatch, by computing $\langle \chi^2_\times \rangle_{\mathrm{data}}$ as follows:
\begin{align}
    \big\langle \chi^2_\times \big\rangle_{\mathrm{data}}
     &= \frac{1}{N_p(N_p-1)} \sum_{\mu \ne \nu} (C_{\mathrm{ana}}^{-1})^{ab} \big\langle \hE_a^\mu \hE_b^\nu \big\rangle
     &  \text{using definition \eqref{eq:chi2_times_def} of $\chi^2_\times$} \nn \\
      &= \frac{1}{N_p(N_p-1)} \sum_{\mu \ne \nu}  (C_{\mathrm{ana}}^{-1})^{ab} \bE_a \bE_b
      & \text{by \cref{eq:EE_patch}} \nn \\
      &= \underbrace{\bE_a \big( C_{\mathrm{ana}}^{-1} \big)^{ab} \bE_b}_{\mathrm{parity\ violation}}
      &
      \label{eq:chi2_times_ev}
\end{align}
As expected, we get the same parity violation term as in our previous calculation \eqref{eq:chi2_tot_ev} of $\langle \chi^2 \rangle$, but without the data--mock mismatch term. 
(As a check, we note that a similar calculation predicts $\langle \chi^2_\times \rangle_{\mathrm{mock}}=0$.
We checked that this is prediction is satisfied, within statistical errors due to the finite number of mocks.)

\begin{figure}
    \centerline{\includegraphics[width=8cm]{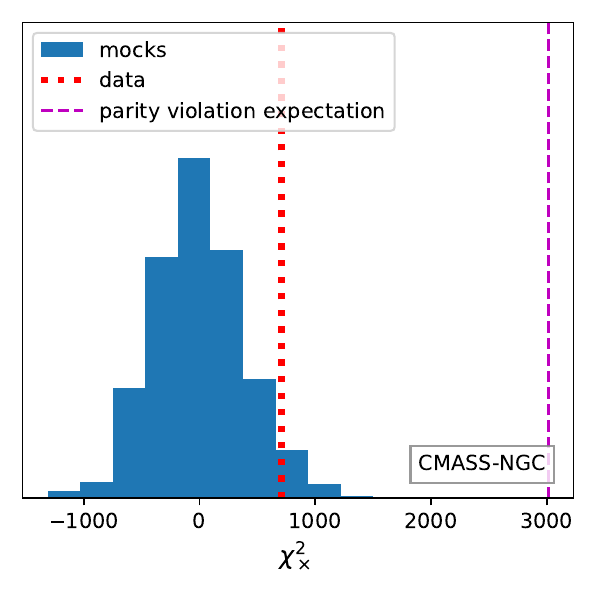} \includegraphics[width=8cm]{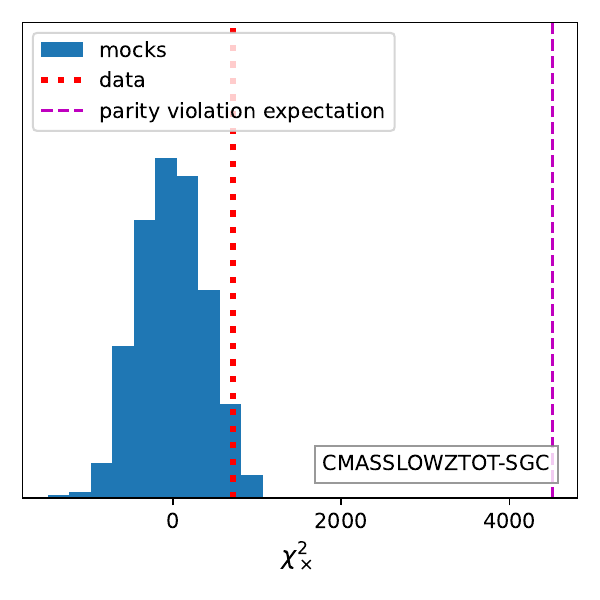}}
    \centerline{\includegraphics[width=8cm]{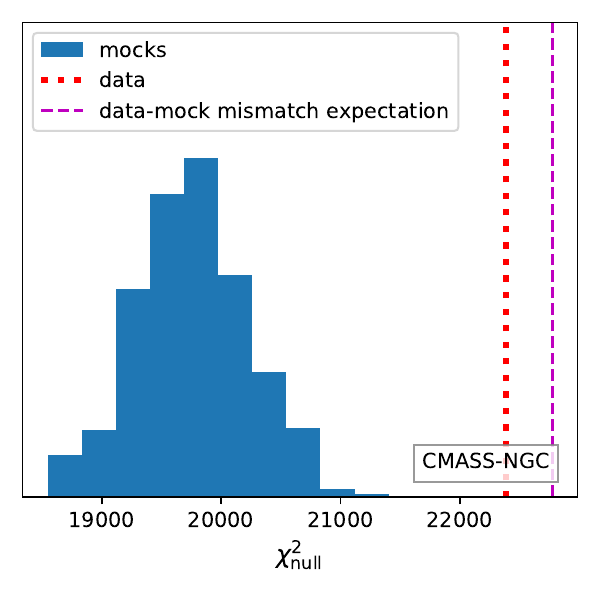} \includegraphics[width=8cm]{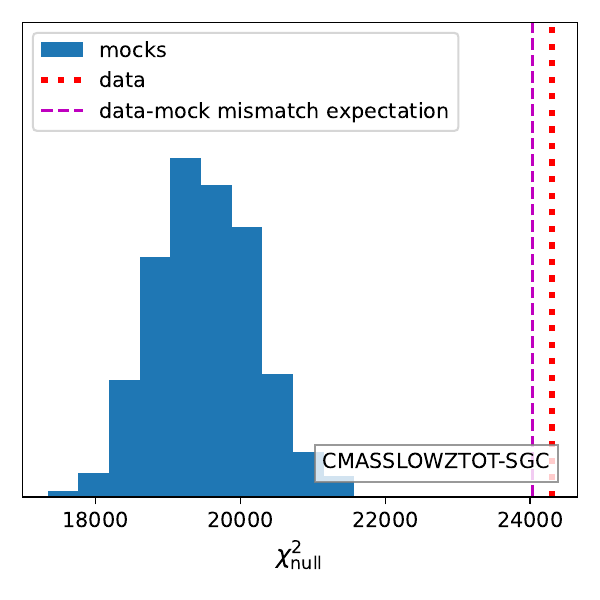}}
    \caption{%
        \textbf{\textit{Top row.}}
        The statistic $\chi^2_\times$ evaluated on mocks (histograms) and data (dotted lines). 
        The values of $\chi^2_\times$ are consistent with zero, as expected if the $\sim 7\sigma$ signal in \cref{fig:chi2} were due to data--mock mismatch.
        We also show (dashed lines) the ``parity violation expectation'', i.\,e.\ expected value of $\chi^2_\times$ if the $\sim 7\sigma$ signal were due to parity violation (\cref{eq:chi2_times_pv}).
        These values of $\chi^2_\times$ are ruled out at \SigNorthCrossTot and \SigSouthCrossTot for \cmassnorth (left) and \cmasssouth (right).
        \textbf{\textit{Bottom row.}}
        The statistic $\chi^2_{\mathrm{null}}$ evaluated on mocks (histogram) and data (dotted lines). 
        The data is inconsistent with mocks, indicating null test failure and evidence for data--mock mismatch.
        We also show (dashed lines) the ``data--mock mismatch expectation'', i.\,e.\ expected value of $\chi^2_{\mathrm{null}}$ if the $\sim 7\sigma$ signal in \cref{fig:chi2} were due to data--mock mismatch (\cref{eq:chi2_null_mm}). The data is consistent with these values, within statistical errors from the mocks.%
    }
    \label{fig:chi2_cross_null}
\end{figure}

In the top panels of \cref{fig:chi2_cross_null}, we evaluate $\chi^2_\times$ on \ac{BOSS} data.
The result is consistent with zero (at \SigNorthCross and \SigSouthCross, for \cmassnorth and \cmasssouth respectively), as would be expected if the $7\sigma$ excess $\chi^2$ (seen previously in \cref{fig:chi2}) were due to data--mock mismatch.
The purple lines in the top panels also show the value of $\chi^2_\times$ that would be expected if the previously-seen excess $\chi^2$ were due entirely to parity violation:
\begin{equation}
    \big( \text{Parity violation expectation for } \chi^2_\times \big)
     =  \chi^2_{\mathrm{data}} - \langle \chi^2 \rangle_{\mathrm{mock}}
      \label{eq:chi2_times_pv}
\end{equation}
These values of $\chi^2_\times$ are inconsistent with the data, at \SigNorthCrossTot and \SigSouthCrossTot for \cmassnorth and \cmasssouth respectively.
This shows that the $7\sigma$ excess $\chi^2$ (\cref{fig:chi2}) cannot be entirely (or even mostly) due to parity violation -- there must be contributions due to data--mock mismatch or systematics.
We emphasize that this conclusion only depends on assumptions 1 and 2 above, which are quite minimal.

One interesting detail: for both \cmassnorth and \cmasssouth, we have $\Var(\chi^2_\times) < \Var(\chi^2)$.
That is, the estimator $\chi^2_\times$ is both more robust than $\chi^2$ (since there is no bias from data--mock mismatch), and more statistically optimal.
This phenomenon surprised us initially, but after some exploration we concluded that it is a finite-volume effect (and the inequality would be reversed for a survey which is larger than \ac{BOSS}).
We explain this in detail in \cref{app:chi2_times}.

\subsection{The statistic \texorpdfstring{$\chi^2_{\mathrm{null}}$}{χ²-null}}
\label{ssec:chi2_null}

The $\chi^2_\times$ results from the previous section are sufficient to conclude that there is no compelling evidence for parity violation in \ac{BOSS}.
However, for the sake of completeness and pedagogy, we can also define a statistic $\chi^2_{\mathrm{null}}$ which is sensitive to data--mock mismatch but not parity violation:
\begin{equation}
    \chi^2_{\mathrm{null}} \equiv \frac{1}{\N} \sum_{\mu\ne\nu} \big( \hE_a^\mu - \hE_a^\nu \big) 
    (C_{\mathrm{ana}}^{-1})^{ab} \big( \hE_b^\mu - \hE_b^\nu \big) 
    \label{eq:chi2_null_def}
\end{equation}
    where the normalization $\N$ is defined by:
\begin{equation}
    \N \equiv 2(N_p-1) \sum_\mu \frac{V_{\mathrm{eff}}}{V_{\mathrm{eff}}^\mu}
    \label{eq:N_def}
\end{equation}
The full-survey effective volume $V_{\mathrm{eff}}$ and single-patch effective volume $V_{\mathrm{eff}}^\mu$ were defined in \cref{eq:Veff}.

In \cref{sec:intro}, we argued that $\chi^2_{\mathrm{null}}$ is sensitive to data--mock mismatch but not parity violation.
Intuitively, this is because $\chi^2_{\mathrm{null}}$ defines a null test: it measures consistency between the four-point function in different parts of the sky.
Parity violation (i.\,e.\ $\bE_a\ne 0$) does not contribute to $\chi^2_{\mathrm{null}}$, since we still expect consistent values of $\hE_a$ in different parts of the sky.

To show this formally, we compute $\langle \chi^2_{\mathrm{null}} \rangle_{\mathrm{data}}$ as follows:
\begin{align}
    \big\langle \chi^2_{\mathrm{null}} \big\rangle_{\mathrm{data}}
     &= \frac{1}{\N} \sum_{\mu\ne\nu} \Big\langle 
      \big( \hE_a^\mu - \hE_a^\nu \big) 
      (C_{\mathrm{ana}}^{-1})^{ab} \big( \hE_b^\mu - \hE_b^\nu \big) 
      \Big\rangle_{\mathrm{data}}
        & \text{using definition \eqref{eq:chi2_null_def} of $\chi^2_{\mathrm{null}}$} \nn \\
      &= \frac{1}{\N} \sum_{\mu\ne\nu}  (C_{\mathrm{ana}}^{-1})^{ab} 
        ( C_{\mathrm{data}}^\mu + C_{\mathrm{data}}^\nu )_{ab} 
        & \text{using \cref{eq:EE_patch}} \nn \\
      &= \frac{2(N_p - 1)}{\N} 
         \sum_\mu \Tr\big( C_{\mathrm{ana}}^{-1} C_{\mathrm{data}}^\mu \big)
\end{align}
Similarly, we have $\langle \chi^2_{\mathrm{null}} \rangle_{\mathrm{mock}} = 2 {\N}^{-1} (N_p-1) \sum_\mu \Tr( C_{\mathrm{ana}}^{-1} C_{\mathrm{mock}}^\mu)$, and therefore:
\begin{equation}
    \big\langle \chi^2_{\mathrm{null}} \big\rangle_{\mathrm{data}} 
      - \big\langle \chi^2_{\mathrm{null}} \big\rangle_{\mathrm{mock}}
     = \frac{2(N_p-1)}{\N} \sum_\mu \Tr\big[ 
       \big( C_{\mathrm{data}}^\mu - C_{\mathrm{mock}}^\mu \big)
       C_{\mathrm{ana}}^{-1} \big]
    \label{eq:chi2_null_ev1}
\end{equation}
which shows that $\chi^2_{\mathrm{null}}$ is sensitive to data--mock mismatch ($C_{\mathrm{data}}^\mu \ne C_{\mathrm{mock}}^\mu$) but not parity violation ($\bE_a \ne 0$).

In the bottom panels of \cref{fig:chi2_cross_null}, we evaluate $\chi^2_{\mathrm{null}}$ on \ac{BOSS}.
The values are not statistically consistent with mocks, indicating null test failure and unambiguous evidence for data--mock mismatch (or systematics), at \SigNorthNull (\SigSouthNull) for \cmassnorth (\cmasssouth).\footnote{%
    We also find that the per-patch $\chi^2$ statistic $\hE_a^\mu (C_{\mathrm{ana}}^{-1})^{ab} \hE_b^\mu$ fluctuates between patches: we get $4.0\sigma$ in \ac{NGC} patch 1, $1.8\sigma$ in \ac{NGC} patch 2, $4.4\sigma$ in \ac{NGC} patch 3, $2.2\sigma$ in \ac{SGC} patch 1, and $8.2\sigma$ in \ac{SGC} patch 2.
    This is a different null test than $\chi^2_{\mathrm{null}}$, since it compares \acp{8PCF} between patches rather than \acp{4PCF}.%
}

We emphasize that our results so far only depend on assumptions 1 and 2 (i.\,e.\ $\hE_a^\mu$ is unbiased, and patches $\mu\ne\nu$ are uncorrelated).
In order to assign a normalization to $\chi^2_{\mathrm{null}}$ and interpret its numerical value, we add a more technical assumption:
\begin{itemize}
    \item \textit{Assumption 3.}
    The single-patch covariance matrices $C_{\mathrm{data}}^\mu, C_{\mathrm{mock}}^\mu$ are approximately proportional to their full-survey counterparts $C_{\mathrm{data}}, C_{\mathrm{mock}}$.
\end{itemize}
This assumption implies the apparently stronger identity:
\begin{equation}
    C_{\mathrm{data}}^\mu \approx \frac{V_{\mathrm{eff}}}{V_{\mathrm{eff}}^\mu} C_{\mathrm{data}}
    \hspace{1.5cm}
    C_{\mathrm{mock}}^\mu \approx \frac{V_{\mathrm{eff}}}{V_{\mathrm{eff}}^\mu} C_{\mathrm{mock}}
    \label{eq:assumption3}
\end{equation}
by the following argument.
Suppose that $C_{\mathrm{data}}^\mu$ and $C_{\mathrm{data}}$ are proportional, i.\,e.\ $C_{\mathrm{data}}^\mu = A C_{\mathrm{data}}$.
Multiplying both sides by $C_{\mathrm{ana}}^{-1}$ and taking traces, we get:
\begin{equation}
    A = \frac{\Tr(C_{\mathrm{ana}}^{-1} C_{\mathrm{data}}^\mu)}{\Tr(C_{\mathrm{ana}}^{-1} C_{\mathrm{data}})}
    = \frac{V_{\mathrm{eff}}}{V_{\mathrm{eff}}^\mu}
\end{equation}
This proves the first half of \cref{eq:assumption3}.
The second half follows by the same argument (replacing $(\cdot)_{\mathrm{data}}$ by $(\cdot)_{\mathrm{mock}}$ everywhere).
Next, plugging \cref{eq:assumption3} into \cref{eq:chi2_null_ev1}, we get:
\begin{equation}
    \big\langle \chi^2_{\mathrm{null}} \big\rangle_{\mathrm{data}} 
      - \big\langle \chi^2_{\mathrm{null}} \big\rangle_{\mathrm{mock}}
     = \underbrace{\Tr\big[ \big( C_{\mathrm{data}} - C_{\mathrm{mock}} \big) C_{\mathrm{ana}}^{-1} \big]}_{\text{data--mock mismatch}}
\end{equation}
i.\,e.\ the data--mock mismatch term is the same as in our previous calculation \eqref{eq:chi2_tot_ev} of $\langle \chi^2 \rangle$, but without the parity violation term. 

Now we can answer the sharper question: is the numerical value of $\chi^2_{\mathrm{null}}$ consistent with the statement that the $\sim 7\sigma$ $\chi^2$ excess from \cref{fig:chi2} is entirely due to data--mock mismatch?
To answer this, we compare $\chi^2_{\mathrm{null}}$ with its ``data--mock mismatch expectation'':
\begin{equation}
    \big( \text{Data-mock mismatch expectation for } \chi^2_{\mathrm{null}} \big)
     = \big\langle \chi^2_{\mathrm{null}} \big\rangle_{\mathrm{mock}}
      + \underbrace{\big( \chi^2_{\mathrm{data}} - \langle \chi^2 \rangle_{\mathrm{mock}} \big)}_{\text{Excess $\chi^2$ from \cref{fig:chi2}}}
    \label{eq:chi2_null_mm}
\end{equation}
In the bottom panel of \cref{fig:chi2_cross_null}, we show that $\chi^2_{\mathrm{null}}$ is statistically consistent with the data--mock mismatch expectation defined by \cref{eq:chi2_null_mm}.
This is expected if the $\sim 7\sigma$ $\chi^2$ excess from \cref{fig:chi2} were entirely due to data--mock mismatch.

\begin{figure}
    \centerline{\includegraphics[width=8cm]{figures/north_2d.pdf} \includegraphics[width=8cm]{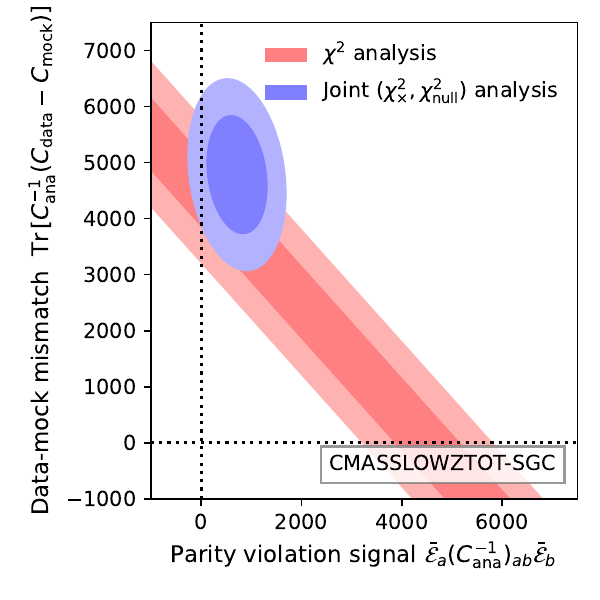}}
    \caption{%
        Summary plots combining previous results from \cref{fig:chi2,fig:chi2_cross_null}.
        The horizontal ($x$) and vertical ($y$) axes represent the parity violation signal $x = \hE_a (C_{\mathrm{ana}}^{-1})^{ab} \hE_b$, and the data--mock mismatch $y = \Tr[ ( C_{\mathrm{data}} - C_{\mathrm{mock}} ) C_{\mathrm{ana}}^{-1} ]$, respectively.
        The statistics $\chi^2$, $\chi^2_\times$, and $\chi^2_{\mathrm{null}}$ measure the parameter combinations $(x + y)$, $x$, and $y$.
        The red regions correspond to the $\chi^2$-analysis in \cref{fig:chi2} and are consistent with either parity violation or data--mock mismatch.
        The blue regions combine $\chi^2_\times$ and $\chi^2_{\mathrm{null}}$ results from \cref{fig:chi2_cross_null}, and show consistency with data--mock mismatch, with no statistically significant evidence for parity violation.
        Throughout this plot, statistical errors are assumed Gaussian, with covariance estimated from mock catalogs.
        Light/dark regions are \SI{68}{\percent} and \SI{95}{\percent} \ac{CL}.%
    }
    \label{fig:2d}
\end{figure}

Finally, in \cref{fig:2d}, we combine our previous results from \cref{fig:chi2,fig:chi2_cross_null} into two summary plots, for \cmassnorth and \cmasssouth.
Each statistic considered so far ($\chi^2$, $\chi^2_\times$, and $\chi^2_{\mathrm{null}}$) can be interpreted as measuring the parity violation signal $\hE_a (C_{\mathrm{ana}}^{-1})^{ab} \hE_b$ on the horizontal axis, and/or the data--mock mismatch $\Tr[ ( C_{\mathrm{data}} - C_{\mathrm{mock}} ) C_{\mathrm{ana}}^{-1} ]$ on the vertical axis.
An analysis based only on the $\chi^2$ statistic shows that the sum of the two signals is nonzero at $\sim 7\sigma$ (red regions).
A joint analysis based on the $\chi^2_\times$ and $\chi^2_{\mathrm{null}}$ statistics separates the two signals.
The joint analysis shows no statistically significant evidence for parity violation (blue regions).

\subsection{Combined NGC + SGC significance}
\label{ssec:combined_ngc_sgc}

In this section we address the question, ``what is the global significance of parity violation, combining \ac{NGC} and \ac{SGC}''?
In particular, since $\chi^2_\times$ is nonzero at $1.8\sigma$ in the \ac{NGC} and $1.7\sigma$ in the \ac{SGC} (\cref{ssec:chi2_cross}), one may wonder whether the global significance is $(1.8 + 1.7)/\sqrt{2} = 2.5\sigma$.

\begin{figure}
    \centerline{\includegraphics[width=12cm]{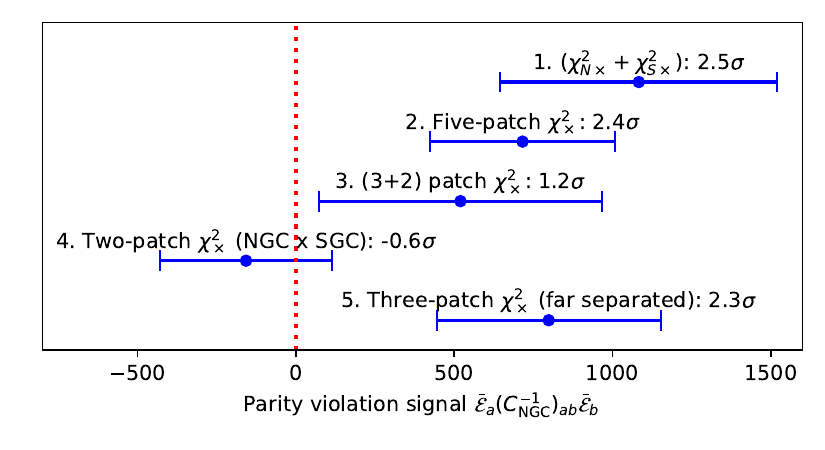}}
    \caption{Statistical significance of parity violation when combining the \ac{NGC} and \ac{SGC}, obtained from five different summary statistics (see descriptions in \cref{ssec:combined_ngc_sgc}).
    In order to combine all five statistics in the same plot, we normalize each statistic so that its expectation value is $\bE_a C_{\mathrm{NGC}}^{-1} \bE_b$, where $C_{\mathrm{NGC}}$ is the 18-bin \ac{NGC} analytic covariance (see \cref{tab:cov_ana_parameters}).}
    \label{fig:ngc_sgc}
\end{figure}

In order to assign global significance, we must choose a summary statistic.
We find that different, well-motivated summary statistics give different results.
In \cref{fig:ngc_sgc}, we compare five summary statistics as follows (from top to bottom):
\begin{enumerate}
    \item $(\chi^2_{\mathrm{N}\times} + \chi^2_{\mathrm{S}\times})$: This summary statistic is defined by simply adding the $\chi^2_\times$ statistics in the \ac{NGC} and \ac{SGC}.
    \label{item:ngc_sgc_combined_1}
    \item Five-patch $\chi^2_\times$: This statistic is defined by treating the three \ac{NGC} patches and the two \ac{SGC} patches (see \cref{fig:regions}) as a single survey with $N_p=5$, and then defining $\chi^2_\times$ as usual (\cref{eq:chi2_times_def}).
    \label{item:ngc_sgc_combined_2}
    \item (3+2)-patch $\chi^2_\times$: This statistic is defined similarly to $\chi^2_\times$, by summing over all $(3\times 2)$ pairs of the form (\ac{NGC} patch) $\times$ (\ac{SGC} patch).
    \label{item:ngc_sgc_combined_3}
    \item Two-patch $\chi^2_\times$ (\ac{NGC} $\times$ \ac{SGC}): this statistic is defined by treating the entire \ac{NGC} as a single patch, and the entire \ac{SGC} as a second patch, then defining $\chi^2_\times$ as usual (\cref{eq:chi2_times_def}) with $N_p=2$. (Statistics \#3 and \#4 are conceptually similar, but non-identical due to edge effects.)
    \label{item:ngc_sgc_combined_4}
    \item Three-patch $\chi^2_\times$ (far separated): This statistic is defined by taking the two most separated patches in the \ac{NGC} (see \cref{fig:regions}), and treating the entire \ac{SGC} as a third patch, then defining $\chi^2_\times$ as usual with $N_p = 3$.\footnote{%
        Since summary statistics 1--5 depend on a choice of $C_{\mathrm{ana}}$, there is a potential ambiguity since one may choose to use either $C_{\mathrm{ana}}^{\mathrm{NGC}}$ or $C_{\mathrm{ana}}^{\mathrm{SGC}}$. However, in the 18-bin case, our covariance matrices $C_{\mathrm{ana}}^{\mathrm{NGC}}$ and $C_{\mathrm{ana}}^{\mathrm{SGC}}$ are proportional, since the two surveys have the same $P_0$ but different $V_{\mathrm{fid}}$ (\cref{tab:cov_ana_parameters}). Therefore, the only ambiguity is the overall normalization of the summary statistic. We have chosen to normalize all five summary statistics to have expectation value $\bE_a C_{\mathrm{NGC}}^{-1} \bE_b$, so that they can be compared to each other in \cref{fig:ngc_sgc}.%
    }
    \label{item:ngc_sgc_combined_5}
\end{enumerate}
Throughout \cref{fig:ngc_sgc}, we assign error bars by evaluating the same summary statistic (e.\,g.\ $(\chi^2_{\mathrm{N}\times} + \chi^2_{\mathrm{S}\times})$ for the top error bar) on both data and mocks.
(In all cases, the mean value of the summary statistic in mocks is consistent with zero, as expected since the mocks are parity-invariant.)

Since summary statistics 1--5 are derived from the same catalogs, they are correlated, and there is more statistical tension in \cref{fig:ngc_sgc} than the error bars would suggest.
In particular, the discrepancy between summary statistic \#\ref{item:ngc_sgc_combined_4} (which has the smallest error bar), and summary statistic \#\ref{item:ngc_sgc_combined_2} (which is a close second) is $3.1\sigma$ relative to mocks.

It is unclear why there is significant tension in \cref{fig:ngc_sgc} relative to mocks.
It may be a symptom of undiagnosed systematics (especially systematics related to \ac{LSS} weights $w_{\mathrm{noz}}, w_{\mathrm{cp}}, w_{\mathrm{sys}}$, see discussion in \cref{ssec:computing_chi2}).
Another possible explanation of the tension in \cref{fig:ngc_sgc} is that error bars have been systematically underestimated, as we now explain in more detail.

The $\chi^2_\times$ statistic has been constructed so that data--mock mismatch does not contribute to $\langle \chi^2_\times \rangle$.
However, we still use mocks to estimate the \emph{variance} of $\chi^2_\times$.
Therefore, data--mock mismatch can still produce a small \emph{multiplicative} bias to $\Var(\chi^2_\times)$.
More quantitatively, it is plausible that we are underestimating error bars on $\chi^2_\times$ by $\sim \SI{20}{\percent}$ throughout the paper, since the results of this paper suggest that the mocks underestimate the \ac{4PCF} covariance $C_{ab}$ by \SI{20}{\percent} (\cref{sec:intro}).
This is much less serious than the original $\chi^2$ statistic, where data--mock mismatch can produce a large \emph{additive bias}, e.\,g.\ a null result could appear to be a $7\sigma$ excess in $\chi^2$.

Nevertheless, the issue of underestimating error bars complicates the interpretation of the \ac{NGC}+\ac{SGC} results in \cref{fig:ngc_sgc}.
The \ac{NGC}+\ac{SGC} significance varies between $(-0.6\sigma)$ and $2.5\sigma$, depending on which summary statistic is used, and the level of tension between summary statistics can be as large as $3.1\sigma$.
Given this level of discrepancy and tension, and the possibility of underestimating error bars, our interpretation of \cref{fig:ngc_sgc} is that there is not compelling evidence for parity violation when combining the \ac{NGC} and \ac{SGC}.

\section{Discussion}
\label{sec:discussion}

The main results of this paper can be summarized as follows:
\begin{itemize}
    \item The $\chi^2$ statistic introduced in \cite{hou_parity,oliver_parity} shows a $\sim 7\sigma$ excess in \ac{BOSS} (\cref{fig:chi2}).
    However, we show in \cref{eq:chi2_tot_ev} that
    $\langle \chi^2 \rangle$ contains two terms: a parity violation term, and a ``data--mock mismatch'' term.
    The data--mock mismatch term is nonzero if the \emph{parity-even} \ac{8PCF} of the mocks does not match the data (which leads to a biased estimate of the parity-odd \ac{4PCF} covariance).
    We define new statistics $\chi^2_\times$ and $\chi^2_{\mathrm{null}}$ (\cref{eq:chi2_times_def,eq:chi2_null_def}) which separate the two terms.

    \item The $\chi^2_\times$ statistic is an improved version of $\chi^2$ which is both \textbf{more robust} and \textbf{more statistically optimal}.
    It is more robust in the sense that $\chi^2_\times$ is sensitive only to parity violation, with no bias from data--mock mismatch (\cref{eq:chi2_times_ev}).
    It is more statistically optimal in the sense that $\Var(\chi^2_\times) \lesssim \Var(\chi^2)$ (due to a finite-volume effect, see \cref{app:chi2_times}).

    When $\chi^2_\times$ is applied to the \ac{NGC} or \ac{SGC} separately, the result is statistically consistent with zero (\cref{fig:chi2_cross_null}, top row).
    This \textbf{rules out parity violation} (at \SigNorthCrossTot and \SigSouthCrossTot for \cmassnorth and \cmasssouth) as the explanation for the $\sim 7\sigma$ excess $\chi^2$ from the previous bullet point.

    \item Conversely, the $\chi^2_{\mathrm{null}}$ statistic defines a null test which is not sensitive to parity violation, but is sensitive to data--mock mismatch (\cref{eq:chi2_null_ev1}).
    If we apply $\chi^2_{\mathrm{null}}$ to \ac{BOSS} data, we get a statistically significant excess $\chi^2_{\mathrm{null}}$ (\cref{fig:chi2_cross_null} bottom row), at \SigNorthNull (\SigSouthNull) for \cmassnorth (\cmasssouth).
    This is a \textbf{failing null test which implies data--mock mismatch (or systematics) with high statistical significance}.

    \item The previous results have made minimal assumptions: the per-patch estimator $\hE_a^\mu$ is unbiased, and patches $\mu\ne\nu$ are uncorrelated (``Assumptions 1 and 2'' in \cref{ssec:chi2_null}).
    Assumption 2 predicts that $\langle \chi^2_{\times} \rangle$ = 0; we have found this to be true on the mocks within the error expected from the finite number of mocks.
    If we add the technical assumption that single-patch covariances are proportional to full-survey covariances (``Assumption~3'' in \cref{ssec:chi2_null}), then we pin down the normalization of $\chi^2_{\mathrm{null}}$.

    When we do this (\cref{ssec:chi2_null} and \cref{fig:2d}), we find that the $\chi^2$ excess from the first bullet point is equal (within statistical errors) to the $\chi^2_{\mathrm{null}}$ excess (which is sensitive to data--mock mismatch but not parity violation).
    This \textbf{further supports the hypothesis that the $7\sigma$ $\chi^2$ excess is due to data--mock mismatch.}

    \item Most results in the paper treat the \ac{NGC} and \ac{SGC} separately.
    In \cref{ssec:combined_ngc_sgc} we consider the combined \ac{NGC}+\ac{SGC} statistical significance.
    We find that different, well-motivated summary statistics give statistical significances between $(-0.6\sigma)$ and $2.5\sigma$, and the statistical tension between summary statistics (relative to mocks) can be as high as $3.1\sigma$.

    Given the level of discrepancy and tension between summary statistics, our interpretation is that there is \textbf{not compelling evidence for parity violation when combining the \ac{NGC} and \ac{SGC}.}
    We speculate that the tension is due to either undiagnosed systematics, or underestimating error bars (\cref{ssec:combined_ngc_sgc}).
\end{itemize}
These results are consistent with the results of \cite{hou_parity,oliver_parity} being due to data--mock mismatch, rather than parity violation.
This does not imply that there are parity-odd systematics in \ac{BOSS} -- it would suffice to have a $\sim \SI{20}{\percent}$ difference between the parity-even \ac{8PCF} of the mocks and data.
Such a difference is plausible since the mocks are primarily intended to model the large-scale \ac{2PCF} \cite{Kitaura:2015uqa,Rodriguez-Torres:2015vqa}.

One issue throughout our analysis is that we still use mocks to estimate the \emph{variance} of our new statistic $\chi^2_\times$.
In \cref{ssec:combined_ngc_sgc} we argue that error bars on $\chi^2_\times$ could plausibly be underestimated by $\sim \SI{20}{\percent}$ throughout the paper.
This complicates the interpretation of results with borderline statistical significance (say $2$--$2.5\sigma$), such as the combined \ac{NGC}+\ac{SGC} results in \cref{ssec:combined_ngc_sgc}.
In future work, it would be interesting to find a better way to assign error bars, e.\,g.\ removing dependence on mocks entirely by estimating $\Var(\chi^2_\times)$ directly from data using a jackknife procedure (see \cite{Adari:2024vkf} for some interesting recent suggestions along these lines).

An extended technical comment on patches.
Since we have defined $\chi^2_\times$, $\chi^2_{\mathrm{null}}$ by splitting the survey into patches, our results will depend on the number and choice of patches.
If needed, we speculate that $\chi^2_\times$ could be defined in a ``patch-free'' way, which we sketch as follows.
We ``promote'' the estimator $\hE_a$ to a 3D field $\smash{\hE_a(\bx)}$ by removing the outer integral $\int d^3\bx \cdots$ in (\cref{eq:Edef}):
\begin{equation}
    \hE_{l_1l_2l_3}^{\beta_1\beta_2\beta_3}(\bx)
     \equiv (-1)^{\sum l_i} \sum_{m_1m_2m_3} 
      \left( \begin{array}{ccc} l_1 & l_2 & l_3 \\ m_1 & m_2 & m_3 \end{array} \right)
      \, \delta_{\mathrm{g}}(\bx) \left( \prod_{i=1}^3 \int d^3\br_i \, W^{\beta_i}_{l_im_i}(\br_i) \, \delta_{\mathrm{g}}(\bx+\br_i) \right)
\end{equation}
and we ``promote'' the estimator $\chi^2$ to a correlation function $\chi^2(\br)$:
\begin{equation}
    \chi^2(\br) \equiv \frac{1}{V_{\mathrm{fid}}} \int d^3\bx\, \hE_a(\bx) \,
     (C^{-1}_{\mathrm{ana}})^{ab} \, \hE_b(\bx+\br)
\end{equation}
In this notation, the original $\chi^2$ statistic is $V_{\mathrm{fid}}^{-1} \int d^3\br \, \chi^2(\br)$.
We can define an alternate statistic $\chi^2_{\mathrm{far}}$ by restricting the integral to large separations (say $|\br| > \SI{100}{Mpc}$).
This is roughly equivalent to splitting the survey into a large number of small patches, and constructing a $\chi^2_\times$-type statistic by cross-correlating patch pairs whose separation is large.
Therefore, $\chi^2_{\mathrm{far}}$ should be sensitive to parity violation but not data--mock mismatch (like $\chi^2_\times$, but without choosing patches explicitly).
We defer exploration of $\chi^2_{\mathrm{far}}$ to future work.
(Adapting the edge correction procedure from \cite{hou_parity,oliver_parity} to $\chi^2_{\mathrm{far}}$ may be nontrivial.)

Despite this null result for parity violation in \ac{BOSS}, it will be interesting to search for parity violation in future datasets such as \ac{DESI} \cite{DESI:2024mwx}.
The $\chi^2$ statistic from \cite{hou_parity,oliver_parity} has the advantage of being template-free (i.\,e.\ no specific model for the parity-odd \ac{4PCF} is assumed), but the disadvantage of being biased by data--mock mismatch.
Therefore, any ``detection'' of parity violation based on $\chi^2$ alone seems likely to remain inconclusive -- parity violation and data--mock mismatch would both be plausible explanations.
In this paper, we have constructed new statistics $\chi^2_\times$, $\chi^2_{\mathrm{null}}$ which separate parity violation from data--mock mismatch, leading to an unambiguous conclusion.

\acknowledgments

AK, SM and KMS have contributed equally to this work and are listed alphabetically in the author list.
We thank the authors of \cite{hou_parity,oliver_parity} for making so much of their code publicly available, and for giving many useful comments on the draft.
We particularly thank Jiamin Hou for sharing the analytic covariance code for the parity-odd four-point function.
AK was supported as a CITA National Fellow by the Natural Sciences and Engineering Research Council of Canada (NSERC), funding reference \#DIS-2022-568580.
SM acknowledges support by the National Science Foundation under Grant No.\ 2108931.
KMS was supported by an NSERC Discovery Grant, by the Daniel Family Foundation, and by the Centre for the Universe at Perimeter Institute.
Research at Perimeter Institute is supported by the Government of Canada through Industry Canada and by the Province of Ontario through the Ministry of Research \& Innovation.

\bibliographystyle{JHEP}
\bibliography{parity_sdss_paper.bib}

\appendix

\section{Deriving the analytic covariance}
\label{app:analytic_covariance}

Recall from \cref{ssec:analytic_covariance} that the ``analytic'' covariance matrix $(C_{\mathrm{ana}})_{ab} = \Cov(\hE_a, \hE_b)$ is the estimator covariance under the approximations that $\delta_{\mathrm{g}}(\bx)$ is a Gaussian field, and the survey geometry is a 3D periodic box with fiducial box volume $V_{\mathrm{fid}}$.
In this appendix, we will derive a closed-form expression for the analytic covariance.
This was first derived in \cite{hou_covariance}, building on \ac{3PCF} results from \cite{Slepian:2015qza}.

Our final expression for $C_{\mathrm{ana}}$ agrees with \cite{hou_covariance}, but we have included a detailed derivation in this appendix for a few reasons.
First, to make the paper self-contained.
Second, because we do the calculation in a different way from \cite{hou_parity}, and an alternate approach may be pedagogically useful.
Third, in order to comment on some details of the numerical implementation.

We checked that the output of our $C_{\mathrm{ana}}$ code agrees with with the output of the \verb|npcf_cov| code from \cite{hou_covariance,hou_parity}, and with the public covariance matrix from \cite{oliver_parity}.
This is an independent cross-check on results from \cite{hou_parity,oliver_parity,hou_covariance}, and also shows that the analysis in our paper is consistent with these previous studies.

\subsection{Special function identities}

Throughout this appendix, $C^{l_1l_2l_3}_{m_1m_2m_3}$ is compressed notation for the Wigner 3j symbol:
\begin{equation}
    C^{l_1l_2l_3}_{m_1m_2m_3} \equiv 
     \left( \begin{array}{ccc} l_1 & l_2 & l_3 \\ m_1 & m_2 & m_3 \end{array} \right)
    \label{eq:C_def}
\end{equation}
We also define the quantity $G_{l_1l_2l_3}$ by:
\begin{equation}
    G_{l_1l_2l_3} \equiv \sqrt{\frac{(2l_1+1)(2l_2+1)(2l_3+1)}{4\pi}} 
      \left( \begin{array}{ccc}
        l_1 & l_2 & l_3 \\ 0 & 0 & 0
      \end{array} \right)
    \label{eq:G_def}
\end{equation}
Using this notation, the integral of three spherical harmonics is given by:
\begin{equation}
    \int d^2\hr \, Y_{l_1m_1}(\hr) \, Y_{l_2m_2}(\hr) \, Y_{l_3m_3}(\hr)
     = G_{l_1l_2l_3} C^{l_1l_2l_3}_{m_1m_2m_3}
    \label{eq:3y_integral}
\end{equation}
An equivalent identity expands the product of two spherical harmonics as a sum of spherical harmonics:
\begin{equation}
    Y_{lm}(\hr) Y_{l'm'}(\hr) 
      = \sum_{LM} G_{ll'L} C^{ll'L}_{mm'M} Y_{LM}^*(\hr)
    \label{eq:expand_yy}
\end{equation}
In \cref{ssec:TU}, we will need the following angular integral involving four spherical harmonics:
\begin{align}
    & \int d^2\hr \, Y_{l_1m_1}^*(\hr) Y_{l'_1m'_1}^*(\hr) Y_{L_2M_2}(\hr) Y_{L_3M_3}(\hr) \nn \\
     & \hspace{2cm} = \int d^2\hr
      \left( \sum_{L_1M_1} G_{l_1l'_1L_1} C^{l_1l'_1L_1}_{m_1m'_1M_1} Y_{L_1M_1}(\hr) \right)
      Y_{L_2M_2}(\hr) Y_{L_3M_3}(\hr) \nn \\
     & \hspace{2cm} = \sum_{L_1M_1} G_{l_1l'_1L_1} G_{L_1L_2L_3} 
       C^{l_1l_1'L_1}_{m_1m_1'M_1} C^{L_1L_2L_3}_{M_1M_2M_3}
    \label{eq:4y_integral}
\end{align}
where we have used \cref{eq:expand_yy} in the first line, and \cref{eq:3y_integral} in the second line.

The following identity is one way of defining the Wigner 9j symbol:
\begin{align}
      \left\{ \begin{array}{ccc}
       l_{11} & l_{12} & l_{13} \\
       l_{21} & l_{22} & l_{23} \\
       l_{31} & l_{32} & l_{33}
      \end{array} \right\}
    &=
     \sum_{m_{ij}}
     \left( \begin{array}{ccc}
      l_{11} & l_{12} & l_{13} \\
      m_{11} & m_{12} & m_{13} 
     \end{array} \right)
     \left( \begin{array}{ccc}
      l_{21} & l_{22} & l_{23} \\
      m_{21} & m_{22} & m_{23} 
     \end{array} \right)
      \left( \begin{array}{ccc}
      l_{31} & l_{32} & l_{33} \\
      m_{31} & m_{32} & m_{33} 
     \end{array} \right) \nn \\
    & \hspace{1cm} \times
     \left( \begin{array}{ccc}
      l_{11} & l_{21} & l_{31} \\
      m_{11} & m_{21} & m_{31} 
     \end{array} \right)
     \left( \begin{array}{ccc}
      l_{12} & l_{22} & l_{32} \\
      m_{12} & m_{22} & m_{32} 
     \end{array} \right)
     \left( \begin{array}{ccc}
      l_{13} & l_{23} & l_{33} \\
      m_{13} & m_{23} & m_{33} 
     \end{array} \right)
    \label{eq:9j_def}
\end{align}

\subsection{Strategy of the calculation}
\label{ssec:strategy}

Restricting to the parity-odd case ($\sum l_i$ odd), we rewrite the estimator \eqref{eq:Edef} as:
\begin{equation}
    \hE_{l_1l_2l_3}^{\beta_1\beta_2\beta_3}
     = -\frac{1}{V_{\mathrm{fid}}} \sum_{m_i} C^{l_1l_2l_3}_{m_1m_2m_3} \int d^3\bx \, \delta_{\mathrm{g}}(\bx) \prod_{i=1}^3 \delta^{\beta_i}_{l_im_i}(\bx)
    \label{eq:estimator_main}
\end{equation}
where we have defined the field:
\begin{equation}
    \delta^\beta_{lm}(\bx) \equiv \int d^3\br \, W_{lm}^\beta(\br) \, \delta_{\mathrm{g}}(\bx+\br)
    \label{eq:delta_blm_def}
\end{equation}
We write the covariance as (defining $\int_{\bx\bx'} = \int d^3\bx \, d^3\bx'\,$):
\begin{align}
    \Cov(\hE_{l_1l_2l_3}^{\beta_1\beta_2\beta_3}, \hE_{l_1'l_2'l_3'}^{\beta_1'\beta_2'\beta_3'*})
     &= -\frac{1}{V_{\mathrm{fid}}^2} \sum_{m_im'_j}
        C^{l_1l_2l_3}_{m_1m_2m_3} C^{l_1'l_2'l_3'}_{m_1'm_2'm_3'} \nn \\
     & \hspace{1cm} \times
        \int_{\bx\bx'} \,
        \bigg\langle 
          \bigg( \delta_{\mathrm{g}}(\bx) \prod_{i=1}^3 \delta^{\beta_i}_{l_im_i}(\bx) \bigg)
          \bigg( \delta_{\mathrm{g}}(\bx') \prod_{j=1}^3 \delta^{\beta'_j}_{l'_jm'_j}(\bx') \bigg)
        \bigg\rangle
    \label{eq:covariance_main}
\end{align}
In this expression, the expectation value $\langle \cdots \rangle$ is an eight-point function in the field $\delta_{\mathrm{g}}$.
Since we are assuming $\delta_{\mathrm{g}}$ is a Gaussian field, we can use Wick's theorem to write the eight-point function as a sum of 24 contractions.
We define $T$ and $U$ to be the following contractions:
\begin{align}    T^{\beta_1\beta_2\beta_3\beta'_1\beta'_2\beta'_3}_{l_1l_2l_3l_1'l_2'l_3'}
     &= -\frac{1}{V_{\mathrm{fid}}^2} \sum_{m_im'_j}
        C^{l_1l_2l_3}_{m_1m_2m_3} C^{l_1'l_2'l_3'}_{m_1'm_2'm_3'}  \nn \\
     & \times \int_{\bx\bx'} \,
     \wick{2345}{\Big( <1 \delta_{\mathrm{g}}(\bx) \,
        <2 \delta^{\beta_1}_{l_1m_1}(\bx) 
        <3 \delta^{\beta_2}_{l_2m_2}(\bx) 
        <4 \delta^{\beta_3}_{l_3m_3}(\bx) \Big)
         \Big( >1 \delta_{\mathrm{g}}(\bx') \,
        >2 \delta^{\beta'_1}_{l'_1m'_1}(\bx') 
        >3 \delta^{\beta'_2}_{l'_2m'_2}(\bx') 
        >4 \delta^{\beta'_3}_{l'_3m'_3}(\bx') \Big)} \label{eq:T_def} \\
U^{\beta_1\beta_2\beta_3\beta'_1\beta'_2\beta'_3}_{l_1l_2l_3l_1'l_2'l_3'}
     &= -\frac{1}{V_{\mathrm{fid}}^2} \sum_{m_im'_j}
        C^{l_1l_2l_3}_{m_1m_2m_3} C^{l_1'l_2'l_3'}_{m_1'm_2'm_3'}  \nn \\
     & \times \int_{\bx\bx'} \,
    \wick{3245}{\Big( <1 \delta_{\mathrm{g}}(\bx) \,
        <2 \delta^{\beta_1}_{l_1m_1}(\bx) 
        <3 \delta^{\beta_2}_{l_2m_2}(\bx) 
        <4 \delta^{\beta_3}_{l_3m_3}(\bx) \Big)
         \Big( >2 \delta_{\mathrm{g}}(\bx') \,
        >1 \delta^{\beta'_1}_{l'_1m'_1}(\bx') 
        >3 \delta^{\beta'_2}_{l'_2m'_2}(\bx') 
        >4 \delta^{\beta'_3}_{l'_3m'_3}(\bx') \Big)}
    \label{eq:U_def}
\end{align}
Additionally, there are 5 terms obtained by permuting indices in $T$, and 17 terms obtained by permuting indices in $U$.
These terms have some nontrivial signs.
To see this, consider the contraction $\bar T$ obtained from $T$ by exchanging the roles of $\delta^{\beta_1}_{l_1m_1}(\bx)$ and $\delta^{\beta_2}_{l_2m_2}(\bx)$:
\begin{align}
    \bar T^{\beta_1\beta_2\beta_3\beta'_1\beta'_2\beta'_3}_{l_1l_2l_3l_1'l_2'l_3'}
     &= -\frac{1}{V_{\mathrm{fid}}^2} \sum_{m_im'_j}
        C^{l_1l_2l_3}_{m_1m_2m_3} C^{l_1'l_2'l_3'}_{m_1'm_2'm_3'}  \nn \\
     & \times \int_{\bx\bx'} \,
     \wick{2345}{\Big( <1 \delta_{\mathrm{g}}(\bx) \,
        <3 \delta^{\beta_1}_{l_1m_1}(\bx) 
        <2 \delta^{\beta_2}_{l_2m_2}(\bx) 
        <4 \delta^{\beta_3}_{l_3m_3}(\bx) \Big)
         \Big( >1 \delta_{\mathrm{g}}(\bx') \,
        >2 \delta^{\beta'_1}_{l'_1m'_1}(\bx') 
        >3 \delta^{\beta'_2}_{l'_2m'_2}(\bx') 
        >4 \delta^{\beta'_3}_{l'_3m'_3}(\bx') \Big)}
\end{align}
Comparing to the definition \eqref{eq:T_def} of $T$, the two contractions are related by the index permutation $(l_1,\beta_1) \leftrightarrow (l_2,\beta_2)$, but we also get an extra minus sign, since $C^{l_2l_1l_3}_{m_2m_1m_3} = (-1)^{\sum l_i} C^{l_1l_2l_3}_{m_1m_2m_3}$, and $\sum l_i$ is odd.
Therefore:
\begin{equation}
    \bar T^{\beta_1\beta_2\beta_3\beta'_1\beta'_2\beta'_3}_{l_1l_2l_3l'_1l'_2l'_3}
      = -T^{\beta_2\beta_1\beta_3\beta'_1\beta'_2\beta'_3}_{l_2l_1l_3l'_1l'_2l'_3}
\end{equation}
Applying similar logic to all 24 contractions, the total covariance can be written as:
\begin{tcolorbox}[ams align]
    \Cov(\hE_{l_1l_2l_3}^{\beta_1\beta_2\beta_3}, \hE_{l_1'l_2'l_3'}^{\beta_1'\beta_2'\beta_3'*})
     &= \Big(
    T^{\beta_1\beta_2\beta_3\beta_1'\beta_2'\beta_3'}_{l_1l_2l_3l'_1l'_2l'_3}
        + U^{\beta_1\beta_2\beta_3\beta_1'\beta_2'\beta_3'}_{l_1l_2l_3l'_1l'_2l'_3}
        + U^{\beta_1\beta_2\beta_3\beta_2'\beta_3'\beta_1'}_{l_1l_2l_3l'_2l'_3l'_1}
        + U^{\beta_1\beta_2\beta_3\beta_3'\beta_1'\beta_2'}_{l_1l_2l_3l'_3l'_1l'_2} 
        \Big) \nn \\
     & \hspace{1cm}
          + \Big( \text{2 even permutations of } (\beta_1,l_1),\ 
       (\beta_2,l_2),\ (\beta_3,l_3) \Big) \nn \\
     & \hspace{1cm}
          - \Big( \text{3 odd permutations of } (\beta_1,l_1),\ 
       (\beta_2,l_2),\ (\beta_3,l_3) \Big) 
    \label{eq:cov_final}
\end{tcolorbox}
\par\noindent
It remains to compute $T, U$.
In \cref{ssec:two_point}, we will compute the following two-point functions, which are needed to evaluate contractions:
\begin{equation}
    \big\langle \delta_{\mathrm{g}}(\bx) \delta_{\mathrm{g}}(\bx') \big\rangle
      \hspace{1.5cm}
    \big\langle \delta_{\mathrm{g}}(\bx) \delta^{\beta'}_{l'm'}(\bx') \big\rangle
      \hspace{1.5cm}
    \big\langle \delta^\beta_{lm}(\bx) \delta^{\beta'}_{l'm'}(\bx') \big\rangle
    \label{eq:two_point_functions}
\end{equation}
Then we will compute $T,U$ in \cref{ssec:TU}.

\subsection{Computing the two-point functions in \texorpdfstring{\cref{eq:two_point_functions}}{eq. (\ref{eq:two_point_functions})}}
\label{ssec:two_point}

The first two-point function in \cref{eq:two_point_functions} is the usual galaxy correlation function $\xi(r) = \big\langle \delta_{\mathrm{g}}(\bx) \, \delta_{\mathrm{g}}(\bx + \br) \big\rangle$.
This is straightforward to compute in Fourier space:
\begin{tcolorbox}[ams align]
    \xi(r) &= \int \frac{d^3\bk}{(2\pi)^3} P_{\mathrm{g}}(k) e^{i\bk\cdot\br} \nn \\
     &= \int_0^\infty dk \, \frac{k^2}{2\pi^2} P_{\mathrm{g}}(k) j_0(kr)
    \label{eq:2pf1}
\end{tcolorbox}
\noindent
The remaining two-point functions in \cref{eq:two_point_functions} can be calculated similarly, but the details are a bit more complicated, since the angular integrals contains spherical harmonics and are nontrivial.
First, we note that $\delta^\beta_{lm}$ can be written as a convolution.
Starting from the definition \eqref{eq:delta_blm_def}:
\begin{align}
    \delta^\beta_{lm}(\bx) &= \int d^3\br \, W_{lm}^\beta(\br) \, \delta_{\mathrm{g}}(\bx+\br) \nn \\
     &= (-1)^l \int d^3\br \, W_{lm}^\beta(-\br) \, \delta_{\mathrm{g}}(\bx+\br) 
      & \text{since } W_{lm}^\beta(\br) = (-1)^l W_{lm}^\beta(-\br) \nn \\
     &= (-1)^l \underbrace{ W_{lm}^\beta \star \delta_{\mathrm{g}}}_{\mathrm{convolution}}
\end{align}
Therefore, $\delta_{lm}^\beta(\bk)$ is given in Fourier space by multiplication:
\begin{equation}
    \delta_{lm}^\beta(\bk) = (-1)^l \, \tW_{lm}^\beta(\bk) \, \delta_{\mathrm{g}}(\bk)
\end{equation}
where $\tW_{lm}^\beta(\bk)$ is the Fourier transform of the function $W^\beta_{lm}(\br)$ defined in \cref{eq:Wr_def} above.
We compute $\tW_{lm}^\beta(\bk)$ as follows:
\begin{align}
    \tW^\beta_{lm}(\bk) 
      &= \int d^3\br\, W^\beta_{lm}(\br) e^{-i\bk\cdot\br} \nn \\
      &= \frac{4\pi}{V_\beta} \int_{r\in\beta} d^3\br\,
       Y_{lm}^*(\hr) \,
       \left[ \sum_{LM} 4\pi (-i)^L j_L(kr)
         Y_{LM}(\hr) Y_{LM}^*(\hk) \right] \nn \\
      &= \frac{4\pi}{V_\beta} \int_{r\in\beta} dr \, 4\pi r^2  (-i)^l j_l(kr) Y_{lm}^*(\bk) \nn \\
      &= 4\pi (-i)^l B^\beta_l(k) Y^*_{lm}(\hk)
\end{align}
where we define the ``binned'' Bessel function $B^\beta_l(k)$ by bin-averaging $j_l(kr)$:
\begin{tcolorbox}[ams equation]
    B^\beta_l(k) = \frac{\int_{r\in\beta} dr \, r^2 j_l(kr)}{\int_{r\in\beta} dr \, r^2}
    \label{eq:B_def}
\end{tcolorbox}
\noindent
See \cref{ssec:binned_bessel} for comments on numerical evaluation of $B_l^\beta(k)$.

Returning to the two-point functions in \cref{eq:two_point_functions}, we can now compute $\langle \delta_{\mathrm{g}}(\bx) \, \delta^{\beta'}_{l'm'}(\bx+\br) \rangle$ as follows:
\begin{align}
    \big\langle \delta_{\mathrm{g}}(\bx) \, \delta^{\beta'}_{l'm'}(\bx+\br) \big\rangle
      &= \left\langle \left( \int \frac{d^3\bk}{(2\pi)^3} \, \delta_{\mathrm{g}}(\bk) e^{i\bk\cdot\bx} \right)
      \left( (-1)^{l'} \int \frac{d^3\bk'}{(2\pi)^3} \, \tW^{\beta'}_{l'm'}(\bk') \, \delta_{\mathrm{g}}(\bk') \, e^{i\bk'\cdot(\bx+\br)} \right) 
      \right\rangle \nn \\
      &= (-1)^{l'} \int \frac{d^3\bk}{(2\pi)^3} \, P_{\mathrm{g}}(k) \, \tW^{\beta'}_{l'm'}(-\bk) \, e^{-i\bk\cdot\br} \nn \\
      &= (-1)^{l'} \int \frac{d^3\bk}{(2\pi)^3} \, P_{\mathrm{g}}(k)
      \bigg( 4\pi (+i)^{l'} B^{\beta'}_{l'}(k) Y^*_{l'm'}(\hk) \bigg) \nn \\
      & \hspace{2.5cm} \times
      \bigg( \sum_{LM} 4\pi (-i)^L j_L(kr) Y_{LM}(\hk) Y_{LM}^*(\hr) \bigg) \nn \\
      &= 4\pi (-1)^{l'} H^{\beta'}_{l'}(k) \, Y_{l'm'}^*(\hr) \label{eq:2pf2}
\end{align}
where the radial function $H_l^\beta(r)$ is defined by:
\begin{tcolorbox}[ams equation]
    H^{\beta}_{l}(r) \equiv \int_0^\infty dk \, \frac{k^2}{2\pi^2} 
           P_{\mathrm{g}}(k) B_{l}^{\beta}(k) j_{l}(kr)
    \label{eq:H_def}
\end{tcolorbox}
\noindent
Similarly, we can compute the third two-point function in \cref{eq:two_point_functions}:
\begin{align}
    \big\langle \delta^\beta_{lm}(\bx) \, \delta^{\beta'}_{l'm'}(\bx+\br) \big\rangle
      &= (-1)^{l+l'} \int \frac{d^3\bk}{(2\pi)^3} \, P_{\mathrm{g}}(k) \, 
       \tW^\beta_{lm}(\bk) \tW^{\beta'}_{l'm'}(-\bk) e^{-i\bk\cdot\br} \nn \\
      &= (-1)^{l+l'} \int \frac{d^3\bk}{(2\pi)^3} \, P_{\mathrm{g}}(k)
      \bigg( 4\pi (-i)^l B^\beta_l(k) Y_{lm}^*(\hk) \bigg) \nn \\
      & \hspace{2.5cm} \times
      \bigg( 4\pi (+i)^{l'} B^{\beta'}_{l'}(k) Y_{l'm'}^*(\hk) \bigg) \nn \\
      & \hspace{2.5cm} \times
      \bigg( \sum_{LM} 4\pi (-i)^L j_L(kr) Y_{LM}^*(\hk) Y_{LM}(\hr) \bigg) 
      \nn \\
    &= (4\pi)^2 \sum_{LM} i^{l-l'-L} 
       \left( \int \frac{k^2\, dk}{2\pi^2} \, P_{\mathrm{g}}(k) B^\beta_l(k) B^{\beta'}_{l'}(k) j_L(kr) \right) \nn \\
    & \hspace{2.5cm} \times 
       \left( \int d^2\hk \, Y_{lm}^*(\hk) Y_{l'm'}^*(\hk) Y_{LM}^*(\hk) \right) 
       Y_{LM}(\hr) \nn \\
    &= (4\pi)^2 \sum_{LM} i^{l-l'-L} F^{\beta\beta'}_{ll'L}(r) 
         G_{ll'L} C^{ll'L}_{mm'M} Y_{LM}(\hr)
    \label{eq:2pf3}
\end{align}
where the radial function $F^{\beta\beta'}_{ll'L}(r)$ is defined by:
\begin{tcolorbox}[ams align]
    F^{\beta\beta'}_{ll'L}(r) &\equiv \int_0^\infty dk \, \frac{k^2}{2\pi^2} \,
     P_{\mathrm{g}}(k) B^\beta_l(k) B^{\beta'}_{l'}(k) j_L(kr)
    \label{eq:F_def}
\end{tcolorbox}

\subsection{Computing \texorpdfstring{$T^{\beta_1\beta_2\beta_3\beta'_1\beta'_2\beta'_3}_{l_1l_2l_3l'_1l'_2l'_3}$}{T(β₁β₂β₃β₁′β₂′β₃′, l₁l₂l₃l₁′l₂′l₃′)} and \texorpdfstring{$U^{\beta_1\beta_2\beta_3\beta'_1\beta'_2\beta'_3}_{l_1l_2l_3l'_1l'_2l'_3}$}{U(β₁β₂β₃β₁′β₂′β₃′, l₁l₂l₃l₁′l₂′l₃′)}}
\label{ssec:TU}

Starting from the definition \eqref{eq:T_def} of $T$, we write the Wick contractions as two-point functions:
\begin{align}
    T^{\beta_1\beta_2\beta_3\beta_1'\beta_2'\beta_3'}_{l_1l_2l_3l'_1l'_2l'_3}
     &= -\frac{1}{V_{\mathrm{fid}}^2} \sum_{m_im_j'} 
       C^{l_1l_2l_3}_{m_1m_2m_3} C^{l_1'l_2'l_3'}_{m_1'm_2'm_3'}  
     \int_{\bx\bx'}
      \big\langle \delta_{\mathrm{g}}(\bx) \delta_{\mathrm{g}}(\bx') \big\rangle \, \prod_{k=1}^3
      \big\langle \delta^{\beta_k}_{l_km_k}(\bx) \delta^{\beta_k'}_{l_k'm_k'}(\bx') \big\rangle
\end{align}
Plugging in the two-point functions in \cref{eq:2pf1,eq:2pf3}, we get:
\begin{align}
    T^{\beta_1\beta_2\beta_3\beta_1'\beta_2'\beta_3'}_{l_1l_2l_3l'_1l'_2l'_3}
    &=  -\frac{(4\pi)^6}{V_{\mathrm{fid}}} \sum_{m_im_j'} 
       C^{l_1l_2l_3}_{m_1m_2m_3} C^{l_1'l_2'l_3'}_{m_1'm_2'm_3'} 
       \nn \\
    & \hspace{1cm} \times \int d^3\br \, \xi(r)
         \prod_{k=1}^3 \bigg( \sum_{L_kM_k} 
           i^{l_k-l'_k-L_k} \,
           F_{l_kl'_kL_k}^{\beta_k\beta'_k}(r) \,
           G_{l_kl'_kL_k} \,
           C^{l_kl'_kL_k}_{m_km'_kM_k} \,
           Y_{L_kM_k}(\hr) \bigg)
    \label{eq:T_intermediate}
\end{align}
We simplify the following subexpression, obtained by bringing the angular integral $(\int d^2\hr)$ and all 9 $m$-sums to the inside:
\begin{align}
    & \sum_{m_im'_j} C^{l_1l_2l_3}_{m_1m_2m_3} C^{l_1'l_2'l_3'}_{m_1'm_2'm_3'} 
      \int d^2\hr \prod_{k=1}^3 \left( 
        \sum_{M_k} C^{l_kl'_kL_k}_{m_km'_kM_k} Y_{L_kM_k}(\hr) \right) \nn \\
    & \hspace{1cm} = 
      G_{L_1L_2L_3} \sum_{m_im'_jM_k}
         C^{l_1l_2l_3}_{m_1m_2m_3} C^{l_1'l_2'l_3'}_{m_1'm_2'm_3'} 
         C^{l_1l_1'L_1}_{m_1m_1'M_1}
         C^{l_2l_2'L_2}_{m_2m_2'M_2}
         C^{l_3l_3'L_3}_{m_3m_3'M_3}
         C^{L_1L_2L_3}_{M_1M_2M_3} \nn \\
    & \hspace{1cm} =
     G_{L_1L_2L_3}
      \left\{ \begin{array}{ccc}
       l_1 & l_2 & l_3 \\
       l'_1 & l'_2 & l'_3 \\
       L_1 & L_2 & L_3
      \end{array} \right\}
    \label{eq:T_subexpression}
\end{align}
In the first line we used \cref{eq:3y_integral}, and in the second line we used definition \eqref{eq:9j_def} of the 9j symbol.
Plugging \cref{eq:T_subexpression} into \cref{eq:T_intermediate}, we get our final expression for $T$:
\begin{tcolorbox}[ams align]
    T^{\beta_1\beta_2\beta_3\beta_1'\beta_2'\beta_3'}_{l_1l_2l_3l'_1l'_2l'_3}
     &= \frac{(4\pi)^6}{V_{\mathrm{fid}}} \sum_{L_1L_2L_3}
        (-i)^{\sum_k(l_k+l'_k+L_k)}
        G_{l_1l_1'L_1}
        G_{l_2l_2'L_2}
        G_{l_3l_3'L_3}
        G_{L_1L_2L_3}
      \left\{ \begin{array}{ccc}
       l_1 & l_2 & l_3 \\
       l'_1 & l'_2 & l'_3 \\
       L_1 & L_2 & L_3
      \end{array} \right\} \nn \\
      & \hspace{1cm} \times \int_0^\infty dr \, r^2 \xi(r) \,
         F^{\beta_1\beta_1'}_{l_1l_1'L_1}(r) \,
         F^{\beta_2\beta_2'}_{l_2l_2'L_2}(r) \,
         F^{\beta_3\beta_3'}_{l_3l_3'L_3}(r)
    \label{eq:T_final}
\end{tcolorbox}

Similarly, starting from the definition \eqref{eq:U_def} of $U$, we write the Wick contractions as two-point functions:
\begin{align}
    U^{\beta_1\beta_2\beta_3\beta_1'\beta_2'\beta_3'}_{l_1l_2l_3l'_1l'_2l'_3}
     &= -\frac{1}{V_{\mathrm{fid}}^2} \sum_{m_im_j'} 
       C^{l_1l_2l_3}_{m_1m_2m_3} C^{l_1'l_2'l_3'}_{m_1'm_2'm_3'} \nn \\
     & \hspace{0.5cm} \times
     \int_{\bx\bx'}
      \big\langle \delta_{\mathrm{g}}(\bx) \delta^{\beta_1'}_{l_1'm_1'}(\bx') \big\rangle \,
      \big\langle \delta^{\beta_1}_{l_1m_1}(\bx) \delta_{\mathrm{g}}(\bx') \big\rangle 
\prod_{k=2}^3
      \big\langle \delta^{\beta_k}_{l_km_k}(\bx) 
           \delta^{\beta'_k}_{l_k'm_k'}(\bx') \big\rangle 
\end{align}
Plugging in the two-point functions in \cref{eq:2pf2,eq:2pf3}, this becomes:
\begin{align}
    U^{\beta_1\beta_2\beta_3\beta_1'\beta_2'\beta_3'}_{l_1l_2l_3l'_1l'_2l'_3}
    &= - \frac{(4\pi)^6}{V_{\mathrm{fid}}} \sum_{m_im_j'} 
       C^{l_1l_2l_3}_{m_1m_2m_3} C^{l_1'l_2'l_3'}_{m_1'm_2'm_3'} \nn \\
    & \hspace{1cm} \times
       \int d^3\br \,
         \Big( (-1)^{l_1'} H^{\beta_1'}_{l_1'}(r) Y_{l_1'm_1'}^*(\hr) \Big)
         \Big( H^{\beta_1}_{l_1}(r) Y_{l_1m_1}^*(\hr) \Big) \nn \\
    & \hspace{1cm} \times
        \prod_{k=2}^3 \bigg( \sum_{L_kM_k} 
           i^{l_k-l'_k-L_k} \, F_{l_kl'_kL_k}^{\beta_k\beta'_k}(r) \,
           G_{l_kl'_kL_k} \, C^{l_kl'_kL_k}_{m_km'_kM_k} \,
           Y_{L_kM_k}(\hr) \bigg)
    \label{eq:U_intermediate}
\end{align}
We simplify the following subexpression, obtained by bringing the angular integral $(\int d^2\hr)$ and all eight $m$-sums to the inside:
\begin{align}
    & \sum_{m_im'_j} C^{l_1l_2l_3}_{m_1m_2m_3} C^{l_1'l_2'l_3'}_{m_1'm_2'm_3'} 
      \int d^2\hr \, Y_{l_1m_1}^*(\hr) Y_{l_1'm_1'}^*(\hr) 
       \prod_{k=2}^3 \left( 
        \sum_{M_k} C^{l_kl'_kL_k}_{m_km'_kM_k} Y_{L_kM_k}(\hr) \right) \nn \\
    & \hspace{1cm}
     = \sum_{m_im'_j} \sum_{M_2M_3} 
       C^{l_1l_2l_3}_{m_1m_2m_3} C^{l_1'l_2'l_3'}_{m_1'm_2'm_3'} 
       C^{l_2l_2'L_2}_{m_2m_2'M_2} C^{l_3l_3'L_3}_{m_3m_3'M_3} \nn \\
    & \hspace{2cm} \times
     \sum_{L_1M_1} G_{l_1l'_1L_1} G_{L_1L_2L_3} 
        C^{l_1l_1'L_1}_{m_1m'_1M_1} C^{L_1L_2L_3}_{M_1M_2M_3}
       \nn \\
     & \hspace{1cm} = \sum_{L_1} G_{l_1l'_1L_1} G_{L_1L_2L_3}
       \left\{ \begin{array}{ccc}
       l_1 & l_2 & l_3 \\
       l'_1 & l'_2 & l'_3 \\
       L_1 & L_2 & L_3
      \end{array} \right\}
    \label{eq:U_subexpression}
\end{align}
In the first line, we used \cref{eq:expand_yy}, and in the second line we used the definition \eqref{eq:9j_def} of the 9j symbol.
Plugging \cref{eq:U_subexpression} back into \cref{eq:U_intermediate}, we get our final expression for $U$:
\begin{tcolorbox}[ams align]
    U^{\beta_1\beta_2\beta_3\beta_1'\beta_2'\beta_3'}_{l_1l_2l_3l'_1l'_2l'_3}
     &= \frac{(4\pi)^6}{V_{\mathrm{fid}}} \! \sum_{L_1L_2L_3} i^{l_2+l_2'-L_2} i^{l_3+l_3'-L_3}
       G_{l_1l'_1L_1} G_{l_2l_2'L_2} G_{l_3l_3'L_3} G_{L_1L_2L_3} 
    \begin{Bmatrix}
       l_1 & l_2 & l_3 \\
       l'_1 & l'_2 & l'_3 \\
       L_1 & L_2 & L_3
    \end{Bmatrix}
      \nn \\
     & \hspace{1cm} \times
      \int_0^\infty dr \, r^2 \, 
       H^{\beta_1}_{l_1}(r) \,
       H^{\beta'_1}_{l'_1}(r) \,
        F^{\beta_2\beta_2'}_{l_2l_2'L_2}(r) \,
        F^{\beta_3\beta_3'}_{l_3l_3'L_3}(r)
    \label{eq:U_final}
\end{tcolorbox}

\subsection{Numerical evaluation of the binned Bessel function \texorpdfstring{$B_l^\beta(k)$}{B(k; β, l)}}
\label{ssec:binned_bessel}

In \cref{eq:B_def}, we defined the binned Bessel function $B^\beta_l(k)$.
Numerical computation of $B^\beta_l(k)$ is nontrivial, and deserves some discussion.
First, we write $B_l^\beta(k)$ as:
\begin{tcolorbox}[ams equation]
    B_l^\beta(k) = 3 \, \frac{ A_l(kR_{\mathrm{max}}) - A_l(kR_{\mathrm{min}}) }{ (kR_{\mathrm{max}})^3 - (kR_{\mathrm{min}})^3}
    \hspace{1.5cm} \text{where } \beta = [R_{\mathrm{min}}, R_{\mathrm{max}}]
\end{tcolorbox}
\noindent
where $A_l(x)$ is the antiderivative of $x^2 j_l(x)$:
\begin{equation}
    A_l(x) \equiv \int_0^x dx' \, x'{}^2 \, j_l(x')
\end{equation}
To compute $A_l(x)$, we use different approaches for small and large $x$. For $x \gtrsim 1$, we use the exact expressions:
\begin{align}
    A_0(x) &= \sin(x) - x \cos(x) \nn \\
    A_1(x) &= 2(1-\cos(x)) - x\sin(x) \nn \\
    A_2(x) &= x\cos(x) - 4 \sin(x) + 3 \, \mathrm{Si}(x) \nn \\
    A_3(x) &= x\sin(x) + 7 \cos(x) - 15 \, \frac{\sin(x)}{x} + 8 \nn \\
    A_4(x) &= -x\cos(x) + 11\sin(x) + \frac{105}{2} \frac{\cos(x)}{x} 
      - \frac{105}{2} \frac{\sin(x)}{x^2} + \frac{15}{2} \mathrm{Si}(x)
    \label{eq:Al_analytic}
\end{align}
where the special function $\mathrm{Si}(x) \equiv \int_0^x dt \, (\sin(t))/t$ can be computed using standard libraries.

For $x \lesssim 1$, the exact expressions in \cref{eq:Al_analytic} are numerically unstable, so to improve numerical accuracy, our code can switch to the series expansion for small arguments:
\begin{equation}
    A_l(x) = \sum_{m=0}^\infty 
     \frac{(-1)^m}{2^m \, m! \, (2m+2l+1)!!} 
      \, \frac{x^{2m+l+3}}{2m+l+3}
    \label{eq:Al_series}
\end{equation}
where $(2n+1)!! \equiv (2n+1)(2n-1)\dots(3)(1)$.
However, we have found that this change does not meaningfully impact the final statistical significances when computing $\chi^2$, $\chi^2_\times$, and $\chi^2_{\mathrm{null}}$.

\subsection{Convergence tests for the analytic covariance matrix}
\label{app:convergence_tests}

As explained in \cref{ssec:analytic_covariance}, the analytic covariance $C_{\mathrm{ana}}$ depends on integration parameters ($k_{\mathrm{max}}$, $N_k$, $r_{\mathrm{max}}$, $N_r$).
Our fiducial choices for these parameters are taken from \cite{hou_covariance}: $k_{\mathrm{max}} = \SI{5}{\hHubble\per\Mpc}$, $N_k = 5000$, $r_{\mathrm{max}} = \SI{1000}{\per\hHubble\Mpc}$, $N_r = 4100$. 
In this section, we will show that these fiducial parameters are chosen conservatively enough that this choice does not affect the rest of the analysis, i.\,e.\ the integrals producing $C_{\mathrm{ana}}$ are converged. 

In order to compare covariance matrices $C_{\mathrm{ana}}$ with different choices of integration parameters, we define the following distance function $J(C_1,C_2)$ between \emph{symmetric positive definite} matrices $C_1,C_2$:
\begin{align}
    J(C_1,C_2) &= \Tr \left( \frac{1}{4} C_1 C_2^{-1} + \frac{1}{4} C_2 C_1^{-1} - \frac{1}{2} I \right) \nn \\
     &= \frac{1}{4} \Tr \Big[ \Delta^2(1+\Delta)^{-1} \Big] 
       \hspace{1.5cm} \text{where } \Delta \equiv C_2^{-1}(C_1-C_2)
    \label{eq:covariance_metric}
\end{align}
where the second line is more numerically stable as $C_1 \rightarrow C_2$.
To motivate this choice of distance function, we state some key properties of $J$ (with proofs omitted for brevity):
\begin{itemize}
    \item $J(C_1,C_2) \ge 0$, with $J(C_1,C_2) = 0$ if and only if $C_1 = C_2$.
    \item Basis independence: if $C_i' = A C_i A^T$ where $A$ is invertible, then $J(C_1',C_2') = J(C_1,C_2)$.
    \item Statistical interpretation: $J(C_1,C_2)$ is the symmetrized KL-divergence $(\mathrm{KL}(\rho_1,\rho_2) + \mathrm{KL}(\rho_2,\rho_1))/2$, where $\rho_i(x) \equiv \mathrm{Det}(2\pi C_i)^{-1/2} \exp(-x^T C_i^{-1} x/2)$ is a multivariate Gaussian \ac{PDF} with covariance $C_i$.
    Thus, $J(C_1,C_2)$ quantifies statistical distinguishability of the \acp{PDF} $\rho_1,\rho_2$: they can be distinguished with $N$ samples if and only if $J(C_1,C_2) \ll (1/N)$.
\end{itemize}
To test convergence of $C_{\mathrm{ana}}$ with respect to an integration parameter (say $k_{\mathrm{max}}$), we compute $C_{\mathrm{ana}}$ twice using different parameter values (say fiducial $k_{\mathrm{max}}^{\mathrm{fid}} = \SI{5}{\hHubble\per\Mpc}$ and ``high'' $k_{\mathrm{max}} = \SI{10}{\hHubble\per\Mpc}$), and verify that the two covariance matrices $C,C'$ satisfy $J(C,C')\ll 1$.
This test shows that changing the parameter from its fiducial value to a conservative value produces a nearly indistinguishable covariance matrix.
This test effectively examines the entire covariance matrix, and is more powerful than testing a few specific matrix entries for convergence.

We apply this convergence test systematically, one integration parameter at a time:
\begin{align}
     J(C_{\mathrm{ana}}^{\mathrm{fid}}, C_{\mathrm{ana}}^{\text{high $k_{\mathrm{max}}$}})
     &= \num{7.3058e-13}
     & \text{where high $k_{\mathrm{max}} = \SI{10}{\hHubble\per\Mpc}$}
     \nn \\
    J(C_{\mathrm{ana}}^{\mathrm{fid}}, C_{\mathrm{ana}}^{\text{high $N_k$})} 
      &= \num{2.7735e-13}
      & \text{where high $N_k = 10000$} \nn \\
    J(C_{\mathrm{ana}}^{\mathrm{fid}}, C_{\mathrm{ana}}^{\text{low $r_{\mathrm{max}}$}})
      &= \num{4.9609e-16}
      & \text{where low $r_{\mathrm{max}} = \SI{500}{\hHubble \: Mpc}$} \nn \\
    J(C_{\mathrm{ana}}^{\mathrm{fid}}, C_{\mathrm{ana}}^{\text{high $N_r$}}) 
      &= \num{1.1658e-9}
      & \text{where high $N_r = 8200$}
\label{eq:convergence_tests}
\end{align}
Since all $J$-values are small, we conclude that our fiducial values of ($k_{\mathrm{max}}$, $N_k$, $r_{\mathrm{max}}$, $N_r$) have been chosen conservatively enough to converge.
Note that when we vary $k_{\mathrm{max}}$, we leave the Gaussian smoothing scale (see \cref{eq:pg_dampening}) fixed at $k_0 = \SI{1}{\hHubble\per\Mpc}$.

\section{Why is \texorpdfstring{$\Var(\chi^2_\times)$}{Var(χ²-×)} smaller than \texorpdfstring{$\Var(\chi^2)$}{Var(χ²)}?}
\label{app:chi2_times}

Our new statistic $\chi^2_\times$ is a more robust version of $\chi^2$.
Empirically, we found that $\chi^2_\times$ is also more statistically optimal than $\chi^2$, in the sense that $\Var(\chi^2_\times) < \Var(\chi^2)$.
We initially found this counterintuitive, since $\chi^2_\times$ uses less sky area than $\chi^2$ (due to gaps between patches), and also throws away information from autocorrelating $\hE_a$ in the same patch.
In this appendix, we will show that this is a non-Gaussian finite-volume effect, i.\,e.\ the inequality $\Var(\chi^2_\times) < \Var(\chi^2)$ would be reversed for a larger survey.

\subsection{A one-parameter model for \texorpdfstring{$\Var(\chi^2)$}{Var(χ²)}}

In this appendix, we use the term ``region'' to mean either the \cmassnorth survey, the \cmasssouth survey, or a patch (in the sense of \cref{ssec:patches}) comprising one of these surveys.
(A total of 7 regions can be defined in this way.)

For any such region, we can define the statistic $\chi^2 = \hE_a (C^{-1}_{\mathrm{ana}})^{ab} \hE_b$.
Recall from \cref{eq:Veff} that we can also define the effective volume $V_{\mathrm{eff}}$ of the region by:
\begin{equation}
    V_{\mathrm{eff}} \equiv \frac{V_{\mathrm{fid}} \, N_{\mathrm{dof}}}{\Tr(C_{\mathrm{ana}}^{-1} C_{\mathrm{mock}})}
    \label{eq:var_chi2_prev_defs}
\end{equation}
Values of $V_{\mathrm{eff}}$ for all 7 regions were given previously in \cref{tab:Veff}.

\begin{figure}
    \centering
    \includegraphics[width=10cm]{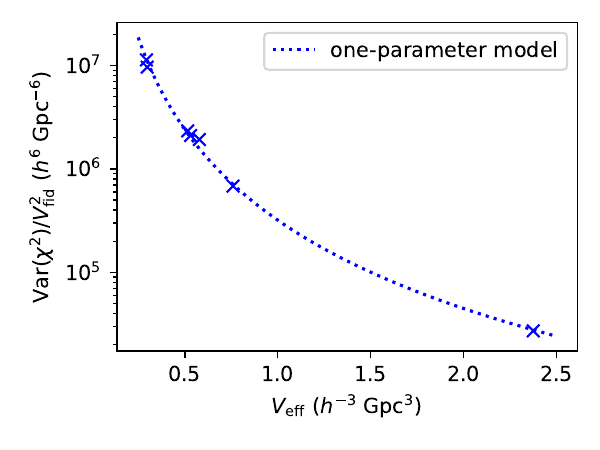}
    \caption{Each cross represents values of $(V_{\mathrm{eff}}, \Var(\chi^2))$ from one region in \cref{tab:Veff}. The dashed line is the one-parameter model in \cref{eq:var_chi2}.}
    \label{fig:veff_variance}
\end{figure}

In \cref{fig:veff_variance} we plot $V_{\mathrm{eff}}$ and the variance $\Var(\chi^2)$ for each region.
The purpose of this section is to describe an analytic model (represented by the dotted curve in \cref{fig:veff_variance}), with a single free parameter $V_{\mathrm{thresh}}$, which does a good job of fitting all 7 points.

Let $T_{abcd}$ be the \emph{connected} four-point function of the estimator $\hE_a$:
\begin{align}
    T_{abcd} &\equiv 
     \big\langle \hE_a \, \hE_b \, \hE_c \, \hE_d \big\rangle^{\mathrm{conn}}_{\mathrm{mock}} \nn \\
     &= \big\langle \hE_a \, \hE_b \, \hE_c \, \hE_d \big\rangle_{\mathrm{mock}}
       - C^{\mathrm{mock}}_{ab} C^{\mathrm{mock}}_{cd} 
       - C^{\mathrm{mock}}_{ac} C^{\mathrm{mock}}_{bd}
       - C^{\mathrm{mock}}_{ad} C^{\mathrm{mock}}_{bc}
\end{align}
Note that $\hE_a$ is a four-point function in the underlying galaxy field $\delta_{\mathrm{g}}$, $C_{\mathrm{mock}}$ is an eight-point function in $\delta_{\mathrm{g}}$, and $T_{abcd}$ is a 16-point(!) function in $\delta_{\mathrm{g}}$.

A short calculation shows that when we compute $\Var(\chi^2)$, we get two terms:
\begin{align}
    \Var(\chi^2) &= \Var\big[ \hE_a (C^{-1}_{\mathrm{ana}})^{ab} \hE_b \big] \nn \\
     &= \underbrace{2 \, \Tr\big[ C_{\mathrm{ana}}^{-1} \, C_{\mathrm{mock}} \, C_{\mathrm{ana}}^{-1} \, C_{\mathrm{mock}} \big]}_{\textrm{ ``Gaussian'' variance}}
      + \underbrace{(C^{-1}_{\mathrm{ana}})_{ab} \, (C^{-1}_{\mathrm{ana}})_{cd} \, T^{abcd}}_{\textrm{ ``non-Gaussian'' variance}}
    \label{eq:var_chi2_1}
\end{align}
We will model these two terms separately.
To model the Gaussian variance, we assume that $C_{\mathrm{mock}}$ is approximately proportional to $C_{\mathrm{ana}}$, which implies:
\begin{equation}
    \Tr\big[ C_{\mathrm{ana}}^{-1} \, C_{\mathrm{mock}} \, C_{\mathrm{ana}}^{-1} \, C_{\mathrm{mock}} \big]
     \approx \frac{1}{N_{\mathrm{dof}}} \Tr\big[ C_{\mathrm{ana}}^{-1} \, C_{\mathrm{mock}} \big]^2
     = N_{\mathrm{dof}} \frac{V_{\mathrm{fid}}^2}{V_{\mathrm{eff}}^2}
    \label{eq:vmodel_g}
\end{equation}
using the definition \eqref{eq:var_chi2_prev_defs} of $V_{\mathrm{eff}}$.
We model the non-Gaussian variance by simply assuming that it is approximately proportional to $1/V_{\mathrm{eff}}^3$ (as expected by mode counting).
We parameterize the proportionality constant as:
\begin{equation}
    (C^{-1}_{\mathrm{ana}})_{ab} \, (C^{-1}_{\mathrm{ana}})_{cd} \, T^{abcd}
     \approx 2 N_{\mathrm{dof}}  \, \frac{V_{\mathrm{thresh}} V_{\mathrm{fid}}^2}{V_{\mathrm{eff}}^3}
    \label{eq:vmodel_ng}
\end{equation}
where $V_{\mathrm{thresh}}$ is a free parameter of our model, with units of volume.

Combining \cref{eq:vmodel_g,eq:vmodel_ng}, our ``bottom-line'' one-parameter model for $\Var(\chi^2)$ is:
\begin{equation}
    \frac{\Var(\chi^2)}{V_{\mathrm{fid}}^2} \approx \frac{2 N_{\mathrm{dof}}}{V_{\mathrm{eff}}^2} \left( 1 + \frac{V_{\mathrm{thresh}}}{V_{\mathrm{eff}}} \right)
    \label{eq:var_chi2}
\end{equation}
We determine the parameter $V_{\mathrm{thresh}}$ by fitting to the points in \cref{fig:veff_variance}, obtaining:
\begin{equation}
    V_{\mathrm{thresh}} = \SI{6.87}{\per\hHubble\cubed\Gpc\cubed}
\end{equation}
Note that all regions in \cref{fig:veff_variance} satisfy $V_{\mathrm{eff}} \lesssim V_{\mathrm{thresh}}$, which implies that $\Var(\chi^2)$ is dominated by the non-Gaussian term in \cref{eq:var_chi2_1}.

\subsection{A zero-parameter model for \texorpdfstring{$\Var(\chi^2_\times)$}{Var(χ²-×)}}

In this section, we will develop an analytic model for $\Var(\chi^2_\times)$, with no free parameters.
Consider a survey which is divided into $N_p$ patches.
We denote the effective volumes of the survey and the patches by $V_{\mathrm{eff}}^{\mathrm{tot}}, V_{\mathrm{eff}}^\mu$.
Following \cref{ssec:chi2_cross}, we assume that for distinct patches $\mu\ne\nu$, the estimators $\hE_a^\mu$ and $\hE_b^\nu$ are uncorrelated.
Then, starting from the definition \eqref{eq:chi2_times_def} of $\chi^2_\times$, a short calculation gives $\Var(\chi^2_\times)$:
\begin{equation}
    \Var(\chi^2_\times) = \frac{2}{N_p^2(N_p-1)^2} 
     \sum_{\mu \ne \nu} \Tr\big[ C_{\mathrm{ana}}^{-1} \, C_{\mathrm{mock}}^\mu \, C_{\mathrm{ana}}^{-1} \, C_{\mathrm{mock}}^\nu \big]
    \label{eq:var_chi2x_1}
\end{equation}
To model the trace on the right-hand side, we assume that $C_{\mathrm{mock}}^\mu$ and $C_{\mathrm{mock}}^\nu$ are approximately proportional to $C_{\mathrm{ana}}$, which implies:
\begin{equation}
    \Tr\big[ C_{\mathrm{ana}}^{-1} \, C_{\mathrm{mock}}^\mu \, C_{\mathrm{ana}}^{-1} \, C_{\mathrm{mock}}^\nu \big]
     \approx N_{\mathrm{dof}} \frac{V_{\mathrm{fid}}^2}{V_{\mathrm{eff}}^\mu \, V_{\mathrm{eff}}^\nu}
    \label{eq:vmodel_x}
\end{equation}
by the same reasoning as in \cref{eq:vmodel_g}.
It will be convenient to define a ``covering fraction'' $f_{\mathrm{cov}}$ by:
\begin{equation}
    f_{\mathrm{cov}} \equiv
    \left( \frac{1}{N_p^3 (N_p-1)} \sum_{\mu\ne\nu} \frac{(V_{\mathrm{eff}}^{\mathrm{tot}})^2}{V_{\mathrm{eff}}^\mu \, V_{\mathrm{eff}}^\nu} \right)^{-1/2}
    \label{eq:fcov_def}
\end{equation}
To motivate this definition, note that if $V_{\mathrm{eff}}^\mu = f V_{\mathrm{eff}}^{\mathrm{tot}}/N_p$, as intuitively expected for equally sized patches which collectively cover fraction $0 < f \le 1$ of the survey area, then $f_{\mathrm{cov}}= f$.
In practice, the value of $f_{\mathrm{cov}}$ will be less than the geometric expectation. 
We find $f_{\mathrm{cov}} = 0.68$ (0.73) for \cmassnorth (\cmasssouth), whereas the geometric expectation is 0.78 (0.88). 

Combining \crefrange{eq:var_chi2x_1}{eq:fcov_def}, we get our final ``bottom-line'' prediction for $\Var(\chi^2_\times)$:
\begin{equation}
    \frac{\Var(\chi^2_\times)}{V_{\mathrm{fid}}^2} \approx 
    \frac{2N_{\mathrm{dof}}}{V_{\mathrm{eff}}^2}
    \left( \frac{N_p}{f_{\mathrm{cov}}^2 (N_p-1)} \right)
    \label{eq:var_chi2x}
\end{equation}
This model for $\Var(\chi^2_\times)$ has no free parameters.
This is because $\Var(\chi^2_\times)$ only has a Gaussian term (\cref{eq:var_chi2x_1}), whereas $\Var(\chi^2)$ has Gaussian and non-Gaussian terms (\cref{eq:var_chi2_1}).
When we compare the model with measurements of $\Var(\chi^2_\times)$ from mocks, it agrees within $\sim \SI{5}{\percent}$ for both surveys (\cmassnorth, \cmasssouth).

\subsection{Comparison between \texorpdfstring{$\Var(\chi^2)$}{Var(χ²)} and \texorpdfstring{$\Var(\chi^2_\times)$}{Var(χ²-×)}}

Now that we have models for $\Var(\chi^2)$ and $\Var(\chi^2_\times)$, we can compare them.
Combining \cref{eq:var_chi2,eq:var_chi2x} we get:
\begin{equation}
    \frac{\Var(\chi^2)}{\Var(\chi^2_\times)} = 
     f_{\mathrm{cov}}^2 \frac{N_p-1}{N_p} 
     \left( 1 + \frac{V_{\mathrm{thresh}}}{V_{\mathrm{eff}}^{\mathrm{tot}}} \right)
\end{equation}
We see that in the large-volume limit $V_{\mathrm{eff}}^{\mathrm{tot}} \gg V_{\mathrm{thresh}}$, we get the inequality $\Var(\chi^2) < \Var(\chi^2_\times)$, as intuitively expected.
However, this inequality reverses in the small-volume limit $V_{\mathrm{eff}}^{\mathrm{tot}} \ll V_{\mathrm{thresh}}$, where the non-Gaussian term in $\Var(\chi^2)$ dominates.

The \ac{BOSS} surveys in this paper are ``small'' surveys, in the sense that $V_{\mathrm{eff}} \lesssim V_{\mathrm{thresh}}$, so we have found the counterintuitive inequality $\Var(\chi^2_\times) < \Var(\chi^2)$ throughout the paper.
The analysis in this appendix has shown that this is a finite-volume effect, i.\,e.\ the inequality would be reversed for a larger survey.

\section{Data and mock investigations}
\label{app:data_mock_investigations}

In an early stage of this paper, we looked for systematic inhomogeneities in the \ac{BOSS} data and mocks which might explain a parity-odd \ac{4PCF}.
We found a previously unknown systematic effect in the PATCHY mocks (\cref{fig:dr_mock}).
We ultimately concluded that the effect was too small to explain the excess $\chi^2$, but in this appendix we describe the systematic effect that we found.

\begin{figure}
    \centerline{%
        \textbf{0.43 < z < 0.70}\hspace{4.8cm}\textbf{0.43 < z < 0.50}%
    } 
    \medskip
    \centerline{%
        \includegraphics[scale=0.35]{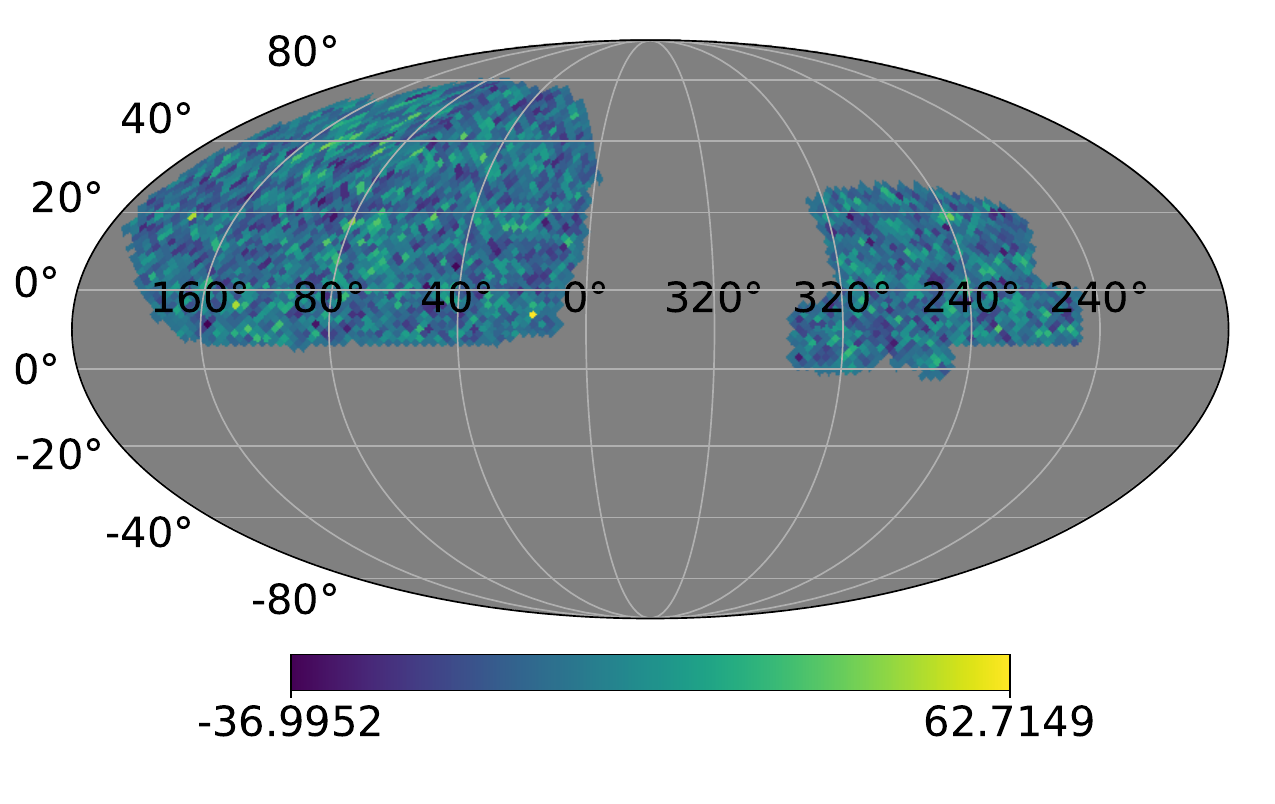}
        \includegraphics[scale=0.35]{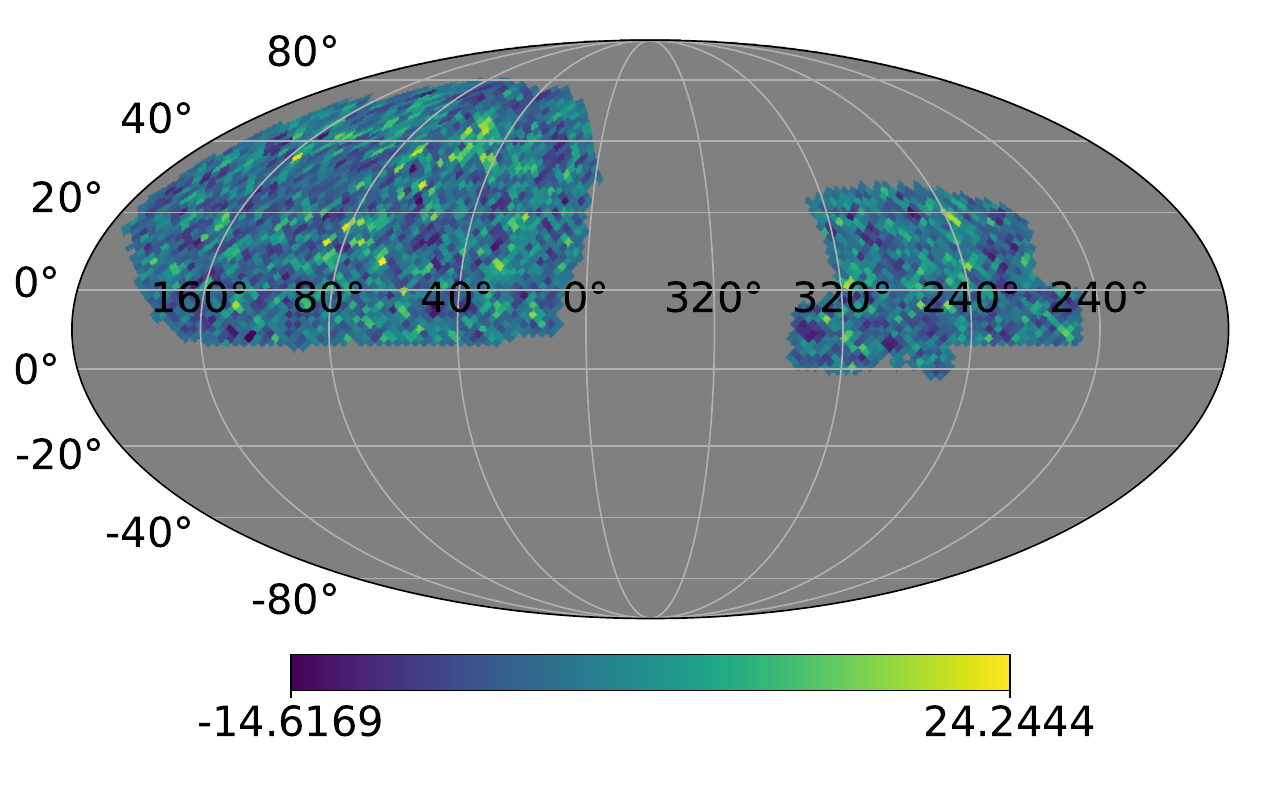}%
    }
    \centerline{%
        \textbf{0.50 < z < 0.60}\hspace{4.8cm}\textbf{0.60 < z < 0.70}%
    }
    \medskip
    \centerline{%
        \includegraphics[scale=0.35]{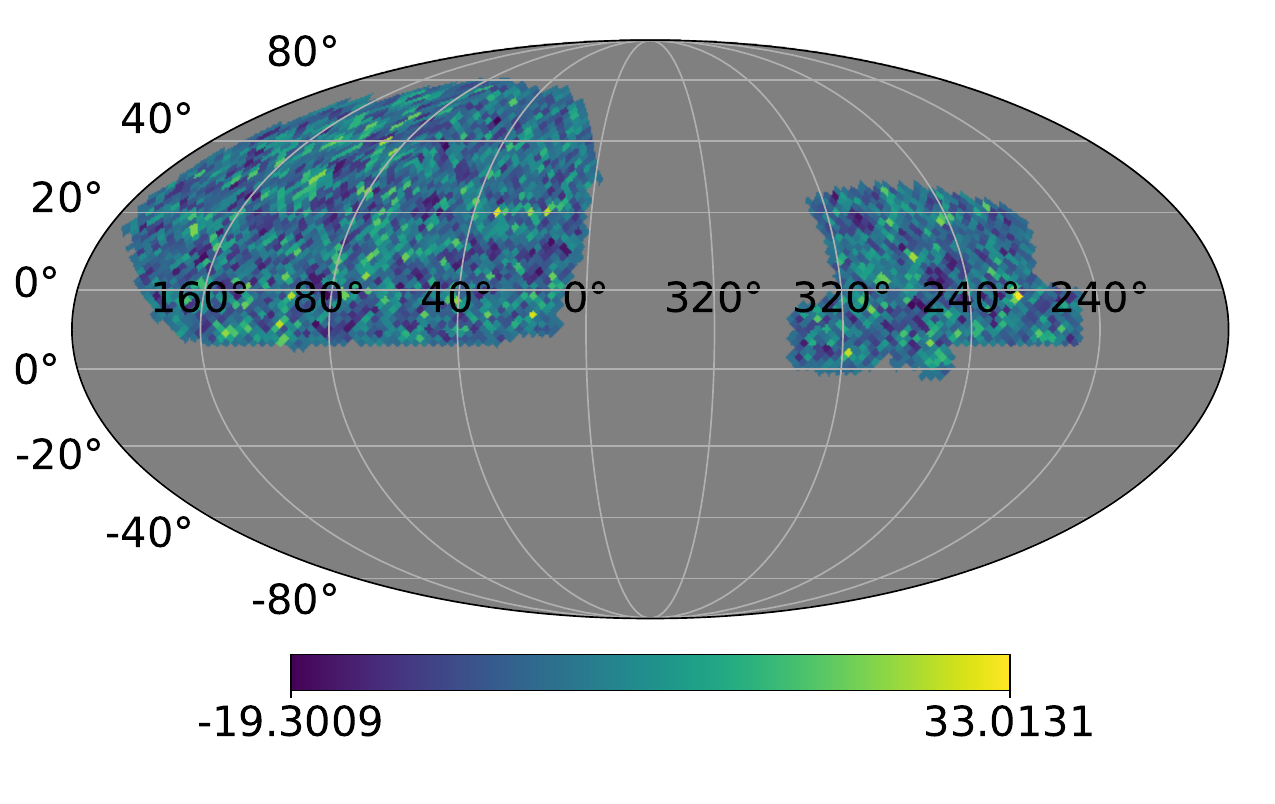}
        \includegraphics[scale=0.35]{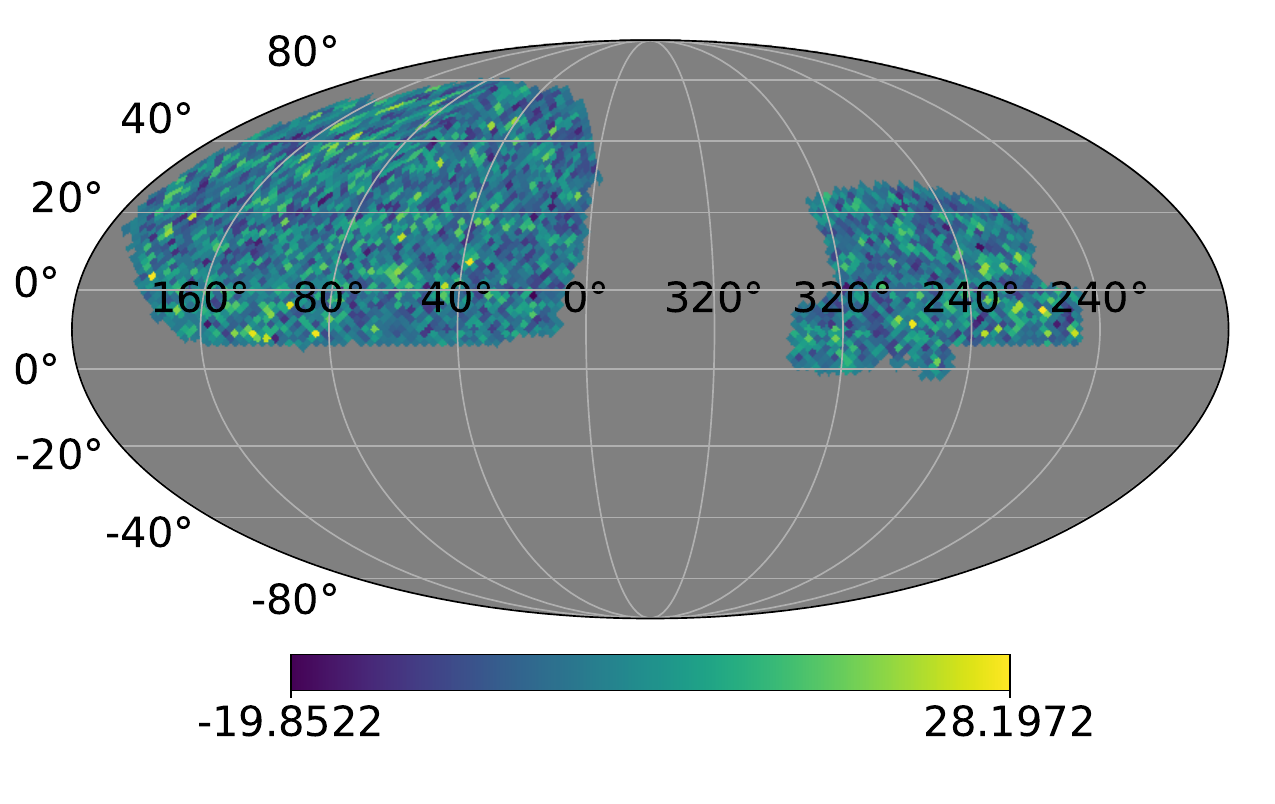}%
    }
    \caption{Sky maps of data minus randoms, normalized to sum to zero, for the \ac{BOSS} data catalog. We show both the entire redshift range as well as three slices in redshift.}
    \label{fig:dr_data}
\end{figure}

\begin{figure}
    \centerline{%
        \textbf{0.43 < z < 0.70}\hspace{4.8cm}\textbf{0.43 < z < 0.50}%
    } 
    \medskip
    \centerline{%
        \includegraphics[scale=0.35]{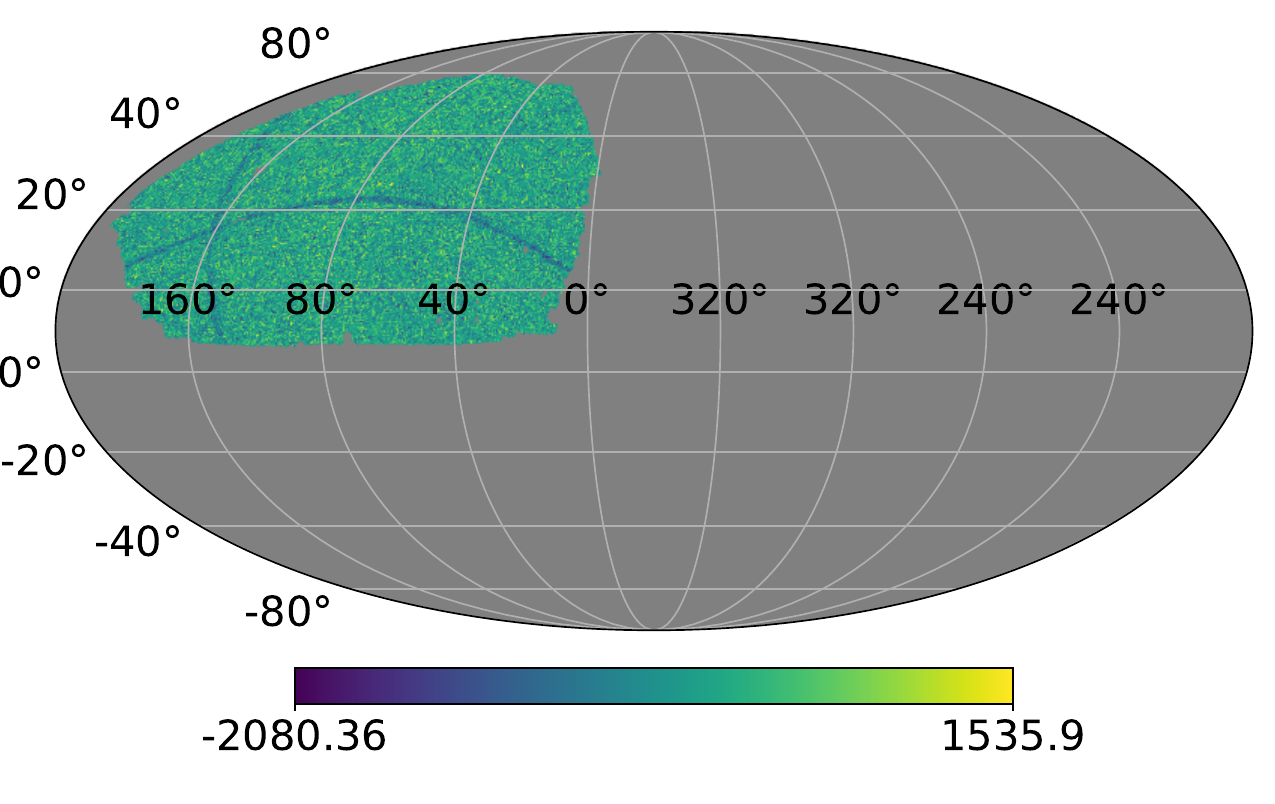}
        \includegraphics[scale=0.35]{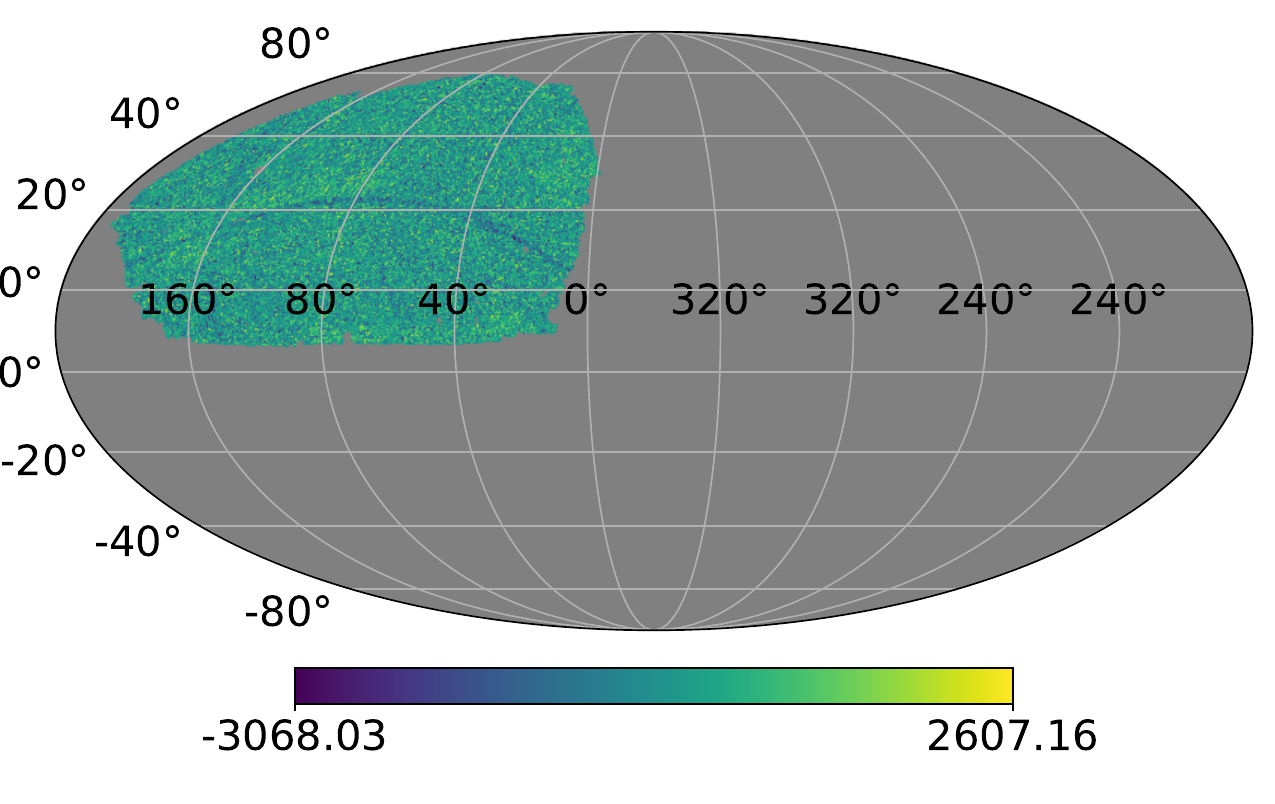}%
    }
    \centerline{\textbf{0.50 < z < 0.60}\hspace{4.8cm}\textbf{0.60 < z < 0.70}}
    \medskip
    \centerline{%
        \includegraphics[scale=0.35]{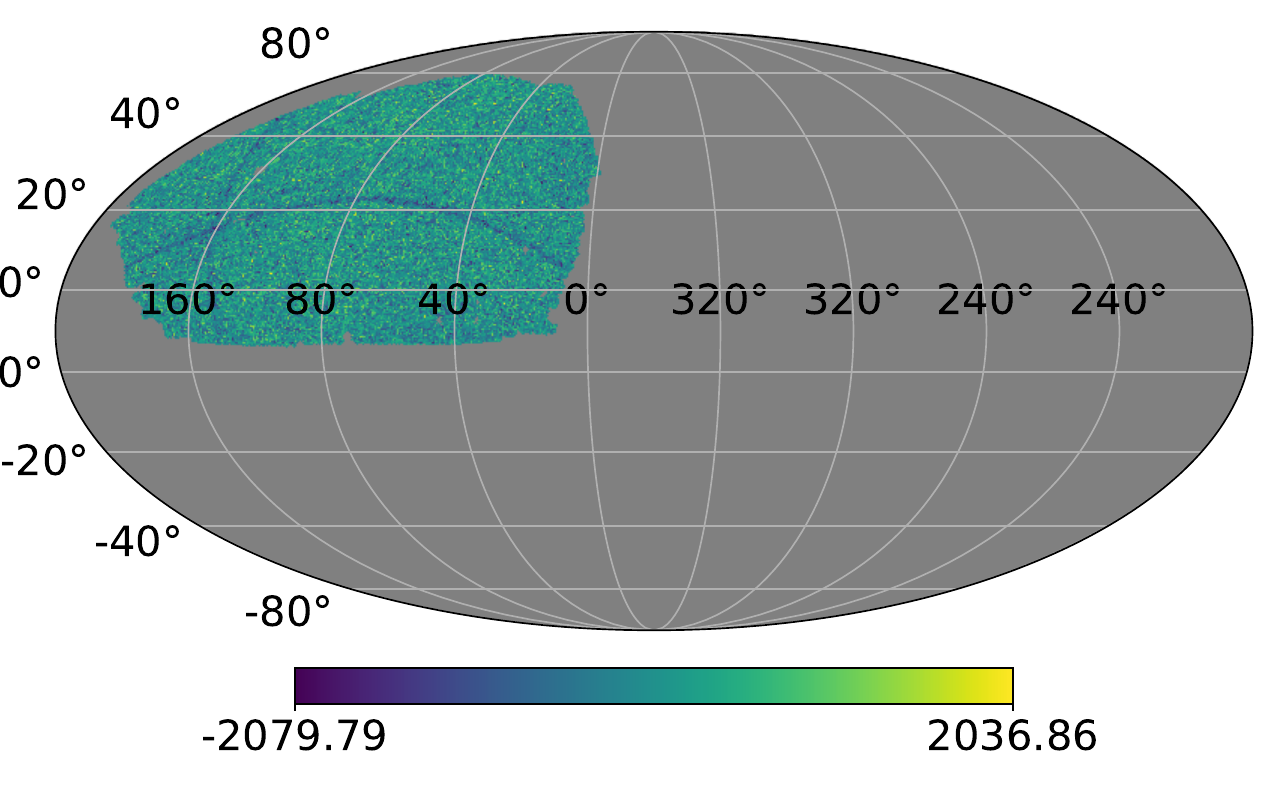}
        \includegraphics[scale=0.35]{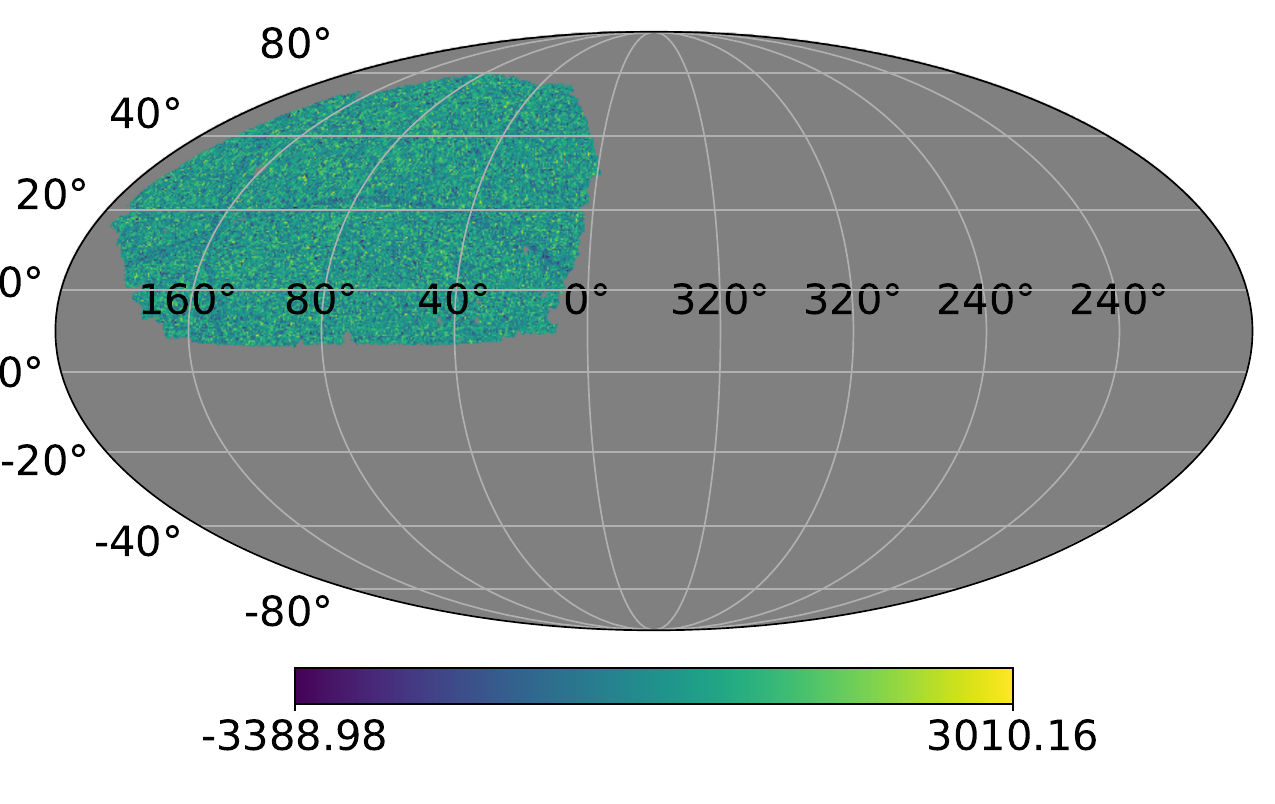}%
    }
    \caption{Sky maps of data minus randoms, normalized to sum to zero, summed over the 2048 Patchy mocks in the \ac{NGC}. Visible striping artifacts are present. We show both the entire redshift range as well as three slices in redshift.}
    \label{fig:dr_mock}
\end{figure}

We searched for angular inhomogeneities in the \ac{BOSS} data and mocks by creating sky maps of the data counts minus the random counts, normalized such that the sum of the map is zero.
In \cref{fig:dr_data}, we show these maps for the data, split into three redshift slices. There are no significant artifacts in these plots, i.\,e.\ the sky distribution of data and randoms matches.

We show the same plot in \cref{fig:dr_mock} for the sum of the 2048 Patchy mocks in the \ac{NGC}.
In contrast to the data, there is a striping artifact apparent in the overall redshift range and in each of the redshift slices. The artifact is caused by a 10\% reduction in mock data counts within these two stripes on the sky.

We create and apply a mask to remove this stripe from the data and mocks.
Testing its impact on the 10-bin results for the first 20 mocks, we find that adding the stripe mask increases $\chi^2$ by 594 for the mocks, on average.
It also increases $\chi^2$ by 670 for the data, and hence increases the detection significance from $2.8\sigma$ to $3.4\sigma$.
We conclude that the striping artifact does not have a significant impact on the parity-odd detection.
Moreover, the striping is only present in \ac{NGC} whereas the parity-odd signal is detected at a similar level in both \ac{NGC} and \ac{SGC}, implying that the striping cannot explain the parity-odd detection.

\section{Condition number and eigenvalues of the analytic covariance}
\label{app:cov_mat_properties}

In an early stage of this paper, we noticed that the analytic covariance matrix $(C_{\mathrm{ana}})_{ab}$ is poorly conditioned.
We wondered whether this could make the statistic $\chi^2 \equiv \hE_a (C_{\mathrm{ana}})^{ab} \hE_b$ susceptible to small effects such as imperfect edge correction, explaining the $\chi^2$ excess.
In this appendix, we show that for the 10-bin and 18-bin cases considered in this paper (see \cref{ssec:hE}) the poor conditioning is unlikely to be an issue.

\begin{figure}
    \hspace{-10pt}
    \includegraphics[scale=0.38]{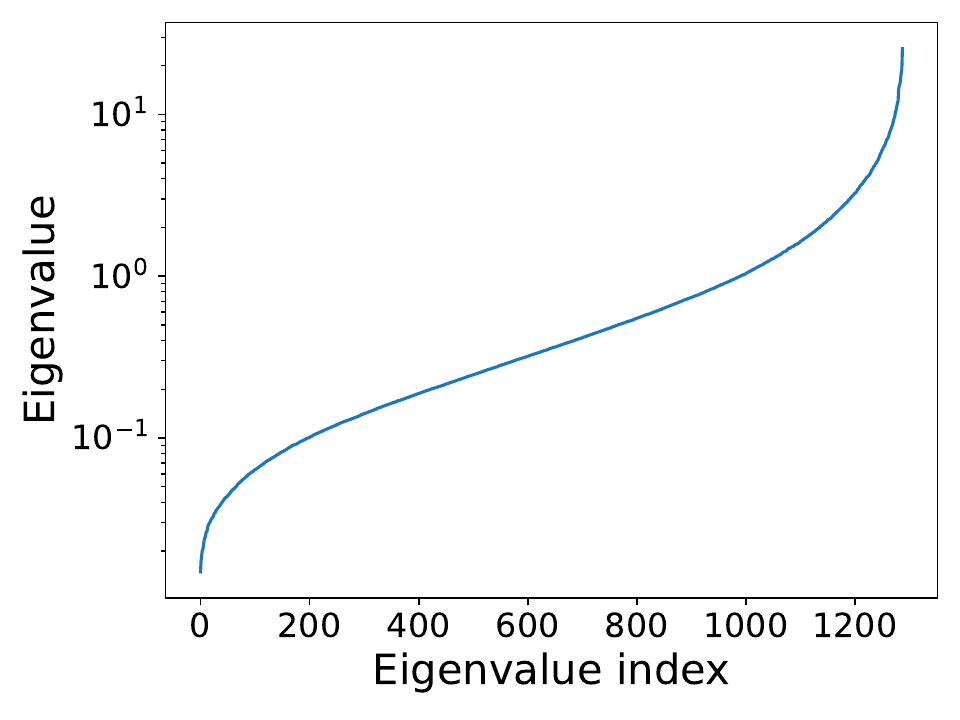}
    \includegraphics[scale=0.38]{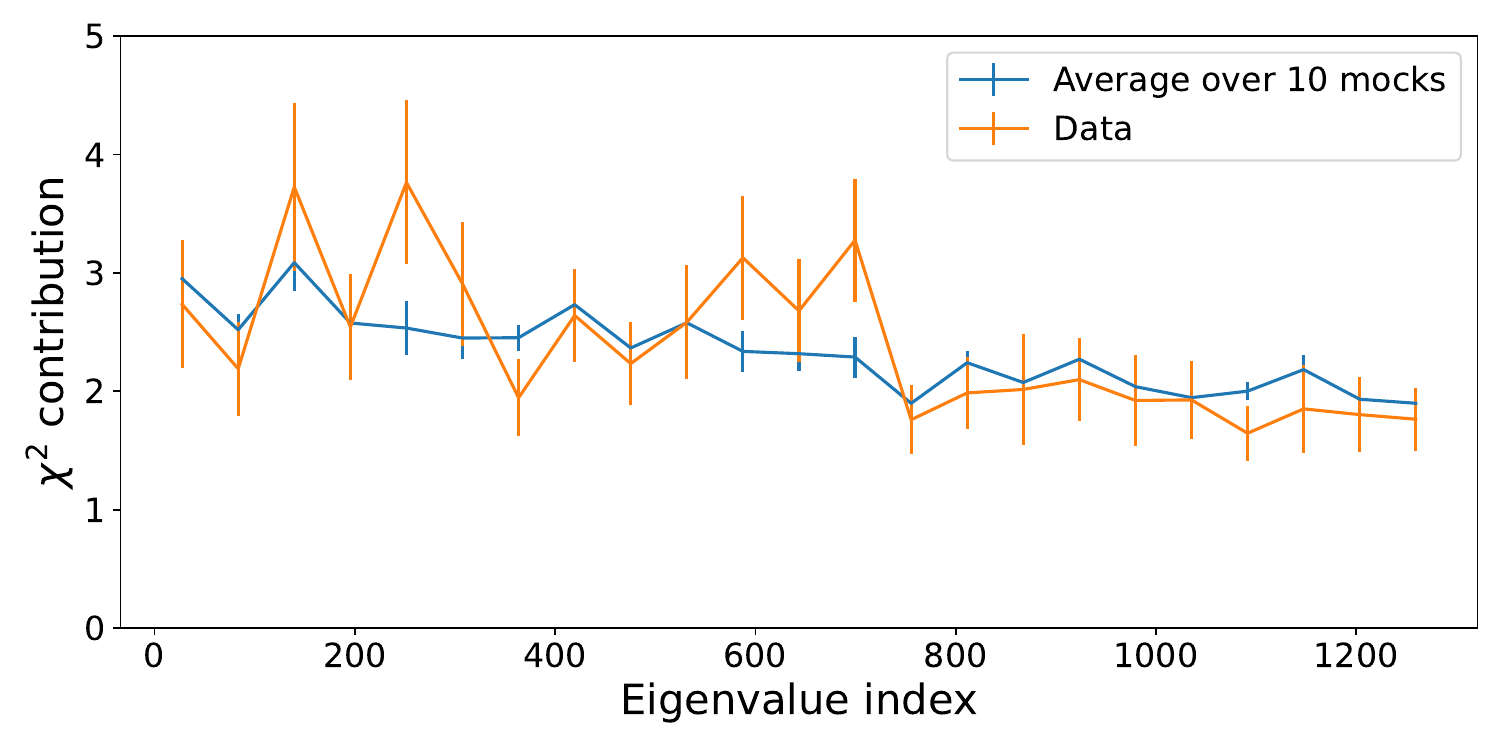}
    \caption{%
        \textit{Left:}
        Distribution of eigenvalues of the correlation matrix for the 10-bin analytic covariance in the \ac{BOSS} \cmassnorth region.
        \textit{Right:}
        Mode-by-mode contribution to the $\chi^2$ for both data (orange) and average over 10 mocks (blue), averaged over blocks of 56 eigenvalues to reduce noise.%
    }
    \label{fig:10_bin_covmat}
\end{figure}

\begin{figure}
    \hspace{-10pt}
    \includegraphics[scale=0.38]{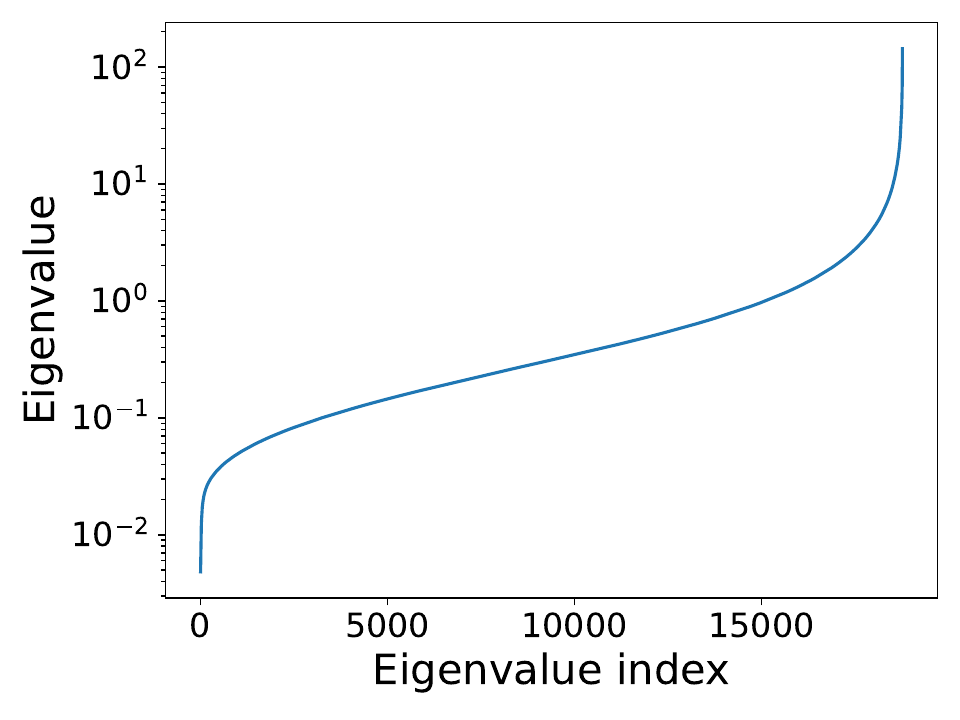}
    \includegraphics[scale=0.38]{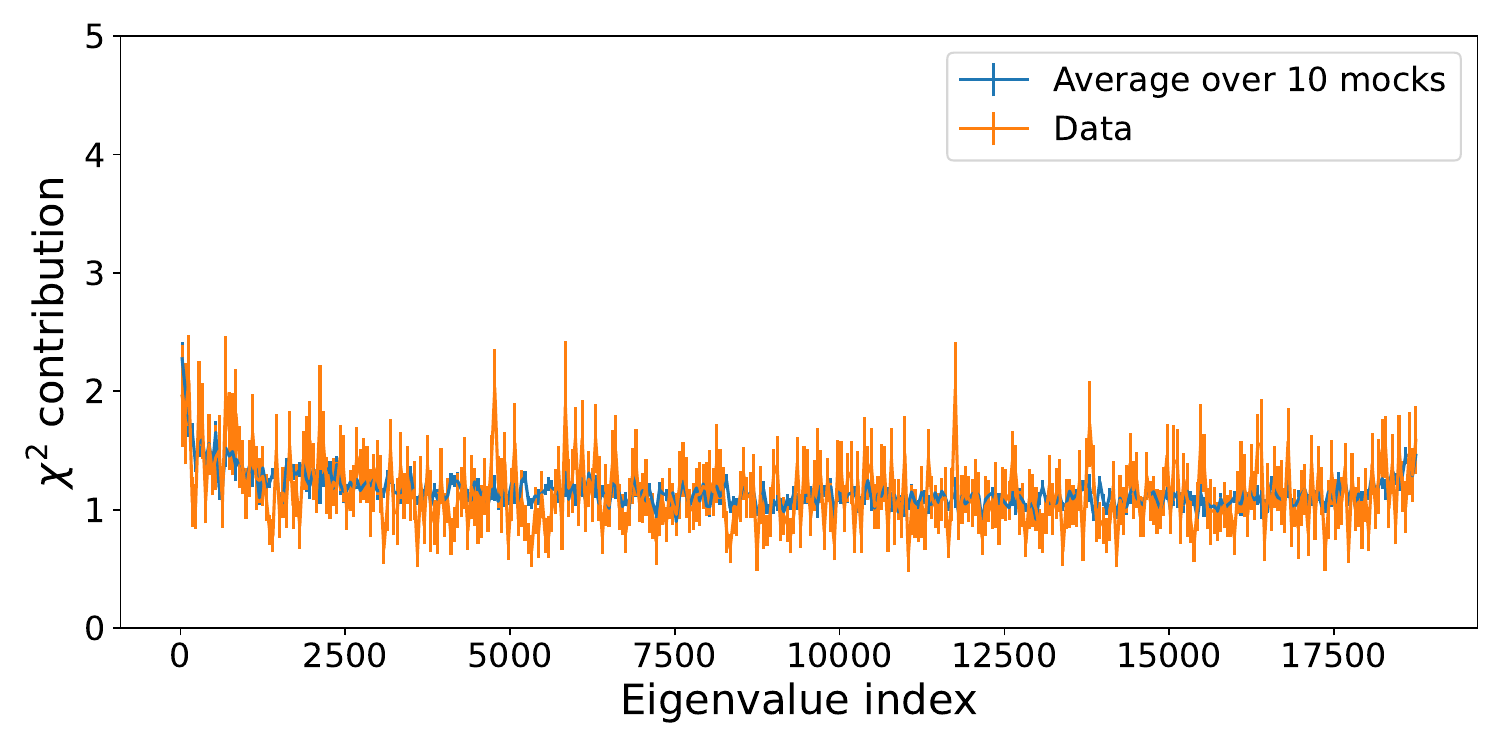}
    \caption{%
        Same as \cref{fig:10_bin_covmat}, but for the 18-bin case. The averaging is done in blocks of 51 eigenvalues.%
    }
    \label{fig:18_bin_covmat}
\end{figure}

In \cref{fig:10_bin_covmat,fig:18_bin_covmat}, we plot the eigenvalues of the correlation matrix for the 10-bin and 18-bin analytic covariance matrices.
We also show the contributions to $\chi^2$ from each eigenmode of the correlation matrix $R_{ab} \equiv (C_{\mathrm{ana}})_{ab}/\sqrt{(C_{\mathrm{ana}})_{aa} (C_{\mathrm{ana}})_{bb}}$.
More precisely, we write $C_{\mathrm{ana}} = D R D$ where $D$ is diagonal, and diagonalize $R = U \Delta U^T$ where $\Delta$ is diagonal and $U$ is unitary.
Then $\chi^2 = \tilde{d} \Delta^{-1} \tilde{d}$ where $\tilde{d} = U^T D^{-1} d$.
Since $\Delta$ is diagonal, we can therefore write $\chi^2$ as a sum and decompose its mode-by-mode contributions.
The mode-by-mode contributions to $\chi^2$ are extremely noisy, so we average them in relatively large blocks for convenience when plotting.
In \cref{fig:10_bin_covmat,fig:18_bin_covmat}, we show the mode-by-mode contributions for both data and mocks, for the 10-bin and 18-bin cases.
While the range in the eigenvalues is quite large (particularly in the 18-bin case), the $\chi^2$ contribution is similar from all eigenmodes, suggesting that the high condition number is not leading to numerical instability in the $\chi^2$ statistic.

\section{Comments on related null tests from \texorpdfstring{\cite{hou_parity,oliver_parity}}{Hou et al., Philcox et al.}}

Our new statistic $\chi^2_{\mathrm{null}}$ is a null test, in the sense that statistical deviation from zero is evidence for systematics or data--mock mismatch (not parity violation).
In \cref{ssec:chi2_null}, we found that this null test failed at \SigNorthNull or \SigSouthNull, for \cmassnorth or \cmasssouth respectively.
The reader may wonder how this high-significance null test failure relates to the null test suites from \cite{hou_parity,oliver_parity}, where no clear null test failure was found.
In this section, we will comment on three null tests from \cite{hou_parity,oliver_parity} that resemble our new statistics $(\chi^2_\times, \chi^2_{\mathrm{null}})$, and explain why these null tests did not detect data--mock mismatch.

\subsection{Matching the 4PCF between data and mocks}

First, we consider the ``realization-dependent amplitude'' from section~VI.C of \cite{oliver_parity}.
In this test, the parity-odd \ac{4PCF} $\hE_a$ is rescaled by an overall constant factor $A$, which is based on the amplitude of the parity-\emph{even} \ac{4PCF}.
We expect that this procedure may ameliorate data--mock mismatch to some extent, since it can compensate for systematic differences in the \ac{4PCF} between data and mocks.
However, the $\chi^2$-statistic is biased by the difference in \acp{8PCF} between data and mocks \cref{eq:intro_2terms}.
Even if the \acp{4PCF} of the data and mocks agree perfectly, there is no guarantee that the \acp{8PCF} are equal.
Therefore, the realization-dependent amplitude test does not conclusively address the issue of data--mock mismatch.

There is a similar test in section~4.2.3 of \cite{hou_parity}, where the covariance is rescaled instead of the \ac{4PCF}.
This reduces the detection significance to $2\sigma$ ($4.6\sigma$) when rescaling the covariance so that the parity-even \ac{4PCF} agrees at $1\sigma$ ($3\sigma$) between mocks and data.
However, this still does not exclude the possibility of an \ac{8PCF} bias driving the residual $\chi^2$ detection.

\subsection{Comparing \texorpdfstring{$\chi^2$}{χ²} in different patches of sky}

In section~6.1.6 of \cite{hou_covariance} and in section~VI.A of \cite{oliver_parity}, there is a null test based on comparing values of $\chi^2$ in different patches of sky.
These patch-based null tests resemble the analysis in our paper, but there is one key difference.
Our null test $\chi^2_{\mathrm{null}}$ is based on comparing the parity-odd four-point function $\hE_a$ in different sky patches, \emph{before} $\hE_a$ is squared to obtain $\chi^2$ (\cref{eq:chi2_null_def}).
On the other hand, the patch-based null tests from \cite{hou_parity,oliver_parity} are based on comparing values of $\chi^2$ (after squaring) in different sky patches.
These are different null tests, and may succeed or fail independently of each other.
We expect that $\chi^2_{\mathrm{null}}$ is more sensitive to covariance mismatch $(C_{\mathrm{data}} \ne C_{\mathrm{mock}})$, whereas the patch-based tests from \cite{hou_parity,oliver_parity} is more sensitive to systematics which break statistical isotropy.

\subsection{The \texorpdfstring{$r_p$}{rp}-statistic (correlating NGC and SGC)}

In section~5.2 of \cite{hou_parity}, there is a statistic $r_p$ which is very closely related to our statistic $\chi^2_\times$.
The $r_p$-statistic is defined for a survey with $N_p = 2$ patches.
In \cite{hou_parity}, the patches are chosen to be the \ac{NGC} and \ac{SGC}.
To define $r_p$, it will be convenient to diagonalize $C_{\mathrm{ana}} = R^T \Lambda R$, and change variables from $\hE_a^\mu$ to the length-$N_{\mathrm{dof}}$ ``data vector'' $d_a^\mu$ defined by:
\begin{equation}
    d_a^\mu = (\Lambda^{-1/2} R)_{ab} \, \hE_b^\mu 
\end{equation}
Then $r_p$ is defined to be the correlation coefficient between the data vectors $d_a^{(1)}$, $d_a^{(2)}$:
\begin{equation}
    r_p \equiv \frac{\sum_a (d_a^{(1)} - \bd^{(1)}) (d_a^{(2)} - \bd^{(2)}) }{
     \sqrt{ \big( \sum_a (d_a^{(1)} - \bd^{(1)})^2 \big)
            \big( \sum_b (d_b^{(2)} - \bd^{(2)})^2 \big) }}
    \hspace{1cm} \mbox{where }
    \bd^{(i)} \equiv \frac{1}{N_{\mathrm{dof}}}
     \sum_{i=1}^{N_{\mathrm{dof}}} d^{(i)}_a
    \label{eq:rp_def}
\end{equation}
In \cite{hou_parity}, the statistic $r_p$ is found to be statistically consistent with zero, but section~5.2 of \cite{hou_parity} argues that this \emph{does not} rule out parity violation.
On the other hand, in this paper, we find that $\chi^2_\times$ is statistically consistent with zero, but we show that this \emph{does} rule out parity violation.
One may wonder how these statements can be consistent, since the statistics $\chi^2_\times$ and $r_p$ are so conceptually similar.

To answer this question, we first note that our $\chi^2_\times$ statistic can be written in ``data vector'' notation as follows:
\begin{equation}
    \chi^2_\times = \sum_a d_a^{(1)} d_a^{(2)}
    \label{eq:chi2_times_datavec}
\end{equation}
Comparing \cref{eq:rp_def,eq:chi2_times_datavec}, we see that the $r_p$-statistic differs from $\chi^2_\times$ in two ways: $r_p$ is defined with $\bd$-subtraction, and $r_p$ is defined with a denominator which ensures $r_p \in [-1, 1]$.

The denominator of \cref{eq:rp_def} is not important (as far as we can tell), but the $\bd$-subtraction in the numerator has an important consequence.
Consider the following toy model of parity violation:
\begin{equation}
    \big\langle d_a^\mu \big\rangle = C
    \hspace{1cm}
    (\text{where $C$ is a constant independent of $a,\mu$})
    \label{eq:C_toy_model}
\end{equation}
In this toy model, the statistic $\chi^2_\times$ is sensitive to the value of $C$, but the statistic $r_p$ is not sensitive to $C$ because of the $\bd$-subtraction in the definition \eqref{eq:rp_def}.

This toy model shows that it is possible for parity violation to make a large contribution to $\chi^2$, but a small (or even zero) contribution to $r_p$.
For this reason, \cite{hou_parity} concluded that a small value of $r_p$ is inconclusive, and does not rule out parity violation.
(The argument in section~5.2 of \cite{hou_parity} is based on a different toy model than \eqref{eq:C_toy_model}, but the principle is the same.)

This issue does not arise for the $\chi^2_\times$ statistic, which is defined without $\bd$-subtraction (\cref{eq:chi2_times_datavec}).
If parity violation makes a statistically significant contribution to $\chi^2$, then it must also make a statistically significant contribution to $\chi^2_\times$.
This follows formally from \cref{eq:chi2_tot_ev,eq:chi2_times_ev}, which show that the statistics $\chi^2$ and $\chi^2_\times$ have the same expectation value $\smash{\bE_a \big( C_{\mathrm{ana}}^{-1} \big)^{ab} \bE_b}$ due to parity violation, plus the statement that $\Var(\chi^2_\times) \lesssim \Var(\chi^2)$.
(See \cref{app:chi2_times} for more discussion of this latter statement.)

\end{document}